\documentclass[10pt]{iopart}
\expandafter\let\csname equation*\endcsname\relax
\expandafter\let\csname endequation*\endcsname\relax
\usepackage{iopams}
\usepackage[utf8x]{inputenc}
\usepackage[UKenglish]{babel}
\usepackage{fancyhdr}
\usepackage{amsmath}
\usepackage{amssymb}
\usepackage{fullpage}
\usepackage{bm}
\usepackage{bbold}
\usepackage{appendix}
\usepackage{graphicx}
\usepackage{color}
\usepackage{tikz}
\usetikzlibrary{calc,decorations.markings}
\newcommand{\ii}{\text{i}}
\newcommand{\p}{\nu}
\newcommand{\be}{\begin{equation}}
\newcommand{\ee}{\end{equation}}
\usepackage{cite}
\numberwithin{equation}{section}
\allowdisplaybreaks[1]
\begin{document}
\title[Subnetwork dynamics in biochemical systems]{Statistical physics approaches to subnetwork dynamics in biochemical systems}
\author{B Bravi$^1$\footnote{Current affiliation: Institute of Theoretical Physics, Ecole Polytechnique F\'ed\'erale de Lausanne (EPFL), CH-1015 Lausanne, Switzerland}
and P Sollich$^1$}
\address{$^1$ Department of Mathematics, King's College London, Strand, London, WC2R 2LS UK}
\ead{barbara.bravi@epfl.ch, peter.sollich@kcl.ac.uk}
\date{}

\begin{abstract}
We apply a Gaussian variational approximation to model reduction in 
large biochemical networks of unary and binary reactions. We focus on a small subset of variables (subnetwork) of interest, e.g.\ because they are accessible experimentally, embedded in a larger network (bulk).
The key goal is to write dynamical equations reduced to the subnetwork but still
retaining the effects of the bulk. As a result, the subnetwork-reduced dynamics contains a memory term and an extrinsic noise term with non-trivial temporal correlations. We first derive expressions for this memory and noise in the linearized (Gaussian) dynamics
and then use a perturbative power expansion to obtain first order 
nonlinear corrections. For the case of vanishing intrinsic noise, our 
description is explicitly shown to be equivalent to projection methods up to quadratic terms, 
but it is applicable also in the presence of stochastic fluctuations in the original dynamics.
An example from the Epidermal Growth Factor Receptor (EGFR) signalling pathway is
provided to probe the increased prediction accuracy and computational efficiency 
of our method.
\end{abstract}

\maketitle
\cleardoublepage
\section{Introduction}

Protein-protein interaction networks, such as the ones building up
signalling pathways, can contain thousands of reacting species and the underlying mechanisms are still largely unclear \cite{kholodenko,seger}, with mathematical descriptions involving large systems of coupled nonlinear differential equations or their stochastic analogues
such as Chemical Langevin Equations \cite{vankampen}. This motivates the search for
approximate descriptions and model reduction techniques (see e.g.~\cite{radulescu,ackermann,okino,apri}). The aim is to simplify the overall analysis, to facilitate
the interpretation of experimental data and ultimately to extract the maximal information from them. Analytical rather than numerical approximations can be particularly useful here by delivering qualitative insights.

The challenge, then, is to find efficient approximations of probability distributions over the temporal
trajectories (paths) of a system with stochastic dynamics. Variational methods achieve this by optimizing an approximating distribution within some chosen family of distributions.
Eyink \cite{eyink} proposed a general scheme to produce approximations by applying a variational principle on a non-equilibrium effective action; see also \cite{sasai,ohkubo} 
 for applications of this variational approach based on Poisson distributions. 

The problem of approximating intractable distributions is also a key one in the area of machine learning techniques \cite{bishop}. Variational approximations are 
there often based on minimizing a Kullback-Leibler divergence \cite{kullback} to obtain the most faithful approximation. Families of approximating distributions include 
the mean field-type (factorized) ansatz \cite{opper_sanguinetti, kupferman_variational} or Gaussian distributions
\cite{opper_gaussian, romanobattistin}, the latter allowing one to retain correlation information beyond mean field. 

The Gaussian assumption can be viewed as a second order moment closure scheme, as first outlined
by Whittle \cite{whittle}. Several precedents exist for its use in modelling biochemical reactions
\cite{bosia,lakatos}; for performance comparisons to other approximations see also \cite{sangui_MA,sangui_MA2,grimaMA}.

The setting we consider in this paper is the following: we consider a network of unary and binary biochemical reactions and focus on a subset of molecular species, the  
``subnetwork''. We view this as embedded in the remainder of the larger network, the ``bulk''. 
The subnetwork is taken as given from the outset. It may be determined by which molecular species can be monitored experimentally, or which are better characterized 
theoretically, or indeed which are more relevant for the biological features of interest.
Our aim is to calculate the time evolution of subnetwork variables without tracking the evolution of the bulk network, 
starting from just some prior information, in the form of the statistics of the bulk initial conditions, and the subnetwork-bulk and bulk-bulk 
interactions. We use a Gaussian Variational Approximation (GVA) of the path distribution
for the stochastic equations describing the entire network, and then add nonlinear corrections. 

We focus on the ``forward" problem of modelling the dynamics of a subnetwork that is part of a known larger network. 
However, this also gives qualitative insights into the situation where only the subnetwork is known and the bulk of the network is thought of 
as providing the significant environmental fluctuations often detected in \emph{in vivo} conditions. 
Deriving models for subsystem dynamics in a \emph{systematic} and \emph{principled} way can then ultimately also help with the ``inference'' 
problem of estimating properties of the environment from observed subnetwork dynamics.

Projection methods are one established approach to model reduction~\cite{chorin,chorin_review,thomasproj,katy}, having been originally introduced in irreversible 
statistical mechanics \cite{zwanzig}. Projection approaches
generically yield a reduced dynamical description for the subnetwork that includes a memory term and extrinsic noise known as ``random force'' (Sec.~\ref{pm}), though finding explicit expressions for these is challenging. Rubin et al.\
\cite{katy,rubin_thesis} achieved this for the setting of unary and binary reactions with mass action kinetics we consider here (with a later extension to Michaelis-Menten kinetics \cite{rubinMM}) in the limit where the dynamics of the entire network is deterministic. These results will provide a useful point of comparison for our approach.

Our method, like that of \cite{katy}, is applicable to \emph{arbitrarily chosen} subnetworks,
while other coarse-graining procedures specifically  tailor the choice of subnetwork, e.g.\ to merge or eliminate fast reactions \cite{sinitsyn}. However, instead of employing projection operators, we derive the reduced subnetwork dynamics
by integrating out or \emph{marginalizing} the bulk degrees of freedom (d.o.f.) from a path integral representation of the full dynamics. 
This marginalization approach has been used also in recent works 
studying fixed or fluctuating random environmental conditions \cite{koeppl_nature,koeppl_plos}. In our treatment, the environment (bulk) is itself a network, with its own structure and dynamics. This makes the approach more flexible, allowing one for example to consider different splits of a given network into subnetwork and bulk as might be relevant for different datasets.

We start in Sec.~\ref{set_up} by introducing the 
model for the network dynamics and the structure of its subnetwork-reduced version expected from 
projection methods. In Sec.~\ref{gva} we outline a variational derivation of a Gaussian approximation over 
paths \cite{opper_gaussian}. We then marginalize to obtain the approximate subnetwork dynamics, including the memory and extrinsic noise due to the interaction with the bulk (Sec.~\ref{linmem}). 
To go beyond the effectively \emph{linearized} dynamics implied by the Gaussian approximation, we write the subnetwork dynamics in terms of an ``effective'' (i.e.\ with the bulk integrated out)
action and perform a perturbative expansion from which the \emph{nonlinear} memory and noise can be read off (Sec.~\ref{nonlinmem0}).
The numerical solution of the resulting subnetwork equations is discussed in Sec.~\ref{num_imp} and is more computationally efficient for large bulk networks than for the projected subnetwork 
equations of~\cite{katy}. Nevertheless, as we show in Sec.~\ref{comp_pm}, the two methods are equivalent in the deterministic limit, to the quadratic order that both treat systematically.

In Sec.~\ref{toy} we use a toy model that is instructive in its simplicity, in order to illustrate
our method and compare its accuracy to existing approximation schemes.
Finally, in Sec.~\ref{egfr}, we apply our model reduction to the protein-protein interaction network around  
Epidermal Growth Factor Receptor (EGFR), where it yields increased prediction accuracy compared to projection methods \cite{katy}.

\section{Set-up}
\label{set_up}
We consider a reaction network of $N$ molecular species such as proteins and protein complexes, with concentrations $\bm{x}=(x_1,\ldots,x_N)$. 
These are assumed to evolve in time according to a set of Chemical Langevin Equations (CLE) \cite{vankampen}
\begin{equation}
\label{eq:steq}
\frac{\partial{x_i(t)}}{\partial{t}}=\Phi_i(\bm{x}(t)) +\xi_i(t) \qquad \quad i=1,\ldots,N
\end{equation}
Here $\xi_i(t)$ is an \emph{intrinsic} noise term from the stochasticity of reaction events and
\begin{equation}\label{phi}
\begin{split}
&\Phi_i(\bm{x}(t))= \sum_{j,l,j \neq l}\left(k^{-}_{l,ij}x_i(t)- k^{+}_{ij,l}x_i(t) x_j(t)\right) + 
\frac{1}{2}\sum_{j,l,j\neq l}\left(k^{+}_{jl,i}x_j(t)x_l(t)-k^{-}_{i,jl}x_i(t)\right)\\
&{}+\sum_{j}\big(\lambda_{ji}x_j(t)-\lambda_{ij}x_i(t)\big)+
\sum_{l}\left(2k^{-}_{l,ii}x_l(t)-k^{+}_{ii,l}x_i(t) x_i(t)\right)+\sum_{j}\left(\frac{1}{2}k^{+}_{jj,i}x_j(t)
x_j(t)-k^{-}_{i,jj}x_i(t)\right)
\end{split}
\end{equation}
These dynamical equations capture the formation of complex $l$ from proteins $i$ and $j$ (with rate constant $k^{+}_{ij,l}$) and its dissociation ($k^{-}_{l,ij}$) and similar reactions where $i$ itself is 
the complex (rate constants $k^{+}_{jl,i}$ and $k^{-}_{i,jl}$), as well as the unary change of species $i$ into species $j$ ($\lambda_{ij}$), e.g.\ by phosphorylation. 
We allow for the formation of both heterodimers and, with the last two terms in \eqref{phi}, homodimers. $\Phi_i(\bm{x}(t))$ encodes the deterministic part of Langevin dynamics \eqref{eq:steq} and can be written as 
$\Phi_i(\bm{x}(t)) = [\bm{S}\bm{f}(\bm{x}(t))]_i$ where $\bm{S}$ denotes the stoichiometric matrix with entries $0,\pm1,\pm2$ and $\bm{f}$ the vector of reaction fluxes.

The CLE-description is equivalent to the Chemical Fokker Planck equation that is obtained by truncating the Kramers-Moyal expansion of the 
underlying Master Equation after the first order in the inverse of the reaction volume $V$ \cite{gardiner}.
By construction of the CLE \cite{vankampen}, the noise is Gaussian distributed
with zero mean. Its covariance matrix
\begin{equation}
\langle \xi_i(t)\xi_j(t') \rangle=\bm{\Sigma}_{ij}(\bm{x})\delta(t-t')
\end{equation}
shows it is uncorrelated in time (white noise) but multiplicative (with $\bm{x}$-dependent correlations, interpreted in the It\^o convention \cite{vankampenito}).
One can show \cite{LNA} that
\begin{equation}
\label{fluct}
 \bm{\Sigma}(\bm{x}(t))= \epsilon\,\bm{S}\,\text{diag}(\bm{f}(\bm{x}(t)))\,\bm{S}^{T}
\end{equation}
The inverse reaction volume $\epsilon=1/V$ sets the amplitude of fluctuations, 
with the noise contribution diminishing as the reaction volume $V$ (typically the cell volume) and hence the number of molecules increases.
Following the approach in \cite{rabello}, we can study the time-dependent statistics of 
protein concentrations by appeal to the Martin--Siggia--Rose--Janssen--De Dominicis (MSRJD) 
formalism \cite{martin,janssen,dedominicis} (see also \cite{kamenev,PathMethods} for a systematic explanation of this formalism).
We discretize time in small steps $\Delta$ and we denote by $P(\bm{x})$ the normalized probability distribution of protein concentration paths, i.e.\ the set of 
temporal trajectories $\bm{x}=\lbrace x_i(t) \rbrace$
satisfying \eqref{eq:steq}; it can be deduced from $P(\bm{\xi})$, the Gaussian distribution of the stochastic fluctuations $\bm{\xi}=\lbrace \xi_i(t) \rbrace$ by
\be
\label{eq:pathprob}
P(\bm{x}) = P(\bm{\xi}(\bm{x}))J(\bm{x})
\ee
where $J(\bm{x}) = |\partial \bm{\xi}/\partial \bm{x}| $ is the Jacobian of this transformation and $J(\bm{x})\equiv 1$ with the It\^o convention \cite{vankampenito,PathMethods}.
The normalization property of \eqref{eq:pathprob} can be rewritten as
\be
\begin{split}
1 \equiv Z &=\int \prod_{it}dx_i(t)d\xi_i(t)\prod_{t} P(\lbrace \xi_i\rbrace(t))\prod_{it} \delta\big( x_i(t+\Delta)-x_i(t)-\Delta\Phi_i(\bm{x}(t))
-\Delta\overline{\xi}_i(t)\big)=\\
&=\int \prod_{it}\frac{dx_i(t)d\hat{x}_i(t)}{2\pi}d\xi_i(t)\prod_{t} P(\lbrace\xi_i\rbrace(t))\prod_{it} e^{\text{i}\hat{x}_i(t)\left[ x_i(t+\Delta)-x_i(t)-\Delta\Phi_i(\bm{x}(t))-\Delta\overline{\xi}_i(t)\right]}
\end{split}
\ee
where we have first expressed $\bm{\xi}$ in terms of $\bm{x}$ by using $\delta-$functions to 
impose~\eqref{eq:steq} and we have then represented these as Fourier integrals over a set of auxiliary variables $\hat{\bm{x}}$.
The noise appears via its average over the time interval $\Delta$, $ \overline{\xi}_i(t)=\frac{1}{\Delta}\int_{t}^{t+\Delta} \xi_i(t')dt'$,
so that $\langle \overline{\xi}_i(t)\overline{\xi}_j(t') \rangle=\Delta^{-1}\bm{\Sigma}_{ij}(\bm{x})\delta(t-t')$. 
The integration over this Gaussian noise can be performed by means of the Hubbard-Stratonovich identity
\begin{equation}
\int D\bm{\xi} P (\bm{\xi}) e^{\pm\text{i}\Delta \hat{\bm{x}} \cdot \overline{\bm{\xi}}} =
e^{-\frac{\Delta}{2}\bm{\hat{x}^{T}}\bm{\Sigma}\bm{\hat{x}}}
\end{equation}
where the measure should be understood as $D\bm{\xi}= \prod_{i}d\xi_i(t)$. 
This brings $Z$ into the final form 
\begin{equation}
\label{norm}
Z = \int D\bm{x}\,D\bm{\hat{x}}\,e^{\mathcal{H}}
\end{equation}
The notation $D\bm{x}\,D\bm{\hat{x}}$ is a shorthand for  $\prod_{it}dx_i(t)d\hat{x}_i(t)/(2\pi)$ and represents a path integral, 
ranging across all possible trajectories for $\bm{x}$ and $\hat{\bm{x}}$. Each trajectory 
$\lbrace x_i(t), \hat{x}_i(t) \rbrace$ is weighted by a factor $\exp(\mathcal{H})$ with the action $\mathcal{H}$ 
\be
\label{action}
\mathcal{H}= \sum_{it}\text{i}\hat{x}_i(t)\left[x_i(t+\Delta)-x_i(t)-\Delta\Phi_i(\bm{x}(t))\right]
+\frac{\Delta}{2}\sum_{ij t} \text{i} \hat{x}_i(t)\Sigma_{ij}(\bm{x}(t)) \text{i} \hat{x}_j(t) 
\end{equation}
Note that $Z \equiv 1$ by construction, so $\exp(\mathcal{H})$ defines a distribution 
(or more precisely a complex-valued measure) over trajectories, including the ones of the auxiliary variable $\hat{\bm{x}}$.
This distribution will be the starting point for our model reduction strategy, which aims to derive 
an accurate description of the dynamics of a chosen \emph{subnetwork} that accounts also for the dynamical effects of the embedding into its environment, the \emph{bulk}.

\subsection{Projection Methods} 
\label{pm}
As explained in the introduction, model reduction to subnetwork dynamics can also be achieved using projection methods~\cite{zwanzig,ritort}. Here one defines an operator 
that projects the dynamics onto observables describing a set of variables of interest. These relevant variables can be slow degrees of freedom, or in the context of optimal 
prediction methods \cite{chorin,chorin_memory,chorin_review} those variables that are better resolved by measurement. 
The contributions of the remaining, less relevant variables, are then isolated into a \emph{memory} kernel and 
a \emph{random force}.

Briefly, if $x_i(t)$ is a subnetwork concentration, with $i=1,\ldots,N^{\rm s}$ where $N^{\rm s}$ is the size of the subnetwork, the projection approach yields
a dynamical equation of the general form (see e.g.~\cite{ritort})
\begin{equation}
\label{eq_proj}
\frac{\partial x_i(t)}{\partial t} = \sum_{j=1}^{N^{\rm s}} x_j(t)\underbrace{\Omega_{ji}}_{\text{Rate matrix}}+
\sum_{j=1}^{N^{\rm s}}\int_0^tdt'x_j(t')\underbrace{M_{ji}(t-t')}_{\text{Memory function}} + \underbrace{r_i(t)}_{\text{Random force }}
\end{equation} 
The rate matrix $\Omega_{ji}$ in the first term on the r.h.s.\ describes processes that are local in time and involving solely subnetwork observables. 
The second term, the memory, 
captures the effects that past subnetwork states at time $t'$ have on the dynamics at time $t$. 
The strength of these effects is given by the memory function, which depends on the time lag $t-t'$ as well as the species involved. The intuition behind the memory 
is that the subnetwork state at $t'$ affects the bulk dynamics, which in turn feeds back into the subnetwork dynamics at $t$.
The final term, the random force, represents the effect of unknown bulk initial conditions, hence a form of extrinsic noise, and is distinct from any stochasticity of the
original dynamics (intrinsic noise).
Note that the projected dynamics formally describes the \emph{conditional average} of the subnetwork concentrations, where the conditioning is on the initial values. Fluctuations due to intrinsic noise are therefore not represented explicitly.
The presence of memory and the random force -- which has non-trivial temporal correlations -- make the projected subnetwork equations non-Markovian, as expected on general grounds given that model reduction to a subnetwork is a form of coarse-graining.
Including these non-Markovian effects
means less information is lost, giving
a more accurate prediction of time courses than any Markovian approximation~\cite{chorin_memory, katy}.
We note finally that by choosing more general observables, e.g.\ products of concentration fluctuations, a nonlinear analogue of \eqref{eq_proj} can be constructed and it is this form that we compare our 
framework to in Sec.~\ref{comp_pm} and~\ref{projvsgva}.

\section{Gaussian Variational Approximation}
\label{gva}
Returning to our probabilistic setup, our model reduction task is in principle to take the full distribution over network trajectories in \eqref{norm} and marginalize 
it, i.e.\ integrate over all possible trajectories of the bulk variables. This is clearly intractable, however, while it would be straightforward if the distribution were Gaussian 
(see Sec.~\ref{linmem} below).
We therefore now construct a Variational Gaussian Approximation (GVA) to the network trajectory distribution. Note that in \eqref{norm} we implicitly assumed the initial 
concentrations $\bm{x}(0)$ to be fixed; this can be relaxed by multiplying by the relevant distribution $P_0(\bm{x}(0))$
\begin{equation}
P(\bm{x},\bm{\hat{x}})=e^{\mathcal{H}(\bm{x},\bm{\hat{x}})}P_0(\bm{x}(0))
\end{equation} 
We approximate $P$ by a variational distribution $Q(\bm{x},\bm{\hat{x}})$ of Gaussian form
\begin{equation}
P(\bm{x},\bm{\hat{x}})
\approx 
Q(\bm{x},\bm{\hat{x}})
= \mathcal{N}(\bm{x},\bm{\hat{x}}| \bm{\mu}_{\text{gen}}, \bm{C}_{\text{gen}})
\end{equation}
This Gaussian is completely determined
by the vector of mean values $\bm{\mu}_{\text{gen}}$ and the covariance matrix $\bm{C}_{\text{gen}}$,
where the ``\text{gen}'' subscript indicates ``generalized'' moments including the auxiliary variables $\hat{\bm{x}}(t)$
\begin{equation}
\bm{\mu}_{\text{gen}}(t)=
\left(
\begin{array}{c}
 \langle \bm{x}(t) \rangle_{Q}\\
-\ii\langle \bm{\hat{x}}(t) \rangle_{Q}\\
\end{array}\right)=\left(
\begin{array}{c}
\bm{\mu}(t)\\
-\ii\bm{\hat{\mu}}(t)\\
\end{array}\right)
\end{equation}
\begin{equation}
 \bm{C}_{\rm gen}(t,t')=
 \left\langle
 \left(
\begin{array}{c}
 \delta\bm{x}(t)\\
-\ii \delta\bm{\hat{x}}(t)\\
\end{array}\right)\big(\delta\bm{x}(t') \quad-\text{i}\delta\hat{\bm{x}}(t')\big) \right\rangle_{Q}=
 \left(
\begin{array}{cc}
\bm{C}(t,t')&\bm{R}(t,t')\\
\bm{R}(t',t)^{T}&\bm{B}(t,t')\\
\end{array}\right) 
\end{equation}
with $\delta\bm{x}(t)=\bm{x}(t)-\bm{\mu}(t)$ and $\delta\bm{\hat{x}}(t)=\bm{\hat{x}}(t)-\bm{\hat{\mu}}(t)$. For the sake of brevity,
we will  write $\langle \rangle_{Q}$ as simply $\langle \rangle$ and the notation used has the following meaning
\begin{equation}
\begin{split}
\bm{C}(t,t')=&\langle \delta\bm{x}(t)\delta\bm{x}^{T}(t') \rangle=
\langle \bm{x}(t)\bm{x}^{T}(t') \rangle-\bm{\mu}(t)\bm{\mu}^{T}(t') \\
\bm{R}(t,t')=&-\ii\langle \delta\bm{x}(t)\delta\bm{\hat{x}}^{T}(t') \rangle=
-\ii\langle\bm{x}(t)\bm{\hat{x}}^{T}(t') \rangle+\ii\bm{\mu}(t)\bm{\hat{\mu}}^{T}(t') \\
\bm{B}(t,t')=&-\langle \delta\bm{\hat{x}}(t)\delta\bm{\hat{x}}^{T}(t') \rangle=-\langle \bm{\hat{x}}(t)\bm{\hat{x}}^{T}(t') \rangle+\bm{\hat{\mu}}(t)\bm{\hat{\mu}}^{T}(t')
\end{split}
\end{equation}
From general results for MSRJD path integrals \cite{kamenev,coolen} it follows that $R_{ij}(t, t')$ has the meaning of
a local response of $x_i(t)$ to a perturbing field $-\text{i}\hat{x}_j(t')$ applied to some other species $j$ at an earlier time $t'$; 
it is therefore non-vanishing only for $t>t'$. Also, $B_{ij}(t,t')$ is expected to vanish for all 
times $t$ and $t'$, as is $\hat\mu_i(t)$: we refer to \cite{fischer} for a derivation of this property from the fact that $Z \equiv 1$.

The optimal Gaussian approximation can be found from the stationary points of the Kullback-Leibler (KL) divergence between $Q(\bm{x},\bm{\hat{x}})$ and
$P(\bm{x},\bm{\hat{x}})$ \cite{kullback}, i.e.\ from the conditions
\begin{equation}
\label{variational1}
\frac{\partial \text{KL}}{\partial{\bm{\mu}_{\text{gen}}}}=0 \qquad \frac{\partial \text{KL}}{\partial{\bm{C}_{\rm gen}}}=0
\end{equation}
These conditions give differential equations for the time-dependent parameters
of the optimal Gaussian $Q(\bm{x},\bm{\hat{x}})$, i.e.\ its first two moments.  
The calculation (see \ref{GaussVar}) is essentially a reformulation in the MSRJD formalism of the variational derivation provided by 
Archambeau et al.~\cite{opper_gaussian}, except that by focussing directly on the stationarity conditions of the KL-divergence we can deal with the concentration dependence
of the diffusion matrix $\bm{\Sigma}(\bm{x}(t))$. We note that, as $P(\bm{x},\bm{\hat{x}})$ and $Q(\bm{x},\bm{\hat{x}})$ are complex-valued, stationarity of the KL-divergence is only
a \emph{heuristic} strategy for finding the optimal approximation, but one that is validated
\emph{a posteriori} by comparison with the results of \cite{opper_gaussian}.

The time evolution of the Gaussian Variational Approximation means is given by
\begin{equation}
\label{meangau0}
\frac{d\mu_i(t)}{ dt} = \langle\Phi_i(\bm{x}(t))\rangle
\end{equation}
so that $\bm{\mu}(t)$ evolves with the average drift of the initial equations of motion \eqref{eq:steq}. The average is over the instantaneous Gaussian distribution with mean $\bm{\mu}(t)$ and covariance $\bm{C}(t,t)$, so that nonlinear terms in $\Phi_i(\bm{x})$ retain information about the fluctuations in the dynamics, 
unlike the Linear Noise Approximation (see below).

In the unapproximated dynamics, the auxiliary variables $\hat{x}_i(t)$ have vanishing moments of any order, and (within the It\^o discretization \cite{PathMethods, vankampenito}) equal-time responses. It turns out that the GVA preserves these properties, giving $\bm{\hat{\mu}}(t)=\bm{B}(t,t') = 0$ and $\bm{R}(t,t)=0$.

It remains to discuss the predictions of the GVA for the correlations $\bm{C}(t,t')$ and the responses $\bm{R}(t,t')$ for $t>t'$. One finds that these are what would be predicted by 
linearization of the original Langevin dynamics \eqref{eq:steq} around the time-dependent means $\bm{\mu}(t)$
\begin{equation}
\label{LanVec3}
\frac{d(\bm{x}(t)-\bm{\mu}(t))}{ dt} =\bm{K}(t)(\bm{x}(t)-\bm{\mu}(t)) + \bm{\xi}(t)
\end{equation}
with an effective rate matrix $\bm{K}(t)$ defined as
\begin{equation}
\label{effdrift}
 K_{ij}(t)=\bigg\langle\frac{\partial \Phi_i(\bm{x}(t))}{\partial x_j(t) }\bigg\rangle
\end{equation}
and white Gaussian additive noise $\bm{\xi}(t)$ with
\be
\label{effdiffusion}
\langle \bm{\xi}(t)\bm{\xi}^{T}(t')\rangle= \langle\bm{\Sigma}(\bm{x}(t))\rangle \delta(t-t')
\ee
Thus, the noise covariance of the GVA is the true one averaged over the approximating Gaussian distribution. 
The result \eqref{effdrift} coincides with that of Archambeau et al.~\cite{opper_gaussian}, while \eqref{effdiffusion} allows one to generalize their approach to the multiplicative noise case in which
the GVA cannot match the true noise covariance instantaneously. The GVA responses and correlations are the ones that follow straightforwardly from \eqref{LanVec3} (see \ref{GaussVar}).

We stress that the GVA differs from the Linear Noise Approximation (LNA), which is obtained within the van Kampen system size expansion of the Master Equation \cite{vankampen}. The LNA also produces a Gaussian approximation, but linearizes around the macroscopic (deterministic) trajectory (see \cite{LNA}) given by
\begin{equation}
\label{meangaulna} 
\frac{d\mu_i(t)}{ dt} = \Phi_i(\bm{\mu}(t))
\end{equation}
The evolution of the means is therefore completely decoupled from any fluctuation effects. In the GVA, on the other hand,
the mean dynamics \eqref{meangau0} is affected by the equal-time fluctuations $\bm{C}(t,t)$, as are the effective rate matrix \eqref{effdrift} and the noise covariance
\eqref{effdiffusion}. GVA and LNA only become equivalent when these fluctuations vanish, i.e.\ in the limit of large reaction volumes.
\section{Subnetwork dynamics within GVA}
\label{linmem} 
We can now exploit the GVA results to perform the marginalization over the bulk dynamics and thus obtain a reduced description of the subnetwork dynamics that still accounts implicitly for the presence of the bulk. It is here that we first go beyond existing work.

We separate the concentrations and auxiliary variables ($x_i$ and $\hat{x}_i$) into subnetwork and bulk. It is easiest to assume they are 
numbered so that $i=1,\ldots,N^{\rm s}$ labels subnetwork species while the remaining $N^{\rm b}=N-N^{\rm s}$ species $i=N^{\rm s}+1,\ldots,N$ are in the bulk. This gives a block structure for $\bm{C}$ and $\bm{R}$
\[\bm{C}=\begin{pmatrix}
\bm{C}^{\rm ss}   & \bm{C}^{\rm sb} \\
\bm{C}^{\rm bs}   & \bm{C}^{\rm bb} \\
\end{pmatrix}\] 
\[\bm{R}=\begin{pmatrix}
\bm{R}^{\rm ss}   & \bm{R}^{\rm sb} \\
\bm{R}^{\rm bs}   & \bm{R}^{\rm bb} \\
\end{pmatrix}\]
where the superscripts (s or b) refer to subnetwork or bulk respectively. To retain bulk effects on the subnetwork dynamics the off-diagonal blocks (sb and bs) are clearly essential. 
Indeed, if one constructs a restricted GVA where these blocks are constrained to vanish, corresponding to a factorized Gaussian approximation, then no memory 
terms appear in the reduced subnetwork dynamics.

Bulk marginalization is now conceptually straightforward: marginals of a joint Gaussian are still Gaussian, with the appropriate mean and (in our case, generalized) covariance. 
The marginal distribution over subnetwork trajectories $\{\bm{x}^{\rm s}(t),\hat{\bm{x}}^{\rm s}(t)\}$ is therefore Gaussian with second moments 
$\bm{C}^{\rm ss}(t,t')$, $\bm{R}^{\rm ss}(t,t')$ and $\bm{B}^{\rm ss}(t,t')\equiv 0$. To write down the distribution itself, one just has to find the 
inverse of this large (generalized) covariance matrix, as this is what features in the exponent of the Gaussian. Finally one has to read off \emph{what dynamical equations this Gaussian distribution over subnetwork trajectories encodes}.

An equivalent and technically simpler route is to write down the dynamics \eqref{LanVec3} implied by the GVA for the full network and eliminate the bulk degrees of freedom explicitly.
We write the subnetwork and bulk parts of $\delta\bm{x}(t)=\bm{x}(t)-\bm{\mu}(t)$ as 
$\delta \bm{x}^{\rm s}(t)$ and
$\delta \bm{x}^{\rm b}(t)$, respectively. Similarly we decompose the rate matrix in equation \eqref{LanVec3} as 
\[\bm{K}(t)=\begin{pmatrix}
\bm{K}^{\rm ss}(t)   & \bm{K}^{\rm sb}(t) \\
\bm{K}^{\rm bs}(t)   & \bm{K}^{\rm bb}(t) \\
\end{pmatrix}\]
Then \eqref{LanVec3} decomposes into
  \begin{eqnarray}
  \label{linearizeddyn}
   \frac{d \delta\bm{x}^{\rm s}}{dt}&= \bm{K}^{\rm ss}(t)\delta\bm{x}^{\rm s} + \bm{K}^{\rm sb}(t)\delta\bm{x}^{\rm b} + \bm{\xi}^{\rm s}\label{eqdxs}\\
   \frac{d \delta\bm{x}^{\rm b}}{dt}&= \bm{K}^{\rm bb}(t)\delta\bm{x}^{\rm b} + \bm{K}^{\rm bs}(t)\delta\bm{x}^{\rm s} + \bm{\xi}^{\rm b} 
  \label{eqdxb}
 \end{eqnarray}
One can now eliminate
$\delta\bm{x}^{\rm b}$ explicitly as
\be
\delta\bm{x}^{\rm b}(t)=
\text{T}\left[e^{\int_{0}^t\bm{K}^{\rm bb}(s)ds}\right]
\delta\bm{x}^{\rm b}(0)+ \int_0^t \,\text{T}\left[e^{\int_{t'}^t\bm{K}^{\rm bb}(s)ds}\right]
\left(\bm{K}^{\rm bs}(t')\delta\bm{x}^{\rm s}(t') + \bm{\xi}^{\rm b}(t')\right)dt'
\ee
where $\delta\bm{x}^{\rm b}(0)$ is the bulk initial condition and the T indicates that the exponentials are time-ordered, with factors of 
$\bm{K}^{\rm bb}(s)$ at each order of the exponential expansion arranged so that time increases from right to left.
Substituting into \eqref{eqdxs} gives the reduced subnetwork dynamics in the form
\begin{equation}
\label{linearreduced}
\frac{d \delta \bm{x}^{\rm s}(t)}{dt}= \bm{K}^{\rm ss}(t) \delta\bm{x}^{\rm s}(t) + 
\int_0^t dt' \bm{M}^{\rm ss}(t,t')\delta \bm{x}^{\rm s}(t') + \bm{\chi}(t)
\end{equation}
All subnetwork reactions are retained directly in the local-in-time term $\bm{K}^{\rm ss}$
while 
the bulk gives a contribution via the additional non-local-in-time terms. The second term in particular contains the memory
function 
  \be
\label{memtemp}
\bm{M}^{\rm ss}(t,t') = \bm{K}^{\rm sb}(t)\,\text{T}\left[e^{\int_{t'}^t\bm{K}^{\rm bb}(s)ds}\right]\,\bm{K}^{\rm bs} (t')
\ee
This memory function has a \emph{boundary} structure~\cite{katy}: memory terms only appear in the reduced dynamical equations for  ``boundary species'', 
defined as those subnetwork species that interact with the bulk and that therefore have at least some nonzero coefficients in the relevant row of $\bm{K}^{\rm sb}$. 
We defer discussion of the effective noise $\bm{\chi}$ in \eqref{linearreduced} to 
\ref{MemNo}; its covariance is given in \eqref{EffNoitem}.

The expression \eqref{memtemp} for the memory function is appealing conceptually -- signals enter the bulk at time $t'$, propagate around the bulk and then return to the subnetwork 
at time $t$ -- and the applicability to time-dependent effective rate matrices $\bm{K}^{\rm bb}(t)$ makes it more general than e.g.\ projection approaches~\cite{katy}. 
For computational purposes a time-dependent $\bm{K}^{\rm bb}(t)$ is inconvenient, however, because of the 
need to evaluate time-ordered exponentials. From now on we therefore consider 
dynamics around a \emph{steady state}, where the effective rate and diffusion matrices become  constant in time, $\bm{K}(t)\equiv \bm{K}$ and
 $\langle\bm{\Sigma}(\bm{x}(t))\rangle\equiv\bm{\Sigma}$.
In this simpler case \eqref{memtemp} becomes
 \begin{equation}
  \label{memlin}
\bm{M}^{\rm ss}(t,t') = \bm{K}^{\rm sb} e^{\bm{K}^{\rm bb}(t-t')}\bm{K}^{\rm bs} 
\end{equation}
This then agrees fully with the memory function in the
reduced description derived via projection methods~\cite{katy}, as summarized in \ref{projlin}.
Given this agreement we can refer to \cite{katy} for a systematic analysis of typical amplitudes and timescales of the memory, which encode its overall effect on dynamics,
in the case of a mass action kinetics such as \eqref{eq:steq}.

In contrast to the memory function, the effective noise term (see \eqref{EffNoi} in \ref{MemNo}) in our marginalization approach differs from its apparent analogue in the projection method, the random force. This is because, as explained in Sec.~\ref{pm}, the projection approach works with conditionally averaged observables and hence eliminates (in a linear theory such as the GVA) all 
intrinsic noise contributions.  
The random force thus contains only the extrinsic noise from the initial uncertainty about the bulk state (see \ref{projvsgva}). Our effective noise, on the other hand, contains in addition the implicit noise effects.

So far no restrictions have been placed
on the relative timescales of bulk and subnetwork. If we specialize to the case of a slowly varying
subnetwork embedded in a fast bulk, equation \eqref{linearreduced} can be shown (see \cite{thesis}) to reduce to the slow-scale LNA introduced by Thomas et al.~\cite{thomas,thomasproj}, and this is a 
useful consistency check for our approach.

\section{Nonlinear corrections by perturbation theory}
\label{nonlinmem}
\subsection{Nonlinear memory}
\label{nonlinmem0}

As explained, our general approach is to derive the reduced dynamics of a subnetwork by marginalizing out bulk degrees of freedom to obtain 
an ``effective'' action that involves \emph{solely} subnetwork variables
\begin{equation}
\label{EffAct}
\mathcal{H}_{\text{eff}}(\bm{x}^{\rm s},\bm{\hat{x}}^{\rm s})= \ln {\int D\bm{x}^{\rm b}D\bm{\hat{x}}^{\rm b} 
e^{\mathcal{H}(\bm{x}^{\rm s},\bm{\hat{x}}^{\rm s},\bm{x}^{\rm b},\bm{\hat{x}}^{\rm b})}}
\end{equation}
where $D\bm{x}^{\rm b}D\bm{\hat{x}}^{\rm b}$ indicates the integration over bulk trajectories and $\mathcal{H}$ is the dynamical action 
\eqref{action} of the whole system. So far we have done this within a Gaussian approximation for $\mathcal{H}$, where the marginalization can be done in closed form; for ease of derivation we sidestepped the dynamical action in this case and eliminated the bulk variables directly from the corresponding dynamical equations.

To go beyond the resulting simple linear expression for the memory function (Sec.~\ref{linmem}) we now use a perturbative expansion to capture nonlinear contributions in the dynamics, specifically cubic terms 
in the effective action.
We decompose the action $\mathcal{H}$ into a ``non-interacting'' part
$\mathcal{H}_0$ containing only the purely Gaussian terms, i.e.\ those that are
quadratic in all variables $\delta\bm{x}$ and $\hat{\bm{x}}$, and an interacting piece $\Delta \mathcal{H}$ containing higher powers.
Assuming formally that the rate constants defining $\Delta\mathcal{H}$ are small, we can expand the effective action \eqref{EffAct} as
\begin{equation}
\label{effac0}
\begin{split}
\mathcal{H}_{\text{eff}}&=  \ln {\int D\bm{x}^{\rm b}D\bm{\hat{x}}^{\rm b} e^{(\mathcal{H}_0 + \Delta \mathcal{H})}}\\
&=\ln {\int D\bm{x}^{\rm b}D\bm{\hat{x}}^{\rm b} e^{\mathcal{H}_0}\left(1+\int D\bm{x}^{\rm b}D\bm{\hat{x}}^{\rm b}
Q_0(\bm{x}^{\rm b},\bm{\hat{x}}^{\rm b}|\bm{x}^{\rm s},\bm{\hat{x}}^{\rm s})
\Delta \mathcal{H}+\mathcal{O}(\Delta \mathcal{H}^2)+ \ldots \right)}
\end{split}
\end{equation} 
Here
\begin{eqnarray}
\label{deltaH0}
&&-\Delta\mathcal{H}= \int_0^{T} dt \bigg\lbrace\text{i}\hat{\bm{x}}^{\rm s}\bigg[\bm{K}^{\rm  s, ss}\big(\delta\bm{x}^{\rm s}\circ
\delta\bm{x}^{\rm s}\big)+
\bm{K}^{\rm  s,sb}\big(\delta\bm{x}^{\rm s}\circ\delta\bm{x}^{\rm b}\big)+
\bm{K}^{\rm  s,bb}\big(\delta\bm{x}^{\rm b}\circ\delta\bm{x}^{\rm b}\big)+\notag\\
&&\qquad\qquad\qquad\quad
\text{i}\hat{\bm{x}}^{\rm b}\bigg[\bm{K}^{\rm b, ss}\big(\delta\bm{x}^{\rm s}\circ\delta\bm{x}^{\rm s}\big)+
\bm{K}^{\rm  b, sb}\big(\delta\bm{x}^{\rm s}\circ\delta\bm{x}^{\rm b}\big)+\bm{K}^{\rm  b,bb}
\big(\delta\bm{x}^{\rm b}\circ\delta\bm{x}^{\rm b}\big)\bigg]\bigg\rbrace
\end{eqnarray}
collects all {\em quadratic} terms in $\delta{\bm{x}}$ in the original mass action dynamics \eqref{phi}: thus the perturbative approach we develop here 
is targeted at this type of nonlinearities, i.e.\ binary reactions. For simplicity we have dropped terms arising from the concentration dependence of the noise 
covariance here; see \ref{pert}. We have written directly the continuous time ($\Delta\to 0$) version and introduced an explicit final time $T$ for trajectories.
We have also introduced convenient shorthands to separate terms that depend on different combinations of subnetwork or bulk variables. 
We use the $\bm{a}\circ\bm{b}$ notation, not to be confused with a Hadamard (elementwise) product, to denote the outer product $\bm{a}\bm{b}^{T}$ rearranged into a single 
(column) vector. This vector then has as its entries all possible componentwise products $a_i b_j$ so that e.g.\ $\delta\bm{x}^{\rm s}\circ\delta\bm{x}^{\rm b}$ is 
a vector of dimension $N^{\rm s}N^{\rm b}$. Where the two vectors are from the same index range (both ``s'' or both ``b'') we use the ordered products only: 
$\delta\bm{x}^{\rm s}\circ\delta\bm{x}^{\rm s}$ is the vector containing the $N^{\rm s}(N^{\rm s}+1)/2$ products $\delta x_i \delta x_j$ with $1\leq i\leq j\leq N^{\rm s}$.
The matrices $\bm{K}^{\rm s,ss}$ etc.\ are then defined to contain the appropriate coefficients to ensure that, for example, 
\be
\big[\bm{K}^{\rm  s, ss}\big(\delta\bm{x}^{\rm s}\circ
\delta\bm{x}^{\rm s}\big)\big]_i=\sum_{j,l=1, j\neq l}^{N^{\rm s}}\bigg(\frac{1}{2}k^{+}_{jl,i}\delta x_j\delta x_l
-k^{+}_{ij,l}\delta x_i\delta x_j\bigg)+
\frac{1}{2}\sum_{j =1}^{N^{\rm s}}k^{+}_{jj,i}\delta x_j\delta x_j
-\sum_{l =1}^{N^{\rm s}}k^{+}_{ii,l}\delta x_i \delta x_i
\ee
Because of the ordering of the $\circ$ product this requires that the elements of $\bm{K}^{\rm  s, ss}$ are defined as
$K_{i,jl} =  k^{+}_{jl,i} - (\delta_{ij}+\delta_{il})\sum_{m=1}^{N^{\rm s}} k^{+}_{jl,m}$ for $j<l$ and 
$K_{i,jj} = \frac{1}{2} k^+_{jj,i} - \delta_{ij}\sum_{m=1}^{N^{\rm s}} k^+_{ii,m}$. These coefficients define the terms in the drift \eqref{phi} for subnetwork variables that depend on quadratic combinations of subnetwork variables.
$\bm{K}^{\rm s,sb}$, $\bm{K}^{\rm b,sb}$, $\bm{K}^{\rm b,bb}$ are used as 
  shorthands for sets of rate constants involved, respectively, in the reaction of a bulk and subnetwork species to give
  a complex in the subnetwork ($\bm{K}^{\rm s,sb}$) or in the bulk ($\bm{K}^{\rm b,sb}$) or two bulk species ($\bm{K}^{\rm b,bb}$) 
  to give another bulk species.
  We will set $\bm{K}^{\rm s,bb}=\bm{K}^{\rm b,ss}\equiv 0$, i.e.\ we do not include in 
  the subnetwork complexes formed by 2 bulk species (and viceversa we do not include in the bulk complexes formed by 2 subnetwork species). This is in line with the spirit of our model reduction setting where subnetwork species are classified 
as the well-characterized ones (e.g.\ because they have been fully resolved biochemically or can be monitored experimentally with accuracy) while bulk species are 
assumed to be less characterized.

As \eqref{effac0} shows, the effective action is calculated by averaging $\Delta\mathcal{H}$ over the distribution of the bulk variables \emph{conditional} on the subnetwork variables.
This average is governed by the Gaussian baseline distribution $Q_0=\exp(\mathcal{H}_0)$, so that also the conditional statistics are Gaussian (see \ref{pert}). 
We could choose $Q_0$ as the steady state GVA distribution, $Q_0=Q$, but this would lead to terms {\em linear} in $\hat{\bm{x}}$ in $\mathcal{H}_0$ that could be interpreted as effective fields. 
These generate nonzero means, which are inconvenient for the conditional Gaussian averaging. To see how such terms arise we write a schematic version of the drift part of the action 
\begin{equation}
 \int_0^{T} dt \,\sum_i\text{i}\hat{x}_i\,\Phi_i(\bm{x})= 
 \int_0^{T} dt \,\sum_i\text{i}\hat{x}_i\,\Phi_i(\bm{\mu}+\delta \bm{x})
 = \int_0^{T} dt \,\sum_i\text{i}\hat{x}_i\,[\Phi_i(\bm{\mu})+\mathcal{O}(\delta\bm{x})+\mathcal{O}(\delta\bm{x}^2)]
\end{equation}
where in the last step we have expanded the drift $\Phi_i$ around the $\bm{\mu}$, up to quadratic terms
in $\delta\bm{x}$. Linear terms appear if we use $\bm{\mu}$ as obtained from the GVA since then $\Phi_i(\bm{\mu})$
 is not zero. It is therefore more convenient to use the (steady state) LNA, see also Sec.~\ref{gva} and \cite{LNA},
 for which $\Phi_i(\bm{\mu})=0$.
The interpretation of $\bm{\mu}$ is then not as a \emph{mean} steady state concentration in a system with intrinsic noise, but rather as the steady state of the corresponding \emph{deterministic} dynamics.
The resulting quadratic action is 
\begin{eqnarray}
\label{H0}
\fl \mathcal{H}_0=&& \int_0^{T} dt \bigg\lbrace\text{i}\hat{\bm{x}}^{\rm s\,\it T}(t)\big[\partial_t\delta\bm{x}^{\rm s}(t)-\bm{K}^{\rm ss} \delta\bm{x}^{\rm s}(t)- \bm{K}^{\rm sb} \delta\bm{x}^{\rm b}(t)\big]+
\frac{1}{2}\text{i}\hat{\bm{x}}^{\rm s\,T}(t)\bm{\Sigma}^{\rm ss}\text{i}\hat{\bm{x}}^{\rm s}(t)+\frac{1}{2}\text{i}\hat{\bm{x}}^{\rm s\,T}(t)\bm{\Sigma}^{\rm sb}\text{i}\hat{\bm{x}}^{\rm b}(t)\notag\\
\fl &&\qquad\text{i}\hat{\bm{x}}^{\rm b\,\it T}(t)\big[\partial_t\delta\bm{x}^{\rm b}(t)-\bm{K}^{\rm bs} \delta\bm{x}^{\rm s}(t)- \bm{K}^{\rm bb} \delta\bm{x}^{\rm b}(t)\big]+
\frac{1}{2}\text{i}\hat{\bm{x}}^{\rm b\,T}(t)\bm{\Sigma}^{\rm bb}\text{i}\hat{\bm{x}}^{\rm b}(t)+\frac{1}{2}\text{i}\hat{\bm{x}}^{\rm b\,T}(t)\bm{\Sigma}^{\rm bs}\text{i}\hat{\bm{x}}^{\rm s}(t)\bigg\rbrace
\end{eqnarray}
The conditional statistics of the bulk variables given the subnetwork ones, which we need for our perturbation expansion, can be worked out from \eqref{H0} using general results for Gaussian conditioning
(see e.g.~\cite{bishop}). The conditional mean of the bulk auxiliary variables is
\begin{equation}
\label{condaux}
 \text{i}\bm{\hat{\mu}}^{\rm b|s} (t)= \int_t^{T} dt' e^{-(\bm{K}^{\rm bb})^{T}(t-t')}(\bm{K}^{\rm sb})^{T}
 \text{i}\bm{\hat{x}}^{\rm s}(t')
\end{equation}
where we use the superscript $\rm{b|s}$ (and $\rm{bb|s}$ below) to indicate bulk averages conditioned on the entire subnetwork trajectories.
This can be viewed as the solution of the differential equation
\begin{equation}
\frac{d\,\text{i}\bm{\hat{\mu}}^{\rm b|s}(t)}{dt}= -(\bm{K}^{\rm bb})^{T}\text{i}\bm{\hat{\mu}}^{\rm b|s}(t) + (\bm{K}^{\rm sb})^{T}
 \text{i}\bm{\hat{x}}^{\rm s}(t)
 \end{equation}
backwards in time, starting from the final boundary condition $\bm{\hat{\mu}}^{\rm b|s} (T)=0$. This is consistent with the theory of conditional Markov processes, 
for which calculating conditional distributions requires information to propagate both forward and backward in time (see also \cite{plefkaobs}).
Note that unlike the marginal auxiliary means, the conditional auxiliary means can be nonzero as long as 
the subnetwork auxiliary variables we condition on are also nonzero.

The conditional means of the bulk concentrations follow from \eqref{H0} and differ from the marginal means by $\delta\bm{\mu}^{\rm b|s} (t)=\bm{\mu}^{\rm b|s} (t)- \bm{\mu}^{\rm b}=
 \bm{\nu}(t)+\bm{\hat{\nu}}(t)$ where
\be
\label{bulkmean}
\bm{\nu}(t) = \int_0^t dt' e^{\bm{K}^{\rm bb}(t-t')}\bm{K}^{\rm bs}\delta\bm{x}^{\rm s}(t') \qquad
\bm{\hat{\nu}}(t) = -
\int_0^{T}dt'\bm{C}^{\rm bb|s}(t,t')(\bm{K}^{\rm sb})^{T}\text{i}\bm{\hat{x}}^{\rm s}(t')
\ee
Here $\bm{\nu}(t)$ is the deterministic part of the conditional time evolution of the bulk concentrations, while $\bm{\hat{\nu}}(t)$ 
involves $\bm{C}^{\rm bb|s}(t,t')$, the conditional correlator of the bulk concentration fluctuations, 
therefore can be interpreted as carrying the ``stochastic'' contributions
 from the bulk, i.e.\ the uncertainty about its initial values and its intrinsic noise. 
 
To complete the description of the conditional bulk statistics one needs the conditional second moments. These can be obtained by setting the subnetwork variables 
in \eqref{H0} to zero (see \cite{thesis}). As such, the conditional correlator of auxiliary variables $\bm{B}^{\rm bb|s}(t,t') \equiv 0$ and the equal-time 
conditional response $\bm{R}^{\rm bb|s}(t,t)\equiv 0$ both vanish, as in the full dynamics. (Causal responses $\bm{R}^{\rm bb|s}(t,t')$ are nonzero, but do not appear in our first order perturbation theory.) 
$\bm{C}^{\rm bb|s}(t,t')$, which also appears in the expression \eqref{bulkmean}, is given by
  \begin{equation}
  \label{bulkcov}
   \bm{C}^{\rm bb|s}(t,t'')=\int_0^{\text{min}(t,t'')}dt'e^{\bm{K}^{\rm bb}(t-t')}\bm{\Sigma}^{\rm bb}e^{(\bm{K}^{\rm bb})^{T}
   (t''-t')}+e^{\bm{K}^{\rm bb}t}\bm{C}^{\rm bb}(0,0) e^{(\bm{K}^{\rm bb})^{T} t''}
  \end{equation}
As is generally the case in Gaussian conditioning, the second order statistics do not depend on the values of the observables (here: the subnetwork concentrations) being conditioned on.

We can now find the reduced subnetwork action from \eqref{effac0} by performing the Gaussian averages over the bulk variables with the conditional statistics as given by
\eqref{condaux}, \eqref{bulkmean} and \eqref{bulkcov}. From this reduced action one can then read off the nonlinear reduced dynamics of the subnetwork. We leave the details
for \ref{pert} and state directly the result
\begin{eqnarray}
\label{nonlingva}
\frac{d\, \delta \bm{x}^{\rm s}(t)}{dt}&=& \bm{K}^{\rm ss} \delta\bm{x}^{\rm s}(t) +
\bm{K}^{\rm s,ss} (\delta\bm{x}^{\rm s}(t) \circ \delta\bm{x}^{\rm s}(t)) +  
\int_0^tdt' \bm{M}^{\rm ss}(t,t')\delta\bm{x}^{\rm s}(t') \notag \\ &+& 
\int_0^t dt'\int_0^{t'}dt'' \bm{M}^{\rm s,ss}(t,t',t'')\big(\delta\bm{x}^{\rm s}(t')\circ\delta\bm{x}^{\rm s}(t'')\big)+\bm{\chi}(t)
\end{eqnarray}
where in $\bm{M}^{\rm s,ss}(t,t',t'')$ the times are ordered, $t>t'>t''$. The coloured noise $\bm{\chi}(t)$ has covariance $\langle \bm{\chi}(t)\bm{\chi}(t')^{T}\rangle=\bm{N}_0^{\rm ss}(t,t')+\bm{N}_1^{\rm ss}(t,t')$ consisting of a 
linear and nonlinear contribution. While $\bm{N}_0^{\rm ss}(t,t')$ is the covariance of a Gaussian coloured noise (namely \eqref{EffNoi}), $\bm{N}_1^{\rm ss}(t,t')$ 
is multiplicative as it is linearly dependent on $\bm{x}^{\rm s}$. 

We see that reactions within the subnetwork contribute to the reduced dynamical equations only via $\bm{K}^{\rm ss}$ and 
$\bm{K}^{\rm s,ss}$ (a compact notation for the rate constants of respectively linear and nonlinear couplings inside the subnetwork): subnetwork reactions are fully reproduced 
in local-in-time terms, which is one of the desiderata of our reduced description. 

While $\bm{M}^{\rm ss}(t,t')$ is the memory function given by \eqref{memlin},
the nonlinear memory function, $\bm{M}^{\rm s,ss}(t,t',t'')$,
can be read off from the coefficients of terms $\sim\bm{\hat{x}}^{\rm s}\delta\bm{x}^{\rm s}\delta\bm{x}^{\rm s}$ in the effective perturbative
action \eqref{effac0}, which themselves 
come from the $\bm{\p}(t)$ part of the conditional mean \eqref{bulkmean}.
The effective noise covariance, $\bm{N}_1^{\rm ss}(t,t')$ is similarly extracted from terms
$\sim \bm{\hat{x}}^{\rm s}\delta\bm{x}^{\rm s}\bm{\hat{x}}^{\rm s}$ that arise from $\bm{\hat{\p}}(t)$.
We refer to \ref{pert} for details of the derivation and just state the final result (for $t>t'>t''$)
\begin{equation}
  \label{eq:effMemoryNLtimes}
   \begin{split}
\bm{M}^{\rm s,ss}(t,t',t'')&\big(\delta\bm{x}^{\rm s}(t')\circ\delta\bm{x}^{\rm s}(t'')\big)=\\
 &+\,\bm{K}^{\rm s,sb}\big(\delta\bm{x}^{\rm s}(t')\circ e^{\bm{K}^{\rm bb}(t-t'')}\bm{K}^{\rm bs}\delta \bm{x}^{\rm s}(t'')\big)\delta(t'-t)\\
 &+\,\bm{K}^{\rm sb} e^{\bm{K}^{\rm bb}(t-t')}\bm{K}^{\rm b,sb}\big(\delta \bm{x}^{\rm s}(t')\circ 
 e^{\bm{K}^{\rm bb}(t'-t'')}\bm{K}^{\rm bs}\delta \bm{x}^{\rm s}(t'')\big)\\
    & +2\int_{t'}^t ds\,\bm{K}^{\rm sb} e^{\bm{K}^{\rm bb}(t-s)} \bm{K}^{\rm b,bb}
  \big(e^{\bm{K}^{\rm bb}(s-t')}\bm{K}^{\rm bs}\delta \bm{x}^{\rm s}(t') \circ e^{\bm{K}^{\rm bb}(s-t'')}\bm{K}^{\rm bs}\delta \bm{x}^{\rm s}(t'')\big)
   \end{split}
  \end{equation}
This form emphasizes that the nonlinear memory that results from our approach involves products of concentration fluctuations 
at \emph{different} times $t'$ and $t''$ in the past. In the projection approach, on the other hand, only equal-time products from the past appear~\cite{katy}.
While our result therefore looks more complicated conceptually, its numerical implementation is in fact more efficient than for the projected subnetwork equations as we discuss next.

\subsection{Numerical implementation}
\label{num_imp}
We will denote by $\bm{\mathcal{M}}(t)$ the total memory appearing in the subnetwork dynamics of $\delta \bm{x}^{\rm s}(t)$. This is given by the 
two integral terms in \eqref{nonlingva}. For numerical purposes, each of these can be represented via the solutions of additional
differential equations. Hence, one can solve the \emph{integro-differential} equations \eqref{nonlingva} using only a
\emph{differential} equation solver but in an enlarged space of variables. To see this,
we first note that $\bm{\mathcal{M}}(t)$ can be expressed in 
terms of the $\bm{\p}$-part of the bulk conditional means as
\be
\label{memfullcond}
\begin{split}
\bm{\mathcal{M}}(t)=\,&\bm{K}^{\rm sb}\bm{\p}(t)+ \bm{K}^{\rm s,sb}\big(\delta\bm{x}^{\rm s}(t)\circ \,\bm{\p}(t)\big)
\\
&{}+\int_0^t dt' \bm{K}^{\rm sb} e^{\bm{K}^{\rm bb}(t-t')}
\left[\bm{K}^{\rm b,sb}\big(\delta\bm{x}^{\rm s}(t')\circ\bm{\p}(t')\big)+ 
\bm{K}^{\rm b,bb} \big(\bm{\p}(t')\circ \bm{\p}(t')\big)\right]
\end{split}
\ee
The bulk means \eqref{bulkmean} are the solution of the $N^{\rm b}$ differential equations
\be
\frac{d \bm{\p}}{dt} = \bm{K}^{\rm bb}\bm{\p}+ \bm{K}^{\rm bs}\delta\bm{x}^{\rm s}
\ee
with $\bm{\p}(0)=0$. 
The $\bm{K}^{\rm b,sb}$ and $\bm{K}^{\rm b,bb}$ pieces of \eqref{memfullcond} contain an additional time integral;
to translate them into differential equations for additional variables 
we need to apply a decomposition into eigenvalues and eigenvectors of $\bm{K}^{\rm bb}$, similarly to the procedure in \cite{rubinMM}. 
This requires $N^{\rm b}$ auxiliary variables in addition to the $\bm{\p}$, giving $2N^{\rm b}$ in total; see \ref{eff_solver} for details.

It is notable here that for subnetwork equations derived by projection methods, a similar conversion into differential equations requires a 
number of additional variables growing \emph{quadratically} with the number of bulk nodes \cite{katy}. The smaller, \emph{linear}, number of auxiliary variables in our method is a significant computational advantage for the typical case where the bulk network is large, e.g.\ because the overall network itself is large while the subnetwork contains only a few, experimentally well resolved, species.

\subsection{Comparison with projection methods}
\label{comp_pm}
The comparison of our perturbatively corrected GVA approach to nonlinear projection methods is not as straightforward as for the linearized equations. 
The reason is that memory functions and the noise correlators derived in the two approaches are not the same 
if taken \emph{separately}. But 
remarkably, we can show that the \emph{combination} of the memory term and the coloured noise from 
equation \eqref{nonlingva} provides an approximation of the subnetwork reduced dynamics that is \emph{equivalent} to the one obtained by
projection techniques \eqref{nonlinproj}. This holds in the \emph{limit of negligible intrinsic noise} and \emph{up to quadratic order} in $\delta\bm{x}$.
The proof of this equivalence is rather non-trivial, both conceptually and algebraically, so we defer it to \ref{comparison}.

\section{Application to biochemical networks}

In this section we illustrate the GVA reduction method by applying it to a simple toy model and to the EGFR biochemical network from \cite{kholodenko}, 
the aim being to assess its accuracy. We solve in parallel the projected equations from \cite{katy}, to verify explicitly the equivalence between the two approaches (under the 
condition specified above of negligible noise) to $\mathcal{O}(\delta x^2)$, and to compare the errors they make at higher order.

Everywhere in what follows we take $\epsilon \rightarrow 0$ (which implies also $\bm{\Sigma}^{\rm bb}=0$) and the bulk is chosen initially at steady state, so
$\bm{C}^{\rm bb|s}(t,t')$, given by \eqref{bulkcov}, vanishes and $\bm{\mu}^{\rm b|s}(t) \equiv \bm{\nu}(t)$
(the conditional bulk means coincide with the deterministic bulk solution of the linearized dynamics). Furthermore, the coloured noise of the GVA is identically zero, as is the random force in the projection methods, at least up to $\mathcal{O}(\delta \bm{x}^2)$
(see \ref{nonlinproj}).

\subsection{Toy model}
\label{toy}
To explain our method step by step we consider first a toy model with five species: 
$x_1$, $x_2$ and $x_3$ belong to the subnetwork, $x_4$ and $x_5$ are in the bulk, and they undergo 
two reversible complex formation/dissociation reactions. Schematically we write
\[x_1+ x_2 \mathrel{\mathop{\rightleftarrows}^{k^+_{12,3}}_{k^-_{3,12}}} x_3 \] 
\[x_1+ x_4 \mathrel{\mathop{\rightleftarrows}^{k^+_{14,5}}_{k^-_{5,14}}} x_5 \]
where $k^+_{12,3}$ and $k^+_{14,5}$ are the complex formation rate constants, while $k^-_{3,12}$ and $k^-_{5,14}$ are the dissociation rates. 
The concentration deviations from the steady states $\delta x_i=x_i-\mu_i$, $i=1,...,5$ follow the mass action kinetics
  \begin{subequations}
  \begin{align}
   \frac{d}{dt}\delta x_1 &= k^-_{3,12}\delta x_3 - k^+_{12,3}(\mu_1 \delta x_2+\mu_2\delta x_1+\delta x_1\delta x_2) \nonumber\\
   &+k^-_{5,14}\delta x_5 - k^+_{14,5}(\mu_1 \delta x_4+\mu_4\delta x_1+\delta x_1\delta x_4)\\
   \frac{d}{dt}\delta x_2 &= k^-_{3,12}\delta x_3 - k^+_{12,3}(\mu_1 \delta x_2+\mu_2\delta x_1+\delta x_1\delta x_2)\label{red2} \ = \ -\frac{d}{dt}\delta x_3 \\
   \frac{d}{dt}\delta x_4 &= k^-_{5,14}\delta x_5 - k^+_{14,5}(\mu_1 \delta x_4+\mu_4\delta x_1+\delta x_1\delta x_4)
\ = \ -\frac{d}{dt}\delta x_5
  \end{align}
  \end{subequations}
There are no constant terms on the r.h.s.\ here as the $\mu_i$ describe a steady state. There are two conservation laws, one for the bulk and one for the subnetwork:
the total concentration of $x_2$ and $x_3$ (similarly for $x_4$ and $x_5$) is constant in time, implying that 
$\delta x_2 = - \delta x_3$ and $\delta x_4 = - \delta x_5$. The subnetwork dynamics is therefore described only in terms of $\delta x_1$ and $\delta x_2$, and the bulk by $\delta x_4$, so that effectively $N^{\rm s}=2$, $N^{\rm b}=1$.

 \subsubsection{Reduced dynamics.}
 There is one boundary species, $\delta x_1$, which interacts with the bulk species $\delta x_4$ and $\delta x_5$.
 Its dynamics is thus the only one  affected by memory effects. By applying the formulas for the memory \eqref{memlin} and \eqref{eq:effMemoryNLtimes}
  we obtain the effective equation for $\delta x_1(t)$ 
 \begin{eqnarray}
 \label{eqdx1}
 \frac{d}{dt} \delta x_1 &=& k^-_{3,12} \delta x_3 - k^+_{12,3}(\mu_1\delta x_2+\mu_2 \delta x_1+ \delta x_1 \delta x_2) +\int_0^t dt' M_{11} (t-t') 
 \delta x_1(t')\notag\\
&+& \int_0^t dt'\int_{t'}^t dt'' M_{11,1} (t,t',t'') \delta x_1(t')\delta x_1(t'') 
 \end{eqnarray}
 with
 \be
 \label{toy_lin}
 M_{11} (t-t')= (k^-_{5,14}+ k^+_{14,5} \mu_1)k^+_{14,5} \mu_4\, e^{-(k^-_{5,14}+ k^+_{14,5} \mu_1)(t-t')}
 \ee
 and
\be
 M_{11,1} (t,t',t'')= k^{+\,2}_{14,5} \mu_4 \,e^{-(k^-_{5,14}+ k^+_{14,5} \mu_1)(t-t'')}[\delta(t-t') 
 - (k^-_{5,14}+ k^+_{14,5} \mu_1)]
 \ee
 As explained in Sec.~\ref{num_imp}, solving the integro-differential equation \eqref{eqdx1} in differential form
requires $2 N^{\rm b}=2$ additional variables. One of these just gives the mean of $\delta x_4$ conditional on $\delta x_1$
\be
\label{eq1}
\frac{d}{dt}\delta \mu_{4|1} = - k^+_{14,5} \mu_4\delta x_1-(k^-_{5,14}+ k^+_{14,5} \mu_1) \delta \mu_{4|1}, \qquad \delta \mu_{4|1}(0)=0
\ee
The other one, $z$, using the procedure outlined in \ref{eff_solver}, is the solution of 
\be
\label{eq2}
\frac{d}{dt} z = -k^+_{14,5}\delta x_1\delta \mu_{4|1} -(k^-_{5,14}+ k^+_{14,5} \mu_1) z, \qquad z(0)=0
\ee
where $-(k^-_{5,14}+ k^+_{14,5} \mu_1)$ is the only nonzero eigenvalue of $\bm{K}^{\rm bb}$.
 These additional equations must be solved jointly with the dynamics of the interior subnetwork species $\delta x_2$, which is given by the original \eqref{red2}, and the version of \eqref{eqdx1} 
 with memory terms expressed in terms of auxiliary variables 
\begin{eqnarray}
\label{eq3}
\frac{d}{dt}\delta x_1& = &k^-_{3,12} \delta x_3 - k^+_{12,3}(\mu_1\delta x_2+\mu_2 \delta x_1+ \delta x_1 \delta x_2) \notag \\
 &&-(k^-_{5,14}+ k^+_{14,5} \mu_1) \delta \mu_{4|1} - k^+_{14,5} \delta x_1 \delta \mu_{4|1} + (k^-_{5,14}+ k^+_{14,5} \mu_1)z
\end{eqnarray}
We solved \eqref{red2}, \eqref{eq1}, \eqref{eq2} and \eqref{eq3} to find the time evolution of $\delta x_1(t)$. Fig.~\ref{fig:toy} (left) shows the corresponding fractional concentration 
deviation from steady state, defined as $\delta\tilde{x}_1=(x_1(t)-\mu_1)/\mu_1$, alongside the baseline obtained from the full reaction equations (black solid line): visually the two are essentially indistinguishable. 
Here and in what follows we refer to the ``full reaction'' equations as the system of Langevin equations describing the evolution of the \emph{entire} 
network (comprising both subnetwork and bulk), taken in the limit of negligible noise $(\epsilon \to 0)$ as for the reduced model.
We contrast this with two simpler approximate subnetwork
reductions for biochemical networks that are Markovian, i.e.\ do not involve memory terms: one in which the subnetwork is considered as isolated, i.e.\ reactions with the bulk are ignored, and one in which the bulk is assumed to be fast, thus always in its steady state given the instantaneous subnetwork concentrations (steady state bulk). 
Fig.~\ref{fig:toy} (left) demonstrates that both of these are distinctly less accurate than our nonlinear GVA reduction. The two simpler approximation schemes are typically justified either by timescale separation between subnetwork and bulk, or by weak subnetwork-bulk coupling. In our case the latter can be achieved by, for example, scaling down  $k^+_{14,5}$ and $k^-_{5,14}$ by a common factor; as this factor is made large, we find indeed that all the approximate schemes
perform well, making identical predictions in the limit. If on the other hand we increase the rates and hence the subnetwork-bulk coupling,
the two alternative approximations, in particular the steady state bulk method, lead to increasingly poor predictions (see inset of Fig.~\ref{fig:toy}, left).
A systematic approach to model reduction, as provided by our method and also by projection approaches~\cite{katy},
gives good agreement with the solution of the full reaction equations \emph{regardless} of any assumption on the strength of
subnetwork-bulk interactions. This makes it more flexible in applications, leaving the choice of subnetwork free according to e.g.\ available data or biological interest in certain molecular species.

 \subsubsection{Quantitative tests.}
 \label{quantitative_tests}
\begin{figure}
\includegraphics[width=0.49\textwidth]{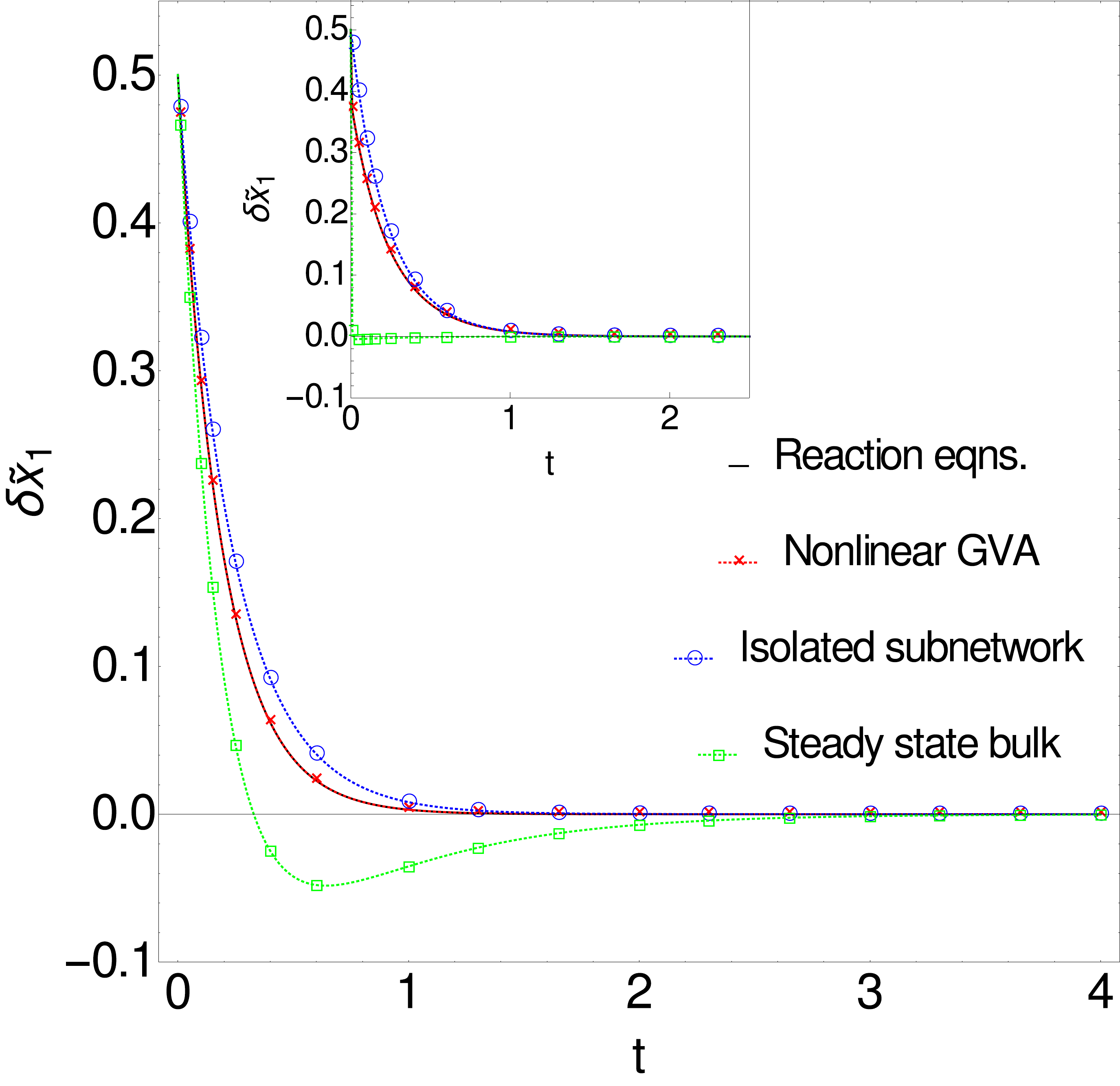}
\includegraphics[width=0.49\textwidth]{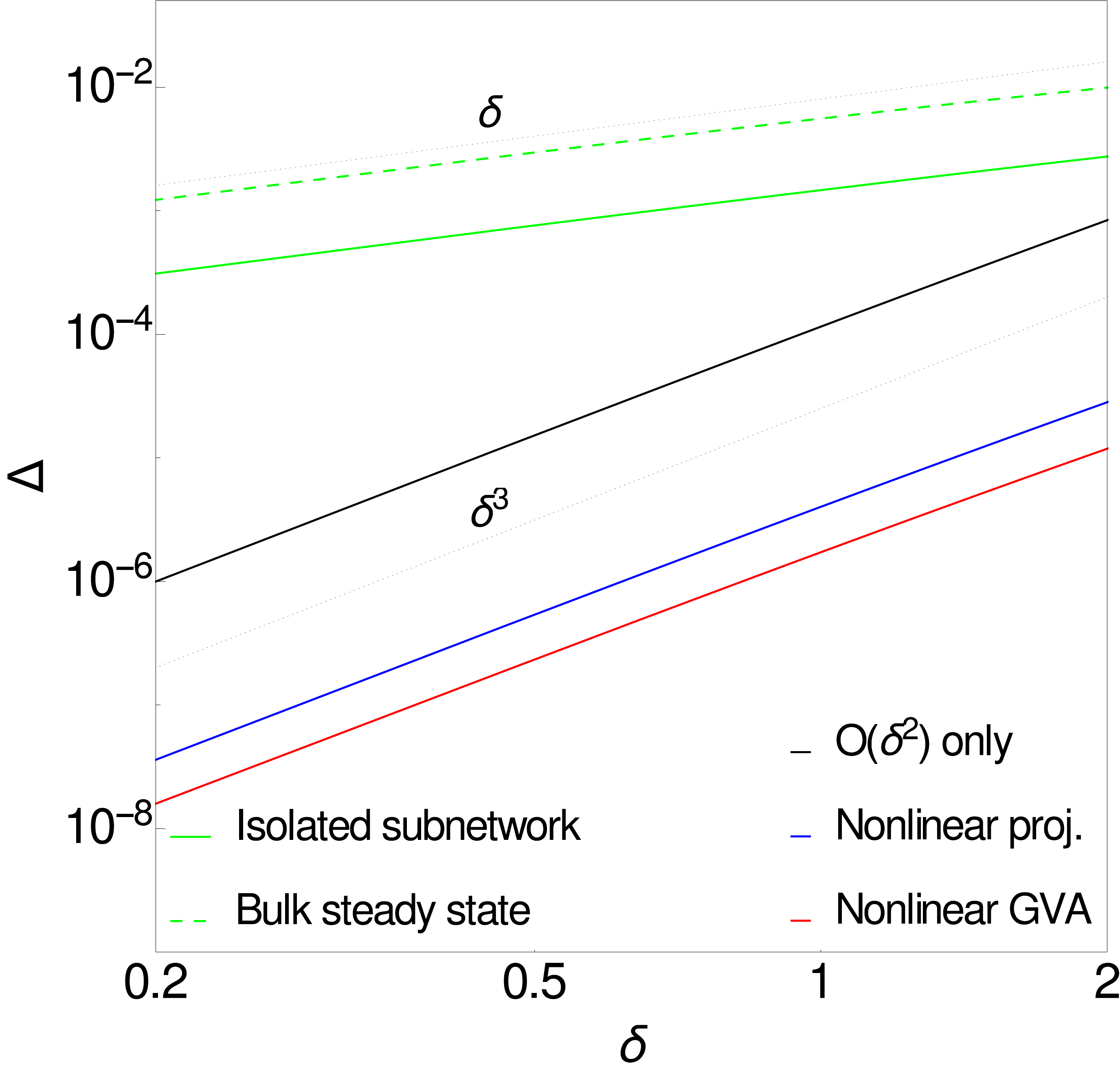}
\caption[Time courses and error of approximation for the toy model.]{{\bf Toy model} (Left) Time courses of $\delta\tilde{x}_1$ given by the nonlinear GVA, the isolated subnetwork and steady state bulk 
approximations compared to the full reaction equations. (Right) Approximation error, $\Delta$, as a function of the initial deviation from steady state
$\delta$; the nonlinear GVA is compared to projection methods (both giving errors scaling as $\delta^3$) and the two simpler approximations 
(whose error grows already as $\delta$). 
 Rates are set to $k^+_{12,3}=k^+_{14,5}=1$, 
$k^-_{3,12}=k^-_{5,14}=2$, giving steady state concentrations $\mu_1=\mu_2=\mu_4=1$, $\mu_3=\mu_5=1/2$. In the inset on the left, the subnetwork-bulk coupling is increased by
setting $k^+_{14,5}=5$, $k^-_{5,14}=10$. Initial conditions are chosen as $\delta\tilde{x}_1(0)=\delta\tilde{x}_2(0)=1/2$, 
$\delta\tilde{x}_3(0)=-1$, $\delta\tilde{x}_4(0)=\delta\tilde{x}_5(0)=0$ (bulk is initially at steady state) for the left figure. On the right we scale these initial conditions 
to get the chosen $\delta$.}
\label{fig:toy}
\end{figure}

 We next look at the error of the various subnetwork reduction methods. As in~\cite{katy} we vary the scale of the initial deviations from steady state, which we characterize by the root mean squared deviation 
 $\delta = (\lbrace\sum_{s}[\delta\hat{x}_s(0)]^2\rbrace/N^{\rm s})^{1/2}$. We quantify the accuracy of the approximation in terms of
 \be
 \label{eqerror}
 \Delta = \frac{1}{T} \int_0^T dt\frac{1}{N^{\rm s}}\sum_{s=1}^{N^{\rm s}}|\delta\tilde{x}_s(t)-\delta\hat{x}_s(t)|
 \ee
where $\delta\hat{x}_s(t)$ is the exact solution. $\Delta$ is
an absolute deviation in the dimensionless concentration of each subnetwork species, averaged over 
species and a time interval of $T=15$ s, chosen here to cover the transient regime of relaxation to the steady state.
From Fig.~\ref{fig:toy} (right) one sees that the errors for the simpler Markovian approximations
are substantially larger in absolute terms, by several orders of magnitude for moderate $\delta$; this emphasizes the importance of memory terms for accurate predictions. 
The errors also grow \emph{linearly} with $\delta$, while they scale \emph{cubically} both for nonlinear projection methods and GVA. The conclusion, supported by the theoretical analysis in 
\ref{projvsgva}, is that by retaining nonlinear memory terms both methods make predictions accurate to $\mathcal{O}(\delta x^2)$. Each method also includes a subset of higher order terms, but not systematically. At $\mathcal{O}(\delta x^3)$ the two methods therefore differ, and this leads to different prefactors in the scaling of their prediction error with $\delta^3$.

In the nonlinear projection method, the error at $\mathcal{O}(\delta x^3)$ is due to the fact that there are cubic contributions to the random force
that do not vanish even if the bulk is initially at steady state.
In the nonlinear GVA, the error arises from the truncation of the perturbative expansion after $\mathcal{O}(\Delta \mathcal{H})$,
while to be systematic in the dynamics at cubic order in $\delta x$ one would have to retain also {\em fourth} order terms $\sim \hat x (\delta x)^3$ in the effective action, which would be affected by the
$\mathcal{O}(\Delta \mathcal{H}^2)$ term in the expansion.
Both methods include \emph{some} higher order terms, and these do improve predictions: the curve labelled ``$\mathcal{O}(\delta x^2)$ only'' in Fig.~\ref{fig:toy} (right) shows that errors increase by more than an order of magnitude if one uses a reduction method that systematically discards all terms beyond quadratic order.
(This is done in practice by evaluating all product terms in the reduced evolution equations from the linearized dynamics only.)

\subsection{Application to EGFR}
  \label{egfr}
 We consider the network of protein-protein interactions around EGFR as in the model by Kholodenko et al. \cite{kholodenko}. This has previously been used as 
 a testbed for projection methods in \cite{katy}; we refer to this work for a characterization
 of the reaction network, including parameters and the choice of initial conditions.
 
 \begin{figure}
 \includegraphics[width=0.493\textwidth]{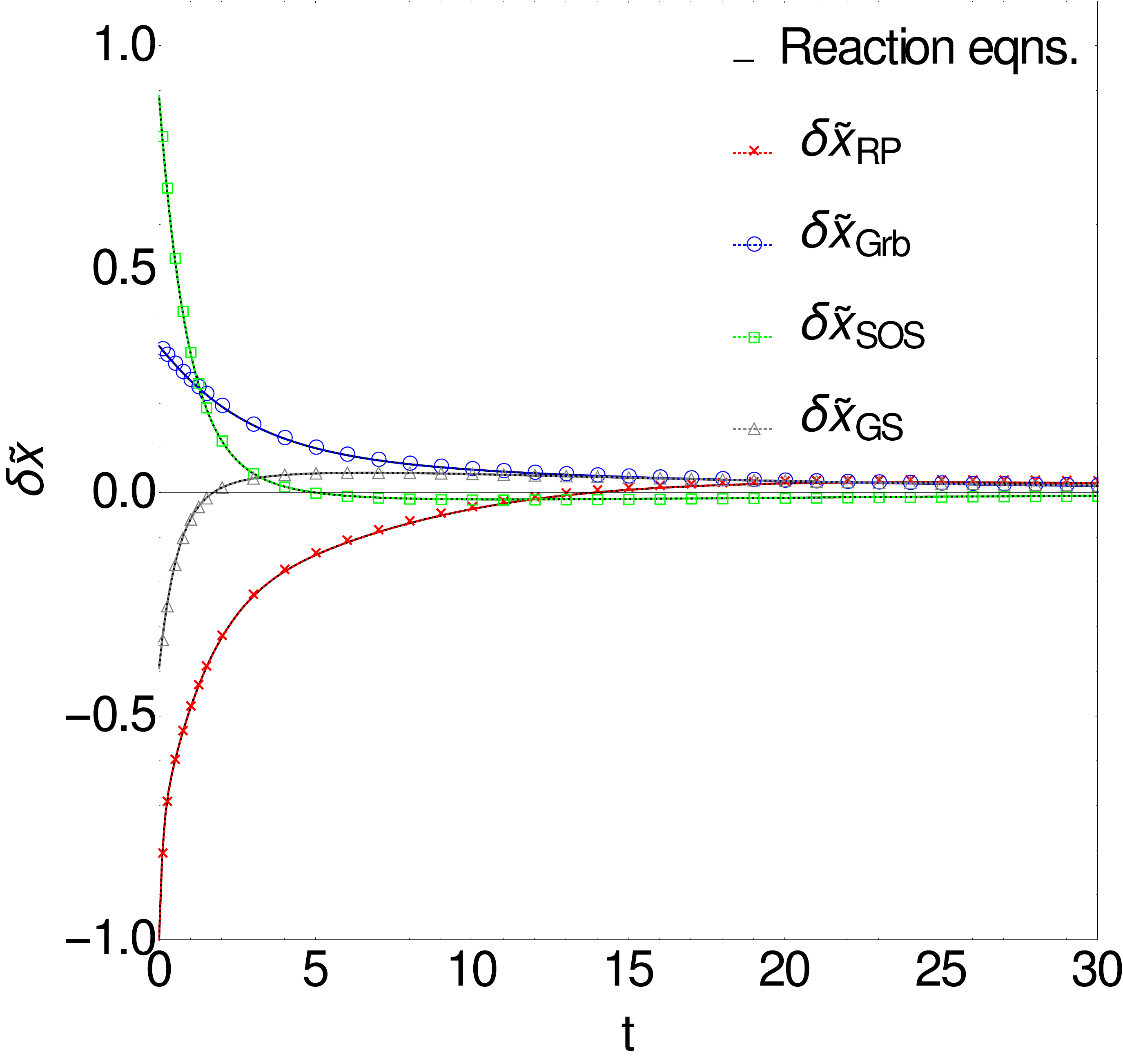}
\includegraphics[width=0.493\textwidth]{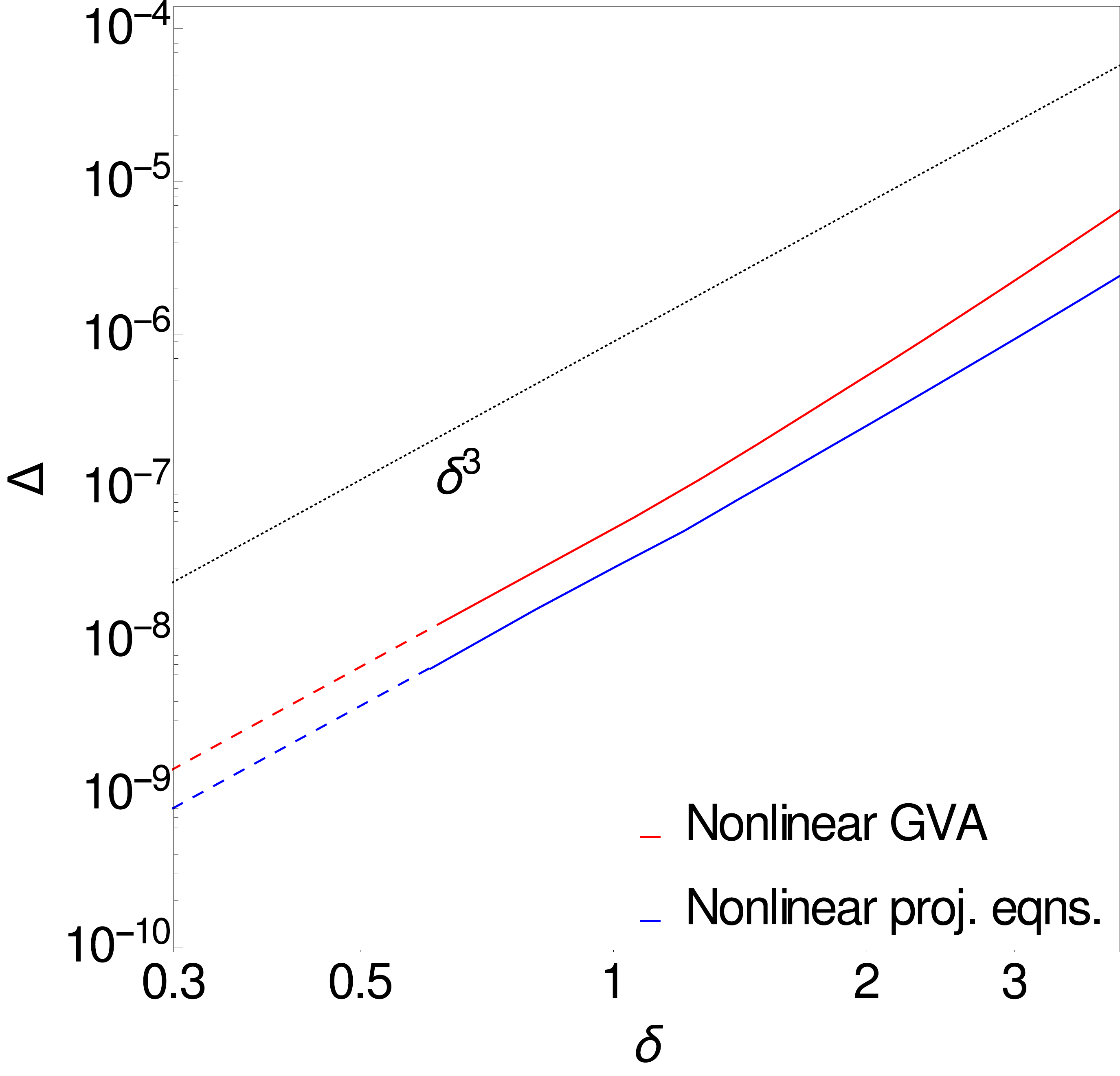}
\caption[Time courses and error of approximation for the EGFR network.]{{\bf EGFR network} (Left) Time courses of the fractional concentration deviations from steady state
for the four subnetwork boundary species (defined as in~\cite{katy}), as predicted by the nonlinear GVA reduction versus the full reaction equations.
 (Right) Approximation error, $\Delta$, as a function of the initial deviation from steady state $\delta$; the error is shown for both the nonlinear GVA and the
nonlinear projected equations; for both the error scales as $\delta^3$ as long as $\delta$ is not too large. 
The dashed lines are an extrapolation into the regime where errors become too small to measure reliably. Rates, steady states and initial conditions are set as 
 in~\cite{katy}.}
\label{fig:egfr}
\end{figure}

\subsubsection{Quantitative tests.}
As before we take the limit of vanishing intrinsic noise, and initial conditions with the bulk at steady state. As shown in Fig. \ref{fig:egfr} (left), a very accurate prediction of subnetwork time courses 
is achieved using the nonlinear GVA reduction method.
For a more quantitative assessment we again vary the initial deviation $\delta$ from steady state and measure the approximation error $\Delta$ 
defined by \eqref{eqerror} (here we consider $T=150$ s as time window for the transient regime as in \cite{katy}). The results are displayed in Fig.~\ref{fig:egfr} (right),
where we compare the approximation error of the the nonlinear GVA and the nonlinear projection methods. 
As expected, $\Delta$ grows as $\delta^3$ for both methods. Again there is a difference in the prefactor, due to different terms being included beyond quadratic order.
For the EGFR network, it is the projected subnetwork equations that are more accurate, by a factor of roughly two, whereas for the toy model it was the nonlinear GVA that produced smaller errors, cf.\ Fig.~\ref{fig:toy} (right). Which method is more accurate is therefore likely to be system-dependent in general. However, the difference in their errors is modest, while both methods are orders of magnitude more accurate than 
simpler Markovian approximations (see data in~\cite{katy} for the EGFR case).
When deciding which method for subnetwork reduction to use one might therefore prioritize the computational advantage offered by the nonlinear GVA over the (potentially, depending on the system)
slightly higher accuracy of projection methods. As an example, for the EGFR case, a numerical implementation of the nonlinear GVA reduction method in terms of differential equations requires only 20 additional variables, as opposed to 255 for the nonlinear projected equations (see \ref{eff_solver} and \cite{katy}).

\subsubsection{Nonlinear memory function.}
\label{mem_nonlin}
\begin{figure}
\includegraphics[width=0.49\textwidth]{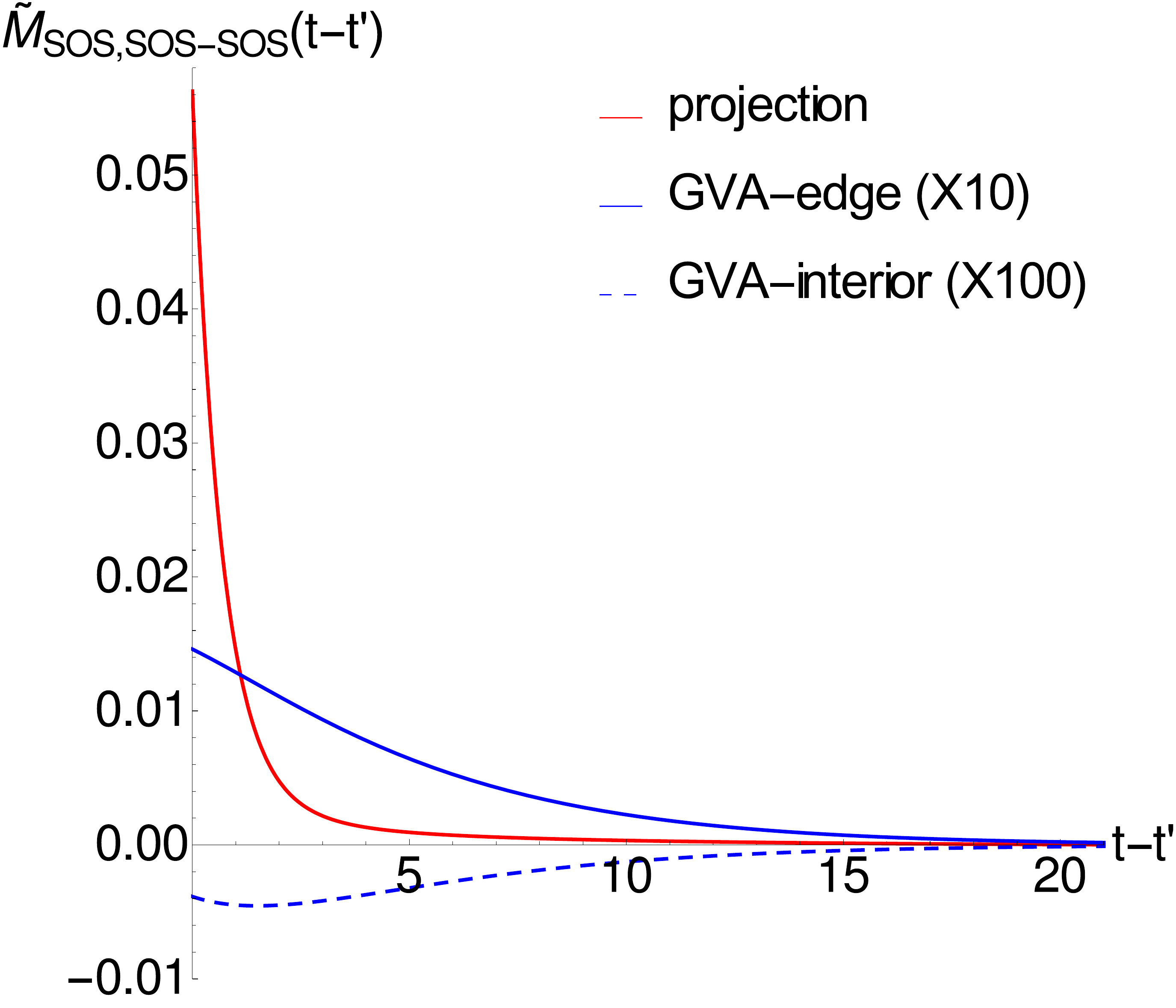}
\hspace{0.3cm}
\includegraphics[width=0.4\textwidth]{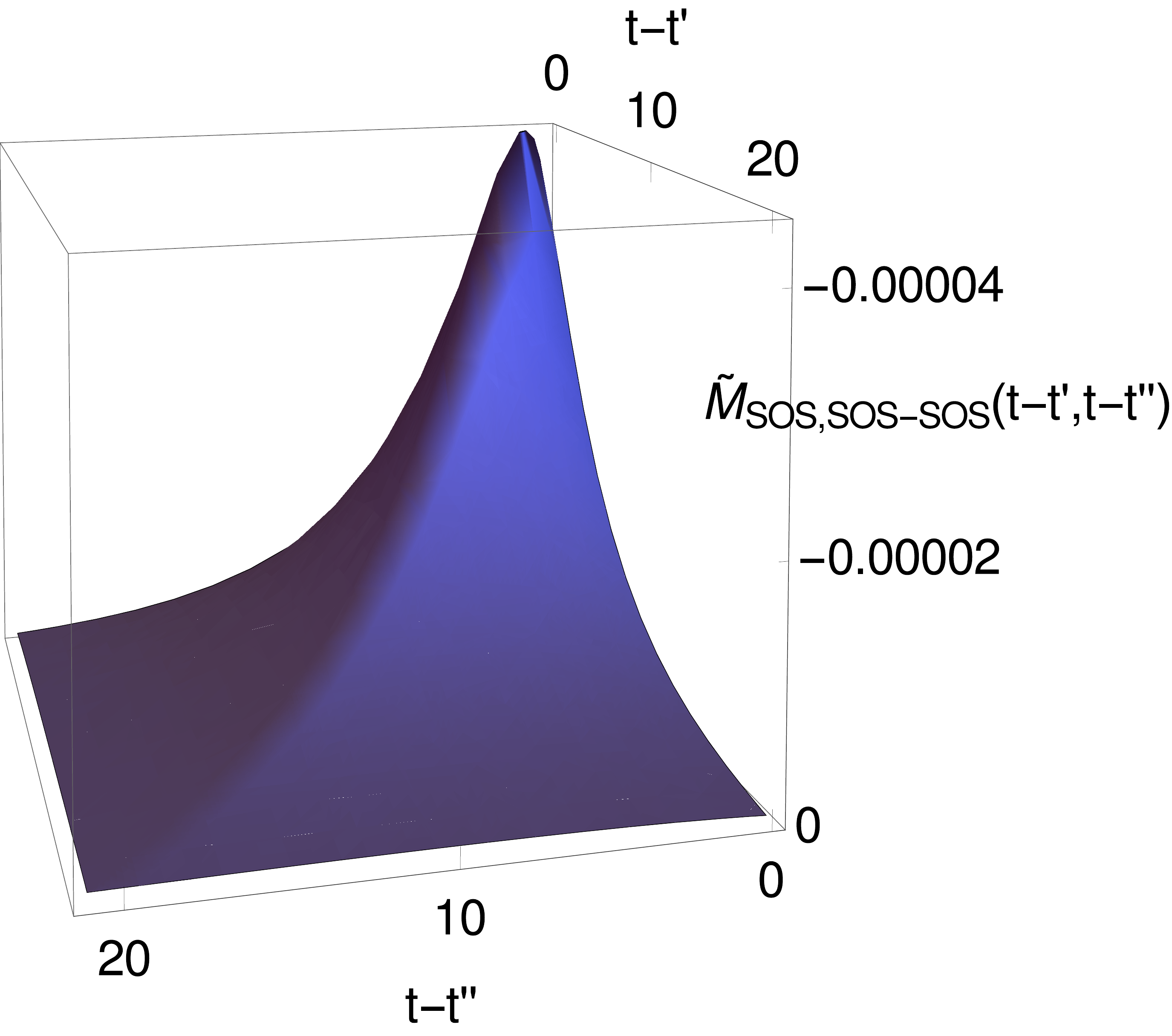}
\caption[Nonlinear self-memory of subnetwork species SOS.]{Nonlinear self-memory of subnetwork species SOS. (Left) Comparison of GVA-edge and GVA-interior memory function terms with nonlinear 
memory function from projection method; 
the GVA-interior term is shown along the diagonal, i.e.\ for $t'=t''$; the edge term is plotted against $t-t'$. 
The GVA-edge and interior terms are multiplied respectively by a factor of 10 and 100 to make them visible on the scale of the projection memory, whose 
amplitude is significantly bigger. Note though that only the projection memory and the GVA-edge memory contribution are comparable, having measurement units of (time)$^{-2}$, 
while the GVA-interior contribution has units (time)$^{-3}$.
(Right) The GVA-interior memory function is plotted in 3D 
to show its shape for generic $t'\neq t''$: the diagonal contribution (i.e.\ for  $t'=t''$) is dominant.}
\label{fig:memorySOS}
\end{figure}

\begin{figure}
\includegraphics[width=0.49\textwidth]{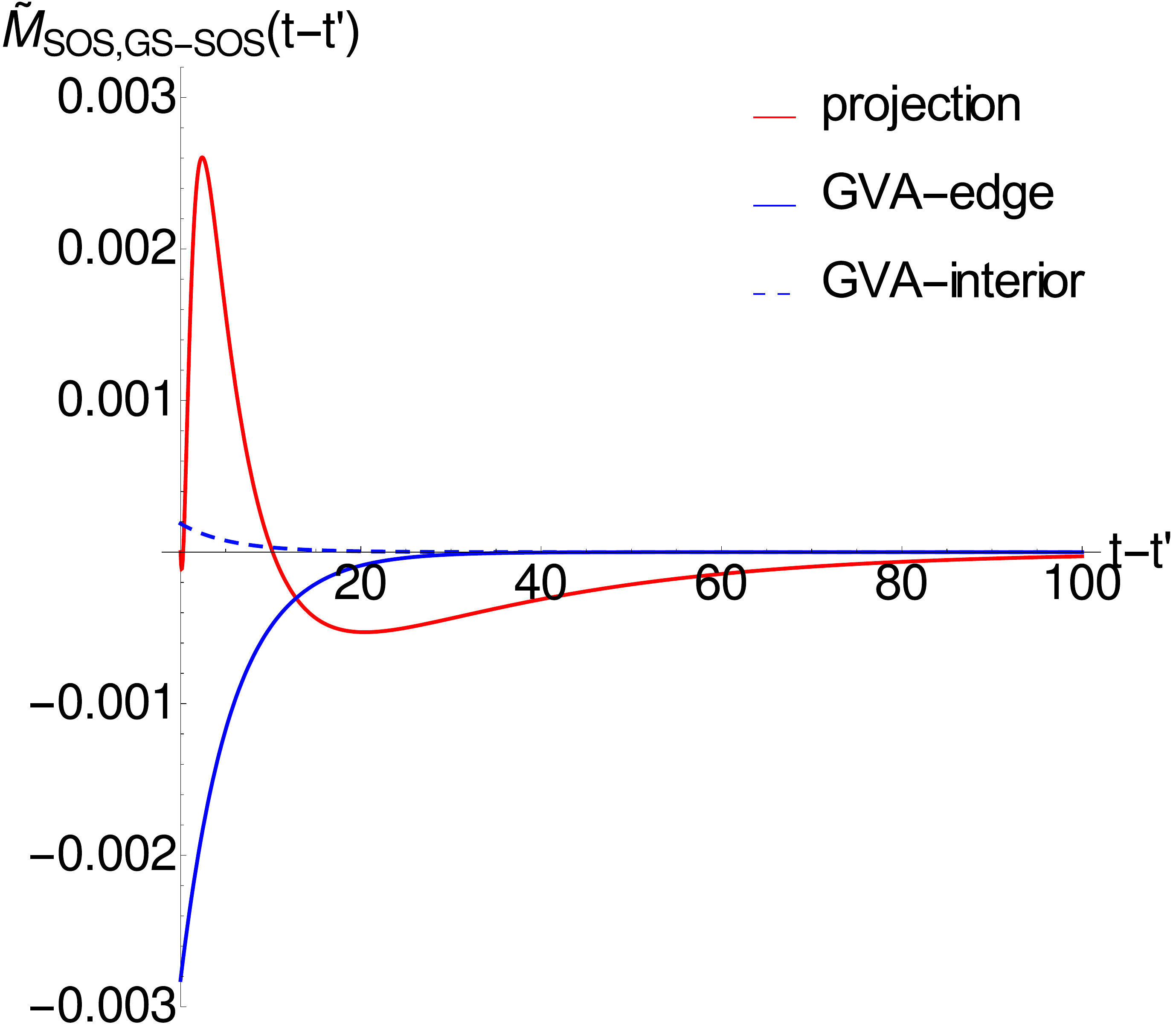}
\hspace{0.3cm}
\includegraphics[width=0.4\textwidth]{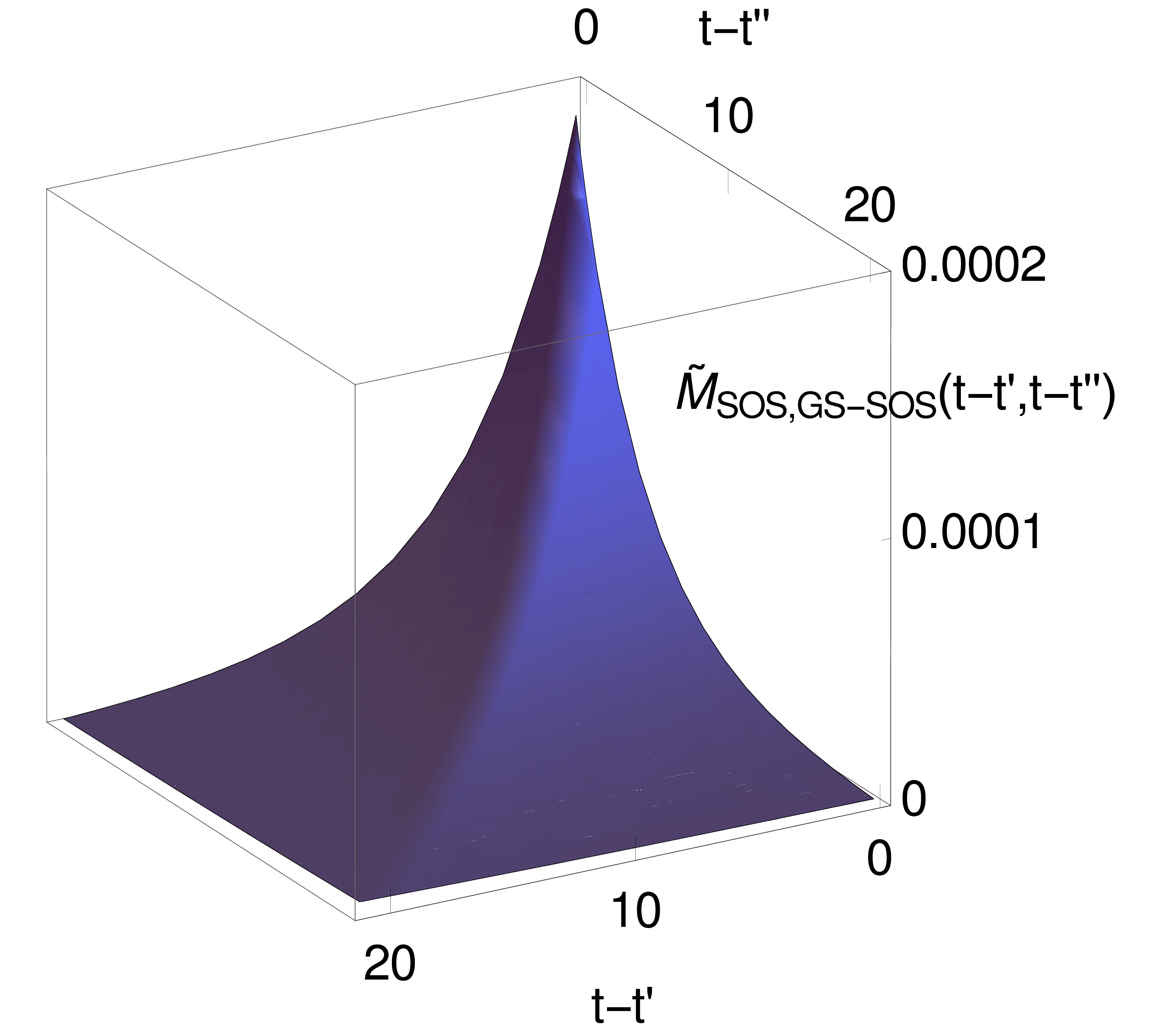}
\caption[Nonlinear cross-memory of SOS for the product GS-SOS.]{Example of nonlinear cross-memory acting on SOS, defined for the product GS-SOS.
(Left) Comparison of GVA-edge and GVA-interior (for $t'=t''$) terms with projection methods memory.
Similarly to $\tilde{M}_{\text{SOS,SOS-SOS}}(t-t')$ in Fig. \ref{fig:memorySOS} (left), 
the GVA nonlinear memory function $\tilde{M}_{\text{SOS,GS-SOS}}(t-t')$ is
smaller in amplitude w.r.t.\ its projection analogue because effectively it accounts 
for what in the projection approach is a sum of terms with positive or 
negative sign - the nonlinear memories on SOS, defined for all subnetwork products but where one could imagine expressing
the interior species in terms of GS and SOS (boundary ones).
(Right) Full GVA-interior term as a function of $t-t'$ and $t-t''$: as in Fig. \ref{fig:memorySOS} (right), it is concentrated
around $t'=t''$.}
\label{fig:memorySOSGS}
\end{figure}

We next analyze the nonlinear memory function $\bm{M}^{\rm s,ss}(t,t',t'')\equiv \bm{M}^{\rm s,ss}(t-t',t-t'')$.
We have defined this in \eqref{eq:effMemoryNLtimes} for ordered times $t'>t''$; for visual purposes it is helpful to plot the symmetrized version, which sets $\bm{M}^{\rm s,ss}(t,t',t'')=\bm{M}^{\rm s,ss}(t,t'',t')$.
The first term in \eqref{eq:effMemoryNLtimes} contains a factor $\delta(t-t')$ and thus gives contributions only along the ``edge'' $t'=t$ of the allowed 
region of times ($t'',t'<t$).
The remaining terms contribute in the ``interior" of this region. We therefore denote these memory function contributions as GVA-edge and GVA-interior, respectively. 
For graphical purposes we plot the edge term simply as a function of $t-t''$, dropping the $\delta$ factor -- see Figs.~\ref{fig:memorySOS} and \ref{fig:memorySOSGS} (left) --
 while examples of the interior terms are shown in the 3D Figs.~\ref{fig:memorySOS} and \ref{fig:memorySOSGS} 
(right). In particular, we consider a version of the nonlinear memory function in \eqref{eq:effMemoryNLtimes}
made dimensionless w.r.t.\ concentrations, as it would appear in the dynamical subnetwork equations for the fractional concentration fluctuations $\delta\tilde{x}$; explicitly,
$\tilde{M}_{i,jk}(t-t',t-t'')=y_i^{-1} M_{i,jk}(t-t',t-t'') y_{j}y_k $.

We compare the GVA nonlinear memory function with the one of the projection approach (see \eqref{memProj} in \ref{nonlinproj}), which
is a function of just one time difference $t-t'$ as it acts only on concentration products at the \emph{same} time $t'$. In Figs.~\ref{fig:memorySOS} and \ref{fig:memorySOSGS} (left) 
we plot two examples from the EGFR network, the self-memory and a cross-memory function for SOS, against the GVA memory terms for the same species 
(where we set $t'=t''$ in the interior piece). In our setting (bulk initially at steady state and vanishing intrinsic noise), we know that 
the overall GVA nonlinear memory acting on a certain species 
is equivalent, to $\mathcal{O}(\delta x^2)$, to the one from the projection method. But the two methods decompose this nonlinear memory differently into sums of contributions
from the possible products of concentrations. In particular, 
the projection memory features products involving \emph{all} subnetwork concentrations, while the GVA memory involves only products of \emph{boundary} species concentrations.
This helps explain the significant difference in amplitude visible in Fig.~\ref{fig:memorySOS} (left), where the GVA memory functions 
are so small that in order to make them visible on the same scale as the projection memory function we had to scale them up by 10 or 100, respectively. 
Inspection of projection functions describing memory to other products of concentrations shows that these have both positive and negative sign and so largely cancel. 
In the GVA memory there are fewer terms, namely those involving products of concentrations of boundary species, hence weaker cancellations: thus the individual memory terms 
can be smaller as we observe. 

\begin{figure}
\includegraphics[width=0.9\textwidth]{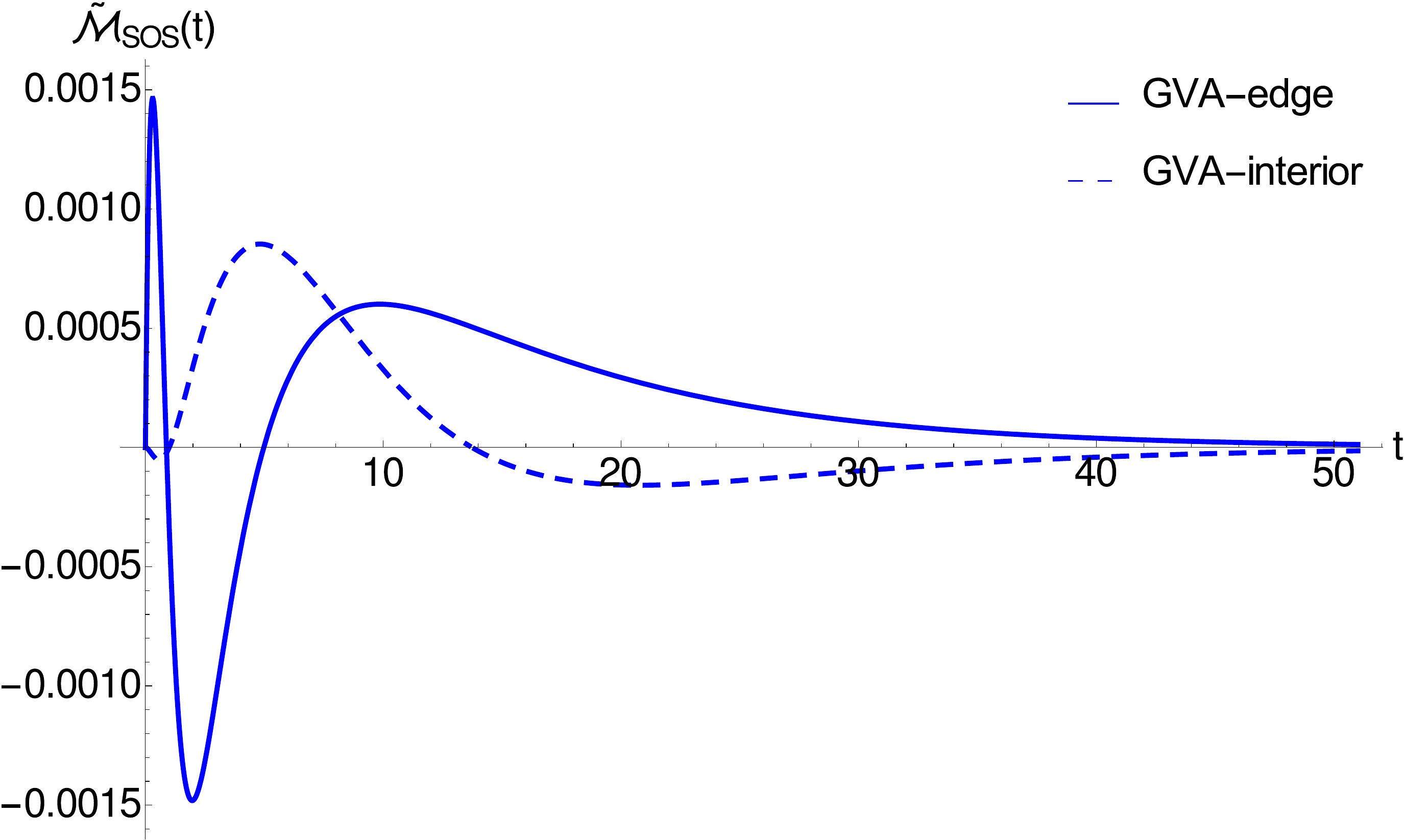}
\caption[GVA-edge and GVA-interior terms in the entire nonlinear memory of SOS.]{The integrals of the GVA-edge and GVA-interior memory terms, which together make up the entire nonlinear memory of SOS, as a function of time $t$ for the time course from Fig.~\ref{fig:egfr} (left). The two contributions are broadly comparable, and which one is dominant actually changes with time.}
\label{fig:memorySOStot}
\end{figure}

Finally, in Fig.~\ref{fig:memorySOStot}, we show the entire (dimensionless) nonlinear memory 
on SOS, $\tilde{\mathcal{M}}_{\rm{SOS}}(t)$, as a function of time along a specific time course, namely the one shown in Fig.~\ref{fig:egfr} (left). We assess the relative importance of the GVA-edge and GVA-interior terms by plotting their 
separate contributions to the memory integral. This shows that either can be dominant, depending on time $t$.


\cleardoublepage

\section{Discussion and conclusions}

In this paper we have developed a model reduction technique 
based on a Gaussian approximation of the Chemical Langevin Equation (CLE) for biochemical networks with unary and binary reactions, describing e.g.\ complex formation and dissociation processes.
We first re-derived a Gaussian Variational Approximation (GVA) via a path integral formalism. This is based on identifying the stationary point of a KL-divergence between complex distributions, and allows one to include the case of multiplicative noise in the CLE, see \eqref{effdiffusion}. One should be able to derive this result also using only real distributions as in \cite{opper_gaussian}: the stationarity conditions for the KL should remain well-defined in the continuous time limit even if the KL itself formally diverges.

The GVA effectively yields a linearization of the dynamics around time-dependent means. 
The evolution of the latter (see \eqref{meangau0}) includes corrections from the intrinsic noise that arises whenever the reaction volume $V$ is finite. 
This gives the GVA a higher accuracy than the Linear Noise Approximation (LNA), which linearizes around the macroscopic, i.e.\ noise-free, time evolution. In particular the error of the GVA on the means can be shown
to scale as $1/V^2$, which is already as small as the error one makes in any case by starting from the CLE rather than a master equation for discrete copy numbers. For the LNA, on the other hand, the error on the means is $\sim 1/V$~\cite{grima,grimaMA}.

From the GVA we derived reduced dynamical equations for a chosen subnetwork, by marginalizing out the other degrees of freedom. The coupling between the subnetwork and the rest of the network, or ``bulk", leads to both a memory term and non-trivial time correlations in the effective noise. Evaluation of these quantities is feasible primarily when at least the bulk means are constant in time, which avoids the numerical calculation of time-ordered matrix exponentials.
While the linearized approximation provided by the GVA may sound too simple, there are fully linear models e.g.\ in signalling pathways where they describe so-called \emph{weakly activated} cascades and are considered of theoretical interest 
for building coarse-grained descriptions \cite{barahona_weakly,del_vecchio}.   

We next developed a perturbative expansion of the effective action to estimate nonlinear corrections to the Gaussian approximation, and were able to derive from this explicit expression for the nonlinear memory functions. As a comparative baseline we used subnetwork equations derived by nonlinear projection techniques. There, nonlinear memory terms involve products of concentration fluctuations taken at the \emph{same} time in the past, while in our new approach products at different time also appear. As we saw in 
two numerical examples in Sec.~\ref{mem_nonlin} the equal-time products nonetheless remain dominant.

We also compared our new nonlinear GVA subnetwork reduction scheme more quantitatively with the projection method. The latter can only be evaluated efficiently in the limit of small intrinsic noise, $\epsilon =1/V\rightarrow 0$, so we focussed on this regime. We were able to show that the nonlinear GVA reaches the same level of
accuracy as the projection approach, giving exact results up to $\mathcal{O}(\delta x^2)$ within an expansion in the overall deviation $\delta x$ of the system from its steady state. Computationally, the new method is rather more efficient: the subnetwork equations can be integrated by mapping them to differential equations using a moderate (linear in the size of the bulk) number of auxiliary variables. The projection method, on the other hand, requires   a number of auxiliary variables scaling quadratically with the number of bulk nodes.

We illustrated the reduction method first on a toy model that can be treated fully analytically. Numerical evaluation showed significantly more accurate predictions of time courses than can be achieved by simpler model reduction schemes (isolated subnetwork and steady state bulk). These work well primarily under appropriate assumptions of timescale separation, whereas 
our systematic derivation of a reduced model does not rely on any such assumption. 
The choice of subnetwork can therefore be adjusted flexibly, as required by whatever specific system one is studying.

Protein interaction networks are a key application for our model reduction strategy. We showed for such a model (the EGFR network from
\cite{kholodenko}) that the agreement with exact time courses is again excellent, the absolute error being slightly higher than for projection methods but still 
extremely small. We regard such improvements in prediction accuracy as important in constraining realistic models of biochemical networks and in helping to interpret experimental data correctly. 

Interesting connections can be established to other approaches \cite{koeppl_nature,koeppl_plos} also based on \emph{marginalizing} out the bulk or ``environment''. A practical limitation of such methods is that their implementation depends on finding suitable approximate schemes 
for the marginal and conditional dynamics, which generally cannot be computed analytically. Here we instead use a Gaussian approximation for the conditional dynamics and evaluate non-Gaussian corrections to the marginal dynamics,
 allowing us to obtain closed form expressions for the effective non-Markovian terms (memory and coloured noise).

Model reduction techniques as developed here can be seen as a tool for partitioning a network consistently and systematically
into separate modules that are easier to study. If memory functions can be represented by sufficiently simple approximations, e.g.\ exponentials, this can lead to significant \emph{computational} savings in predicting the subnetwork dynamics, compared to simulating the entire network. At the same time there are \emph{conceptual} 
benefits in studying how a cellular subsystem couples to the rest of the cellular environment.
It has been recently shown, for example, that equilibrium-like domains can arise in non-equilibrium systems \cite{saad} in the form of 
generic bipartite graphs, a scenario akin to our subnetwork--bulk split.

In future work we plan to study the implications of our approach for noise modelling in biochemical systems and how these compare to e.g.\ projection methods. Here we can exploit the fact that in the covariance of the effective noise, see \eqref{EffNoi}, we obtain an explicit
decomposition, i.e.\ intrinsic noise, inherent to the random timing 
of biochemical events, and extrinsic noise due to the interaction with the environment \cite{elowitz,swain,shahrezaei}. 

One would also like to lift some of the restrictions in our approximations. E.g.\ for biochemical networks that exhibit bi- or multistable states or oscillations a Gaussian approximation with its single peak may be poor.
Gaussian mixture models  \cite{barahona_GMM} could then be deployed instead, or one could use the 
general variational formalism of Eyink \cite{eyink}. For systems where small copy number fluctuations are important, 
we are studying the application of our methods to path integral representations of the chemical master equation (e.g.\ \cite{doi2,doi1,peliti}, as discussed in \cite{thesis}).

Also, there are types of nonlinearities in reaction networks that are not captured by unary and binary reactions, e.g. Michaelis-Menten (enzymatic) kinetics 
and Hill functions. To extend our formalism to such cases one could follow a procedure similar to the one used to treat Michaelis-Menten kinetics by projection methods 
\cite{rubinMM}: introduce fast variables and binary reactions that in the limit of large rates reproduce the desired kinetics; apply the reduction method to this larger system;
and finally take the large rate limit to get back to a reduced description for the original kinetics.

As is clear from this work, a path integral representation of biochemical dynamics has a 
 structure that allows the application of variational principles and is convenient for formal manipulations such as the
 marginalization over the bulk; furthermore it offers a flexible framework for developing a perturbative theory for nonlinear problems.
More generally, mathematical tools borrowed from field theories in physics, such as path integrals \cite{sinitsyn,gaffney} and Feynman diagrams \cite{grima_diagrams},  
can yield more accurate prediction methods and an understanding of dynamical behaviours grounded in detailed descriptions of
fundamental components. These results are promising and could motivate a systematic transfer of 
approaches from theoretical physics to systems biology that is still to be fully explored.

Finally, as a natural continuation of our analysis, one could address the inverse problem: given observed time courses for the subnetwork, can
we use the insight into 
reduced subnetwork dynamics that we have established to infer properties of the bulk?
If the structure of the bulk is known, one could aim to infer the bulk dynamics from the observed subnetwork behaviour. In our approach this is exactly what is being done by conditioning on the subnetwork degrees of freedom, with bulk dynamics estimated by Gaussian conditioning; an interesting challenge would be to incorporate nonlinear corrections.
Other approximate approaches to this problem, 
based on variational methods \cite{opper_sanguinetti,kupferman_variational,opper,opper_gaussian} or system size expansions \cite{opper_ruttor}, have turned out to be
 clearer conceptually than MCMC sampling inference methods for intracellular stochastic kinetics.

\section*{Acknowledgements}
This work was supported by the Marie Curie Training Network NETADIS (FP7, grant 290038).
We thank Reimer K\"{u}hn for insightful comments on possible future directions,
Luca Dall'Asta and Heinz K\"{o}ppl for suggesting several related papers.
We are indebted to Katy Rubin and Edgar Herrera Delgado for making available 
their code which helped us with the numerical implementation of projection methods. 
PS acknowledges the stimulating research environment provided by the EPSRC Centre for Doctoral Training in Cross-Disciplinary Approaches to 
Non-Equilibrium Systems (CANES, EP/L015854/1). BB acknowledges the Simons Foundation Grant No. 454953.

\cleardoublepage
\bibliography{Phdbib.bib}

\providecommand{\newblock}{}
\begin{thebibliography}{10}
\expandafter\ifx\csname url\endcsname\relax
  \def\url#1{{\tt #1}}\fi
\expandafter\ifx\csname urlprefix\endcsname\relax\def\urlprefix{URL }\fi
\providecommand{\eprint}[2][]{\url{#2}}

\bibitem{kholodenko}
{Kholodenko} B~N, {Demin} O~V, {Moehren} G and {Hoek} J~B 1999 {\em J. Biol.
  Chem.\/} {\bf 274} 30169--30181

\bibitem{seger}
{Plotnikov} A, {Zehorai} E, {Procaccia} S and {Seger} R 2011 {\em Biochimica et
  Biophysica Acta\/} {\bf 1813} 1619--1633

\bibitem{vankampen}
{Van Kampen} N~G 2007 {\em Stochastic Processes in Physics and Chemistry\/}
  (Elsevier, 3rd Edition)

\bibitem{radulescu}
{Radulescu} O, {Gorban} A, {Zinovyev} A and {Noel} V 2012 {\em Front. Genet.\/}
  {\bf 3}

\bibitem{ackermann}
{Ackermann} J, {Einloft} J, {Nothen} J and {Koch} I 2012 {\em J. Theor.
  Biol.\/} {\bf 315} 71–80

\bibitem{okino}
{Okino} M and {Mavrovouniotis} M 1998 {\em Chem. Rev.\/} {\bf 98} 391--408

\bibitem{apri}
{Apri} M, {De Gee} M and {Molenaar} J 2012 {\em J. Theor. Biol.\/} {\bf 304}
  16--26

\bibitem{eyink}
{Eyink} G 1996 {\em Phys. Rev. E\/} {\bf 54} 3419--3435

\bibitem{sasai}
{Sasai} M and {Wolynes} P 2003 {\em PNAS\/} {\bf 100} 2374--2379

\bibitem{ohkubo}
{Ohkubo} J 2007 {\em JSTAT\/} {\bf P09017}

\bibitem{bishop}
{Bishop} C~M 2006 {\em Pattern Recognition and Machine Learning\/} (Springer)

\bibitem{kullback}
{Kullback} S and {Leibler} R~A 1951 {\em Ann. Math. Stat.\/} {\bf 22} 79–86

\bibitem{opper_sanguinetti}
{Opper} M and {Sanguinetti} G 2007 {\em Proceedings of NIPS\/}

\bibitem{kupferman_variational}
{Cohn} I, {El-Hay} T, {Friedman} N and {Kupferman} R 2009 {\em Proceedings of
  NIPS\/}

\bibitem{opper_gaussian}
{Archambeau} C, {Cornford} D, {Opper} M and {Shawe-Taylor} J 2007 {\em JMLR:
  Workshop and Conference Proceedings\/} {\bf 1} 1--16

\bibitem{romanobattistin}
{Bachschmid-Romano} L, {Battistin} C, {Opper} M and {Roudi} Y 2016 {\em J.
  Phys. A: Math. Theor.\/} {\bf 49} 434003

\bibitem{whittle}
{Whittle} P 1957 {\em J.R. Stat. Soc.\/} {\bf 19} 268–281

\bibitem{bosia}
{Bosia} C, {Pagnani} A and {Zecchina} R 2013 {\em Plos One\/} {\bf 8} 1--13

\bibitem{lakatos}
{Lakatos} E, {Ale} A, {Kirk} P~D~W and {Stumpf} M~P~H 2015 {\em J. Chem.
  Phys.\/} {\bf 143} 094107

\bibitem{sangui_MA}
{Schnoerr} D, {Sanguinetti} G and {Grima} R 2014 {\em J. Chem. Phys.\/} {\bf
  141}

\bibitem{sangui_MA2}
{Schnoerr} D, {Sanguinetti} G and {Grima} R 2015 {\em J. Chem. Phys.\/} {\bf
  143}

\bibitem{grimaMA}
{Grima} R 2012 {\em J. Chem. Phys.\/} {\bf 136}

\bibitem{chorin}
{Chorin} A~J, {Hald} O~H and {Kupferman} R 2000 {\em PNAS\/} {\bf 97}
  2968–2973

\bibitem{chorin_review}
{Chorin} A~J, {Hald} O~H and {Kupferman} R 2006 {\em J. Scient. Comput.\/} {\bf
  28} 245--261

\bibitem{thomasproj}
{Thomas} P, {Grima} R and {Straube} A 2012 {\em Phys. Rev. E\/} {\bf 86} 041110

\bibitem{katy}
{Rubin} K~J, {Lawler} K, {Sollich} P and {Ng} T 2014 {\em J. Theor. Biol.\/}
  {\bf 357} 245--267

\bibitem{zwanzig}
{Zwanzig} R 1961 {\em Phys. Rev.\/} {\bf 124} 983--992

\bibitem{rubin_thesis}
{Rubin} K 2014 {\em Dynamics of protein interaction subnetworks\/} Ph.D. thesis
  King's College London

\bibitem{rubinMM}
{Rubin} K~J and {Sollich} P 2016 {\em J. Chem. Phys.\/} {\bf 144} 174114

\bibitem{sinitsyn}
{Sinitsyn} N, {Hengartner} N, {Nemenman} I and {Press} W 2009 {\em PNAS\/} {\bf
  106} 10546--10551

\bibitem{koeppl_nature}
{Zechner} C, {Unger} M, {Pelet} S, {Peter} M and {Koeppl} H 2014 {\em Nature
  Methods\/} {\bf 11} 197--202

\bibitem{koeppl_plos}
{Zechner} C and {Koeppl} H 2014 {\em PLoS Comput. Biol.\/} {\bf 10} e1003942

\bibitem{gardiner}
{Gardiner} C~W 1985 {\em Handbook for Stochastic Methods\/} (Springer, 2nd
  edition)

\bibitem{vankampenito}
{Van Kampen} N~G 1982 {\em J. Stat. Phys.\/} {\bf 24} 175--187

\bibitem{LNA}
{Elf} J and {Ehrenberg} M 2003 {\em Genome Res.\/} {\bf 13} 2475–2484

\bibitem{rabello}
{Coolen} A and {Rabello} S 2009 {\em JPCS\/} {\bf 197} 012006

\bibitem{martin}
{Martin} P, {Siggia} E and {Rose} H 1973 {\em Phys. Rev. A\/} {\bf 8} 423--436

\bibitem{janssen}
{Janssen} H~K 1976 {\em Z. Phys. B: Cond. Mat.\/} {\bf 23} 377--380

\bibitem{dedominicis}
{De Dominicis} C 1978 {\em Phys. Rev. B\/} {\bf 18} 4913--4919

\bibitem{kamenev}
{Kamenev} A 2011 {\em Field theory of non-equilibrium systems\/} (Cambridge
  University Press)

\bibitem{PathMethods}
{Hertz} J~A, {Roudi} Y and {Sollich} P 2017 {\em J. Phys. A: Math. Theor\/}
  {\bf 50} 033001

\bibitem{ritort}
{Ritort} F and {Sollich} P 2003 {\em Advances in Physics\/} {\bf 52} 219 —
  342

\bibitem{chorin_memory}
{Chorin} A~J, {Hald} O~H and {Kupferman} R 2002 {\em Physica D\/} {\bf 166}
  239--257

\bibitem{coolen}
{Coolen} A~C~C 2001 {\em Handbook of Biological Physics\/} vol~4
  (Elsevier-Academic Press, Amsterdam) chap Statistical Mechanics of Recurrent
  Neural Networks II Dynamics

\bibitem{fischer}
{Fischer} K~H and {Hertz} J~A 1991 {\em Spin Glasses\/} (Cambridge Studies in
  Magnetism)

\bibitem{thesis}
{Bravi} B 2016 {\em Path integral approaches to subnetwork dynamics and
  inference\/} Ph.D. thesis King's College London

\bibitem{thomas}
{Thomas} P, {Straube} A and {Grima} R 2012 {\em BMC Syst. Biol.\/} {\bf 6}

\bibitem{plefkaobs}
{Bravi} B and {Sollich} P 2017 {\em JSTAT, in press\/}

\bibitem{grima}
{Grima} R, {Thomas} P and {Straube} A 2011 {\em J. Chem. Phys.\/} {\bf 135}

\bibitem{barahona_weakly}
{Diaz} M, {Desikan} R and {Barahona} M 2016 {\em J. R. Soc. Interface\/} {\bf
  13}

\bibitem{del_vecchio}
{Herath} N, {Hamadeh} A and {Del Vecchio} D 2015 {\em Proceedings of ACC\/}

\bibitem{saad}
{Saad} D and {Mozeika} A 2013 {\em Phys. Rev. E\/} {\bf 87} 032131

\bibitem{elowitz}
{Elowitz} M~B, {Levine} A~J, {Siggia} E~D and {Swain} P~S 2002 {\em Science\/}
  {\bf 297} 1183--1186

\bibitem{swain}
{Swain} P, {Elowitz} M~B and {Siggia} E~D 2002 {\em PNAS\/} {\bf 99}
  12795--12800

\bibitem{shahrezaei}
{Shahrezaei} V, {Olivier} J~F and {Swain} P 2008 {\em Mol. Syst. Biol.\/} {\bf
  4}

\bibitem{barahona_GMM}
{Chang} H, {Hemberg} M, {Barahona} M, {Ingber} D and {Huang} S 2008 {\em
  Nature\/} {\bf 453} 544--547

\bibitem{doi2}
{Doi} M 1976 {\em J. Phys. A: Math. Gen.\/} {\bf 9} 1465--1477

\bibitem{doi1}
{Doi} M 1976 {\em J. Phys. A: Math. Gen.\/} {\bf 9} 1479--1495

\bibitem{peliti}
Peliti L 1985 {\em Journal de Physique\/} {\bf 46} 1469--1483

\bibitem{gaffney}
{Santos} F~A~N, {Gadelha} H and {Gaffney} E~A 2015 {\em Phys. Rev. E\/} {\bf
  92} 062714

\bibitem{grima_diagrams}
{Thomas} P, {Fleck} C, {Grima} R and {Popovic} N 2014 {\em J. Phys. A: Math.
  Theor.\/} {\bf 47}

\bibitem{opper}
{Opper} M and {Archambeau} C 2009 {\em Neural Comput.\/} {\bf 21} 786--92

\bibitem{opper_ruttor}
{Ruttor} A and {Opper} M 2009 {\em Phys. Rev. Lett.\/} {\bf 103} 230601

\bibitem{kleinert}
{Kleinert} H 2006 {\em Path {Integrals} in {Quantum} {Mechanics}, {Statistics},
  {Polymer} {Physics}, and {Financial} {Markets}\/} 4th ed (World scientific
  publishing)

\end{thebibliography}
\title{Supplementary material}
\maketitle
\appendix
\section{Gaussian Variational Approximation}
\label{GaussVar}
We derive in this appendix the stationarity conditions for the KL divergence that define the Gaussian Variational Approximation (GVA), leading in particular to 
\eqref{meangau0}-\eqref{effdiffusion} in the main text.

We denote $\bm{y}=(\bm{x},-\ii\bm{\hat{x}})$ in such a way that the joint ``distribution'' of $\bm{x}$ and $\hat{\bm{x}}$ can be written
\begin{equation}
P(\bm{y})=\frac{e^{\mathcal{H}(\bm{y})}}{Z}P_0(\bm{x}(0))
\end{equation}
with $P_0(\bm{x}(0))$ being the unknown initial probability distribution. We then proceed with the variational approximation by assuming 
that the approximating distribution is Gaussian
\begin{equation}
P(\bm{y})=\frac{e^{\mathcal{H}(\bm{y})}}{Z}P_0(\bm{x}(0))\approx \mathcal{N}(\bm{y}| \bm{\mu}_{\rm gen}, \bm{C}_{\rm gen})=Q(\bm{y})
\end{equation}
The Gaussian $\mathcal{N}(\bm{y}| \bm{\mu}_{\rm gen}, \bm{C}_{\rm gen})$ is completely determined by two sets of 
parameters, the vector of mean values $\bm{\mu}_{\rm gen}=\langle \bm{y} \rangle_Q$ and the covariance matrix 
$\bm{C}_{\rm gen}=\langle \bm{y} \bm{y}^{T}\rangle_Q - \bm{\mu}_{\rm gen} \bm{\mu}_{\rm gen}^T$.
We define the Kullback-Leibler (KL) Divergence \cite{kullback} between $Q$ and $P$
\begin{equation}
\label{eq:KL}
\begin{split}
\text{KL}(Q||P)&=\int D\bm{y}\, Q(\bm{y})\ln {\frac{Q(\bm{y})}{P(\bm{y})}} =\\
&=\left\langle -\frac{1}{2}(\bm{y}-\bm{\mu}_{\rm gen})^{T}\bm{C}_{\rm gen}^{-1}(\bm{y}-\bm{\mu}_{\rm gen}) 
-\frac{1}{2}\ln{(2\pi)^d \det(\bm{C}_{\rm gen})} + \mathcal{H}(\bm{y}) + \ln{Z}-\ln{P_0(\bm{x}(0))} \right\rangle_{Q}=\\
&=-\frac{d}{2} -\frac{1}{2}\ln{\big[(2\pi)^d \det(\bm{C}_{\rm gen})\big]} + \ln{Z}-\big\langle \mathcal{H}(\bm{y})+\ln{P_0(\bm{x}(0))} \big\rangle_{Q}
\end{split}
\end{equation}
where $d$ is the dimension of the vectors involved, i.e. $d=N(2 T/\Delta+1)$ and $\Delta$ is the elementary time step.
With real measures, $\text{KL}(Q||P)\geq0$ with equality if and only if
$P(\bm{y})\equiv Q(\bm{y})$, so it can be seen as a measure of the dissimilarity 
(a ``distance'') of the distributions $Q(\bm{y})$ and 
$P(\bm{y})$. This leads one to choose the variational parameters by minimizing the KL divergence, or more generally in the case of complex measures to make it stationary
\begin{subequations}
\label{variational}
 \begin{align}
\frac{\partial \text{KL}}{\partial{\bm{\mu}_{\rm gen}}}=&0\label{firstvar} \\
\frac{\partial \text{KL}}{\partial{\bm{C}}_{\rm gen}}=&0\label{secondvar} 
\end{align}
\end{subequations}
Given the KL expression (\ref{eq:KL}), the set of equations \eqref{firstvar} for the components 
of $\bm{\mu}_{\rm gen}$ reduces to
\begin{equation}
\nabla_{\bm{\mu}_{\rm gen}} \left(\big\langle \mathcal{H}(\bm{y})+\ln{P_0(\bm{x}(0))} \big\rangle_{Q}\right)=0 
\end{equation}
while the equation for the inverse correlation matrix \eqref{secondvar} is
\begin{equation}
\label{eq:correlvar}
\frac{1}{2}\nabla_{\bm{C}_{\rm gen}}\ln{(\det\bm{C}_{\rm gen}})+\nabla_{\bm{C}_{\rm gen}}\left\langle 
\mathcal{H}(\bm{y}) \right\rangle_{Q}+\nabla_{\bm{C}_{\rm gen}}\left\langle \ln{P_0(\bm{x}(0))} \right\rangle_{Q}=0
\end{equation}
For the first term in (\ref{eq:correlvar}), one can exploit the properties
\begin{equation}
\nabla_{\bm{C}_{\rm gen}}\ln{(\det\bm{C}_{\rm gen}})=\nabla_{\bm{C}_{\rm gen}}\operatorname{Tr}(\ln{\bm{C}_{\rm gen}})
\end{equation}
\begin{equation}
\frac{\partial\operatorname{Tr}(\ln{\bm{C}_{\rm gen}})}{\partial (\bm{C}_{\rm gen})_{ijtt'}} = (\bm{C}_{\rm gen}^{-1})_{jit't}
\end{equation}
and the following identities (see \cite{opper} for a proof)
\begin{subequations}
\label{identity}
 \begin{align}
\nabla_{\bm{\mu}_{\rm gen}} \left\langle \mathcal{H}(\bm{y}) \right\rangle_{Q}=& \left\langle \nabla_{\bm{y}}\mathcal{H}(\bm{y})\right\rangle_{Q}\label{firstid}\\
\nabla_{\bm{C}_{\rm gen}} \left\langle \mathcal{H}(\bm{y}) \right\rangle_{Q}=&\frac{1}{2}\left\langle \nabla_{\bm{y}}\nabla_{\bm{y}}\mathcal{H}(\bm{y})\right\rangle_{Q}\label{secondid}
\end{align}
\end{subequations}
To apply these, one has to calculate the derivatives of $\mathcal{H}(\bm{x},\bm{\hat{x}})$ with respect 
to the variables $\bm{x}$ and $-\ii\bm{\hat{x}}$
\begin{equation}
\label{eq:deriv1}
\frac{\partial \mathcal{H}(\bm{x},\bm{\hat{x}})}{\partial (\text{i}\hat{x}_i(t))}= x_i(t+\Delta)-x_i(t)-
\Delta\Phi_i(\bm{x}(t)) +\Delta\sum_{j} \text{i}\hat{x}_j(t)\Sigma_{ji}(\bm{x}(t)) 
\end{equation}
\begin{equation}
\label{eq:deriv2}
\begin{split}
\frac{\partial \mathcal{H}(\bm{x},\bm{\hat{x}})}{\partial x_i(t)}&=\text{i}\hat{x}_i(t-\Delta)-\text{i}\hat{x}_i(t)
-\Delta\text{i}\hat{x}_i(t)\frac{\partial \Phi_i(\bm{x}(t))}{\partial x_i(t)}+
\frac{\Delta}{2}\sum_{jk}\text{i}\hat{x}_j(t)\frac{\partial\Sigma_{jk}(\bm{x}(t))}{\partial x_i(t)}\text{i}\hat{x}_k(t)
\end{split}
\end{equation}
with
\be
\frac{\partial \Phi_i(\bm{x}(t))}{\partial x_i(t)} = \sum_{j,l,j \neq l}k^{+}_{ij,l}x_j(t) + \frac{1}{2}\sum_{j,l,j\neq l}k^{-}_{i,jl}+\sum_{j}\lambda_{ij}+\sum_{l}2k^{+}_{ii,l}x_i(t)+\sum_{j}k^{-}_{i,jj}
\ee
For the second derivatives we need a number of different combinations
\begin{subequations}
\label{secondder}
\begin{align}
\frac{\partial}{\partial x_j(t)}\frac{\partial \mathcal{H}(\bm{x},\bm{\hat{x}})}{\partial x_i(t)} &= 
-\Delta\text{i}\hat{x}_i(t)\frac{\partial}{\partial x_j(t)}\frac{\partial \Phi_i(\bm{x}(t))}{\partial x_i(t)}+
\frac{\Delta}{2}\sum_{lk}\text{i}\hat{x}_l(t)\frac{\partial}{\partial x_j(t)}\frac{\partial\Sigma_{lk}(\bm{x}(t))}
{\partial x_i(t)}\text{i}\hat{x}_k(t)\\
\begin{split}
\frac{\partial}{\partial (\text{i}\hat{x}_i(t))}\frac{\partial \mathcal{H}(\bm{x},\bm{\hat{x}})}{\partial x_i(t)} &=
-1-\Delta\frac{\partial \Phi_i(\bm{x}(t))}{\partial x_i(t) }  +
\Delta \sum_{k}\frac{\partial\Sigma_{ik}(\bm{x}(t))}{\partial x_i(t)}\text{i}\hat{x}_k(t)
\end{split}\\
\frac{\partial}{\partial (\text{i}\hat{x}_j(t))}\frac{\partial \mathcal{H}(\bm{x},\bm{\hat{x}})}{\partial
(\text{i}\hat{x}_i(t))} &=  \Delta \Sigma_{ij}(\bm{x}(t))\\
\frac{\partial}{\partial {x}_j(t)}\frac{\partial \mathcal{H}(\bm{x},\bm{\hat{x}})}{\partial (\text{i}\hat{x}_i(t))}&=
-\Delta\frac{\partial \Phi_i(\bm{x}(t))}{\partial x_j(t)} + \Delta\sum_{k}\text{i}
\hat{x}_k(t)\frac{\partial \Sigma_{ki}(\bm{x}(t))}{\partial x_j(t)} \qquad i\neq j\\
\frac{\partial}{\partial {x}_i(t+\Delta)}\frac{\partial \mathcal{H}(\bm{x},\bm{\hat{x}})}{\partial (\text{i}\hat{x}_i(t))} &=
\frac{\partial}{\partial (\text{i}\hat{x}_i(t-\Delta))}\frac{\partial \mathcal{H}(\bm{x},\bm{\hat{x}})}{\partial {x}_i(t)}=1
\end{align}
\end{subequations}
For all other time differences the second derivatives are zero.

We can now obtain the explicit equations for the means. Considering the $\bm{x}$ and $\bm{\hat{x}}$ variables separately, one can cast the equations \eqref{firstvar} 
in the form
\begin{equation}
\label{meaninter}
\left\langle \frac{\partial \mathcal{H}(\bm{x},\bm{\hat{x}})}{\partial x_i(t)}+
\frac{\partial(\ln{P_0(\bm{x}(0)})}{\partial x_i(t)}\right\rangle=0 
\qquad \left\langle \frac{\partial \mathcal{H}(\bm{x},\bm{\hat{x}})}{\partial (\ii\hat{x}_i(t))}+\frac{\partial(\ln{P_0(\bm{x}(0))})}{\partial (\ii\hat{x}_i(t))}\right\rangle=0
\end{equation}
The initial distribution $P_0(\bm{x}(0))$ only contributes in the left equation, for $t=0$. We make the natural choice of assuming it to be Gaussian,
$P_0(\bm{x}(0))= \mathcal{N}(\bm{x}|\bm{\mu}_0, \bm{C}_0)$ so that
\begin{equation}
 \bigg\langle\frac{\partial \ln{P_0}}{\partial x_i(0)}\bigg\rangle= -\sum_j(\bm{C}_0^{-1}(0))_{ij}(\bm{\mu}(0)-\bm{\mu}_0(0))_j
\label{P0_derivative}
\end{equation}
Then given the derivatives (\ref{eq:deriv1}) and (\ref{eq:deriv2}) and carrying out 
the average in \eqref{meaninter} one obtains
\begin{equation}
\label{hatmu}
\text{i}\hat{\mu}_i(t)-\text{i}\hat{\mu}_i(t-\Delta)+\Delta \left\langle\text{i}\hat{x}_i(t)\frac{\partial \Phi_i(\bm{x}(t))}{\partial x_i(t)}\right\rangle
-\frac{\Delta}{2}\bigg\langle\sum_{jk}\text{i}\hat{x}_j(t)\frac{\partial\Sigma_{jk}(\bm{x}(t))}{\partial x_i(t)}\text{i}\hat{x}_k(t)\bigg\rangle=0
\end{equation}
where the term (\ref{P0_derivative}) has to be added on the l.h.s.\ for $t=0$. For the ordinary means, on the other hand,
one has
\begin{equation}
\label{mu}
\mu_i(t+\Delta)-\mu_i(t)-\Delta \langle \Phi_i(\bm{x}(t)) \rangle +\Delta \bigg\langle\sum_{j} \text{i}\hat{x}_j(t)
\Sigma_{ji}(\bm{x}(t))\bigg\rangle=0
\end{equation}
with
\begin{equation}
\begin{split}
&\langle \Phi_i(\bm{x}(t)) \rangle =\Phi_i(\bm{\mu}(t), \bm{C}(t,t))=\sum_{j,l,j \neq l}\left(k^{-}_{l,ij}\mu_l(t)- k^{+}_{ij,l}(\mu_i(t) \mu_j(t)+{C}_{ij}(t,t))\right) +\\
&+\frac{1}{2}\sum_{j,l,j\neq l}\left(k^{+}_{jl,i}( \mu_j(t)\mu_l(t)+{C}_{jl}(t,t))-k^{-}_{i,jl}\mu_i(t)\right)
+\sum_{j}\big(\lambda_{ji}\mu_j(t)-\lambda_{ij}\mu_i(t)\big)+\\
&+\sum_{l}\left(2k^{-}_{l,ii}\mu_l(t)-k^{+}_{ii,l}(\mu_i(t) \mu_i(t)+{C}_{ii}(t,t))\right) + \sum_{j}\left(\frac{1}{2}k^{+}_{jj,i}({C}_{jj}(t,t)+\mu_j(t) \mu_j(t))-k^{-}_{i,jj}\mu_i(t)\right)
\end{split}
\end{equation}
The ``initial'' (boundary) condition for \eqref{hatmu} at $t=T$, given 
by $\langle \partial \mathcal{H}/\partial x_i(T)\rangle=0$, yields $\hat{\mu}_i(T-\Delta)=0$. 
If we also assume that the approximate solution is causal (i.e.\ $R_{ij}(t,t)=-\ii\langle \hat{x}_i(t)x_j(t)\rangle = 0)$ and that $-\langle \hat{x}_i(t)\hat{x}_j(t) \rangle\equiv0$, then the two averages in \eqref{hatmu} vanish term by term in a Taylor expansion in $\bm{x}(t)$, and solving 
backwards in time we have that $\hat{\mu}_i(t)=0$ $\forall t$. At the initial time this then requires $\bm{\mu}(0)=\bm{\mu}_0(0)$ to make also 
the term \eqref{P0_derivative} zero, which is intuitively reasonable.

Note that in the exact trajectory distribution, the above assumptions on averages clearly hold: averages of any product of the $\hat{x}_i(t)$ vanish, as a consequence of the normalization of the overall trajectory distribution (as shown in \cite{fischer}), while the vanishing of the response function $-\ii\langle \hat{x}_i(t)x_j(t') \rangle$ for $t> t'$ follows from causality and, at $t=t'$, from the fact that we are using an It\^o discretization \cite{PathMethods}. It then makes sense that a Gaussian approximation to the trajectory distribution would maintain these properties.

In equation \eqref{mu}, the vanishing of the equal-time response again makes the last term zero.
Rearranging and taking the limit $\Delta \rightarrow 0$ one obtains
\begin{equation}
\frac{d\mu_i(t)}{dt} = \Phi_i(\bm{\mu}(t),\bm{C}(t,t))
\end{equation}
This differs from the deterministic equation of motion for the mean values by additional 
terms stemming from equal time (co-)variances $\bm{C}(t,t)$.

Moving on to the variational equations for the (generalized) covariance matrix, one starts from (\ref{eq:correlvar}) and uses the property \eqref{secondid} as well as the explicit 
second derivatives \eqref{secondder} of $\mathcal{H}(\bm{x},\bm{\hat{x}})$. This shows that 
nonzero inverse correlation elements, for $t\neq 0$, are given by
\begin{subequations}
\label{correl}
 \begin{align}
 (\bm{C}_{\rm gen}^{-1})_{jitt}&= \Delta\ii\hat{\mu}_i(t)\sum_{l}k^{+}_{ij,l}-\frac{\Delta}{2}\bigg
 \langle\sum_{lk}\text{i}\hat{x}_l(t)\frac{\partial}{\partial x_j(t)}\frac{\partial\Sigma_{lk}(\bm{x}(t))}{\partial x_i(t)}
 \text{i}\hat{x}_k(t)\bigg \rangle \qquad i\neq j\label{correlfirst}\\
 (\bm{C}_{\rm gen}^{-1})_{iitt}&=\Delta\ii\hat{\mu}_i(t)\sum_{l}2k^{+}_{ii,l}-
 \frac{\Delta}{2}\bigg\langle\sum_{jk}\text{i}\hat{x}_j(t)\frac{\partial}{\partial x_i(t)}
 \frac{\partial\Sigma_{jk}(\bm{x}(t))}{\partial x_i(t)}\text{i}\hat{x}_k(t)\bigg\rangle\label{correlsecond}\\
 (\bm{C}_{\rm gen}^{-1})_{\hat{\imath}itt}&= -1-\Delta K_{ii}(t)
+\left\langle \Delta \sum_{k}\frac{\partial\Sigma_{ik}(\bm{x}(t))}{\partial x_i(t)}\text{i}\hat{x}_k(t)\right\rangle\\
 (\bm{C}_{\rm gen}^{-1})_{\hat{\imath}\hat{\jmath}tt}&= -\Delta \langle\Sigma_{ij}(\bm{x}(t))\rangle\\
 (\bm{C}_{\rm gen}^{-1})_{j\hat{\imath}tt}&= -\Delta K_{ij}(t)+
 \left\langle \Delta\sum_{k}\text{i}
\hat{x}_k(t)\frac{\partial \Sigma_{ki}(\bm{x}(t))}{\partial x_j(t)}\right\rangle \qquad i\neq j\\
 (\bm{C}_{\rm gen}^{-1})_{i\hat{\imath}t+\Delta t}&=(\bm{C}_{\rm gen}^{-1})_{\hat{\imath}it-\Delta t}=1
 \end{align}
 \end{subequations}
 where the indices $\hat{\imath}$ and $\hat{\jmath}$ are introduced to refer to $\hat{x}_i$ and $\hat{x}_j$. 
 For $t=0$, there would be an extra term on the r.h.s.\ of \eqref{correlfirst} from
\begin{equation}
 \bigg\langle\frac{\partial}{\partial x_j(0)}\frac{\partial \ln{P_0(\bm{x}(0))}}{\partial x_i(0)}\bigg\rangle= -(\bm{C}^{-1}(0))_{ij}
\end{equation}
which just sets the initial covariance matrix to be the exact one.
 We have defined the quantity $K_{ij}(t)$ as follows
 \begin{equation}
 \label{defK}
  K_{ij}(t)=\bigg\langle\frac{\partial \Phi_i(\bm{x}(t))}{\partial x_j(t) } \bigg\rangle
 \end{equation}
From the expressions for the inverse correlation matrix we now want to recover differential equations for the temporal evolution of the 
correlations. This can be done by imposing the inverse relation
\begin{equation}
\label{id_corr}
\sum_{jt''}\big[(\bm{C}_{\rm gen})_{kjtt''}(\bm{C}_{\rm gen})^{-1}_{jit''t'} + 
(\bm{C}_{\rm gen})_{k\hat{\jmath}tt''}(\bm{C}_{\rm gen})^{-1}_{\hat{\jmath}it''t'}\big]=\delta_{ki}\delta_{tt'}
\end{equation}
If we consider explicitly only the nonzero terms in the summation over $t''$, the sum becomes
\begin{eqnarray}
\label{id_corr_new}
&&\sum_{j\neq i}(\bm{C}_{\rm gen})_{kjtt'}(\bm{C}_{\rm gen})^{-1}_{jit't'}+
 (\bm{C}_{\rm gen})_{kitt'}(\bm{C}_{\rm gen})^{-1}_{iit't}+\sum_{j\neq i}(\bm{C}_{\rm gen})_{k\hat{\jmath}tt'}
 (\bm{C}_{\rm gen})^{-1}_{\hat{\jmath}it't}+\notag\\
&&+(\bm{C}_{\rm gen})_{k\hat{\imath}tt'}(\bm{C}_{\rm gen})^{-1}_{\hat{\imath}it't}+
 (\bm{C}_{\rm gen})_{k\hat{\imath}tt'-\Delta}(\bm{C}_{\rm gen})^{-1}_{\hat{\imath}it'-\Delta t'}=\delta_{kitt'}
\end{eqnarray}
As before, we can use the fact that we the solution will be causal and all 
averages of products of auxiliary variables to vanish. Substituting the expressions \eqref{correl} into \eqref{id_corr_new}, dividing by $\Delta$ and taking the limit 
for $\Delta\rightarrow 0$ then yields the differential equations 
\begin{equation}
\label{diffCorrel4}
\frac{\partial R_{ki}(t,t')}{\partial t'} = -\delta_{ki}\delta(t-t')-\sum_{j}R_{kj}(t,t')K_{ji}(t') 
\end{equation}
\begin{equation}
\label{diffCorrel3}
\frac{\partial R_{ki}(t,t')}{\partial t} = \delta_{ki}\delta(t-t') + \sum_{j\neq k}K_{kj}(t)R_{ji}(t,t')
\end{equation}
\begin{equation}
\label{diffCorrel2}
\frac{\partial C_{ki}(t,t')}{\partial t'}=\sum_{j}C_{kj}(t,t')K_{ij}(t')+\sum_{j}R_{kj}(t,t')
\langle\Sigma_{ji}(\bm{x}(t'))\rangle
\end{equation}

As discussed in the main text, the results of the GVA can be understood as an effective linearization of the
Langevin dynamics \eqref{eq:steq} around time dependent means. Let us show in more detail that the equations
\eqref{diffCorrel4}, \eqref{diffCorrel3}, \eqref{diffCorrel2} 
can be mapped into this case. We shall start from the fluctuations about the mean, which obey
\begin{equation}
\label{LanVec2}
\frac{d(\bm{x}(t)-\bm{\mu}(t))}{ dt} = \bm{K}(t)(\bm{x}(t)-\bm{\mu}(t)) + \bm{\xi}(t)
\end{equation}
$\bm{K}(t)$ represents a time dependent rate matrix (as given by the linearization around time-dependent means \eqref{defK}) 
and the noise satisfies $\langle \bm{\xi}(t)\bm{\xi}^{T}(t')\rangle= \langle \bm{\Sigma}(\bm{x}(t))\rangle \delta(t-t')$.
The general solution reads
\begin{equation}
\bm{x}(t)= \bm{\mu}(t)+\bm{R}(t,0)(\bm{x}(0)-\bm{\mu}(0))+\int_0^t ds\,\bm{R}(t,s)\bm{\xi}(s)
\end{equation}
in terms of the response function
\begin{equation}
\label{expon}
\bm{R}(t,t')= \theta(t-t')\,\text{T}\left[e^{\int_{t'}^{t}\bm{K}(t'')dt''}\right]
\end{equation}
Here the time-ordering operator T indicates that in the expansion of the exponential the matrices $\bm{K}(t'')$ are always ordered in order of ascending time from right to left. 
Applying the definition of correlations one then also has 
\begin{equation}
\label{eq:GenCor}
\begin{split}
 \bm{C}(t,t')&= \bm{R}(t,0)\bm{C}(0,0)\bm{R}^T(t',0)
+\int_0^{\min(t,t')} ds\,\bm{R}(t,s)\langle\bm{\Sigma}(\bm{x}(s))\rangle \bm{R}^T(s,t')
\end{split}
 \end{equation}
It is straightforward to check that these response and correlation functions satisfy the equations \eqref{diffCorrel4}, \eqref{diffCorrel3}, \eqref{diffCorrel2} that we had obtained variationally.

\section{Memory and coloured noise in the linearized dynamics}
\label{MemNo}
\subsection{Non-stationary case}
We start from the Langevin equations for concentrations deviations from the means $\delta\bm{x}^{\rm s}(t)$ and $\delta\bm{x}^{\rm b}(t)$ given 
by \eqref{eqdxs} and \eqref{eqdxb} of the main text, obtained via a linearization of the dynamics around possibly time-dependent means $\bm{\mu}^{\rm s}(t)$ and $\bm{\mu}^{\rm b}(t)$.
For simplicity, we set the subnetwork-bulk blocks of the initial conditions and of fluctuations 
to zero, i.e.\ $\bm{C}^{\rm sb}(0,0)=\langle \delta\bm{x}^{\rm s}(0)\delta\bm{x}^{\rm b\, T}(0)\rangle=\mathbb{0}$ (and similarly for $\bm{C}^{\rm bs}(0,0)$), 
$\langle\bm{\Sigma}^{\rm sb}(\bm{x}(t))\rangle=\langle \bm{\xi}^{\rm s}(t)\bm{\xi}^{\rm b\, T}(t)\rangle=\mathbb{0}$ (and similarly for 
$\langle\bm{\Sigma}^{\rm bs}(\bm{x}(t))\rangle$): this assumption can be easily lifted at the price of longer expressions.

We want to read off the structure of the reduced dynamics for $\delta \bm{x}^{\rm s}(t)$, which can be cast as
\begin{equation}
\label{linearreducedA}
\frac{d \delta \bm{x}^{\rm s}(t)}{dt}= \bm{K}^{\rm ss}(t) \delta\bm{x}^{\rm s}(t) + 
\int_0^t\bm{M}^{\rm ss}(t,t')\delta \bm{x}^{\rm s}(t') + \bm{\chi}(t)
\end{equation}
as we discussed in Sec.~\ref{linmem}. The memory kernel $\bm{M}^{\rm ss}(t,t')$ and the covariance of the effective noise $\bm{N}_0^{\rm ss}(t,t')=\langle \bm{\chi}(t)\bm{\chi}(t')^{T}\rangle$
can be derived by direct elimination of the bulk variables from \eqref{eqdxs}, which is equivalent to substituting the solution for $\delta\bm{x}^{\rm b}$, 
\be
\delta\bm{x}^{\rm b}(t)=
\text{T}\left[e^{\int_{0}^t\bm{K}^{\rm bb}(s)ds}\right]
\delta\bm{x}^{\rm b}(0)+ \int_0^t \,\text{T}\left[e^{\int_{t'}^t\bm{K}^{\rm bb}(s)ds}\right]
\left(\bm{K}^{\rm bs}(t')\delta\bm{x}^{\rm s}(t') + \bm{\xi}^{\rm b}(t')\right)dt'
\ee
(where T is the time-ordering operator). In this way one straightforwardly finds the explicit expression for the effective dynamics 
\eqref{linearreducedA}. One reads off that the memory function is given by \eqref{memtemp}, namely
\be
\label{memtempapp}
\bm{M}^{\rm ss}(t,t') = \bm{K}^{\rm sb}(t)\,\text{T}\left[e^{\int_{t'}^t\bm{K}^{\rm bb}(s)ds}\right]\,\bm{K}^{\rm bs} (t')
\ee
The effective noise covariance has the expression 
\begin{eqnarray}
\fl &&\bm{N}_0^{\rm ss}(t,t')= 
 \bm{K}^{\rm sb}(t)\,\text{T}\left[e^{\int_{0}^t\bm{K}^{\rm bb}(s)ds}\right]
 \bm{C}^{\rm bb}(0,0)\,\text{T}\left[e^{\int_{0}^{t'}(\bm{K}^{\rm bb}(s))^{T}ds}\right](\bm{K}^{\rm sb})^{T}(t')
+\langle\bm{\Sigma}^{\rm ss}(\bm{x}(t))\rangle\delta(t-t')
\\ \fl &&
+\bm{K}^{\rm sb}(t)\int_0^{\text{min}(t,t')} dt'' 
 \,\text{T}\left[e^{\int_{t''}^{t}\bm{K}^{\rm bb}(s)ds}\right]
 \langle\bm{\Sigma}^{\rm bb}(\bm{x}(t''))\rangle\,\text{T}\left[e^{\int_{t'}^{t''}(\bm{K}^{\rm bb})^{T}(s)ds}\right]
 (\bm{K}^{\rm sb})^{T}(t'')\notag
\end{eqnarray}
which can be also rewritten as
\begin{equation}
\label{EffNoitem}
\begin{split}
 \bm{N}_0^{\rm ss}(t,t')=& 
 \bm{K}^{\rm sb}(t)\bm{R}^{\rm bb}(t,0)\bm{C}^{\rm bb}(0,0)\bm{R}^{\rm bb\, \it{T}}(t',0)(\bm{K}^{\rm sb})^{T}(t')+\langle\bm{\Sigma}^{\rm ss}(\bm{x}(t))\rangle\delta(t-t')\\
 &+\bm{K}^{\rm sb}(t)\int_0^{\text{min}(t,t')} dt'' 
 \bm{R}^{\rm bb}(t,t'') \langle\bm{\Sigma}^{\rm bb}(\bm{x}(t''))\rangle\bm{R}^{\rm bb\, \it{T}}(t',t'')(\bm{K}^{\rm sb})^{T}(t'')
 \end{split}
\end{equation}
by using the definition \eqref{expon} for the marginal bulk responses.

\subsection{Stationary case}

If we consider a dynamics linearized around stationary means (with the diffusion matrix and the dynamical matrix both
evaluated then at steady state, i.e.\ $\langle\bm{\Sigma}(\bm{x}(t))\rangle\equiv\bm{\Sigma}$ and $\bm{K}(t)\equiv \bm{K}$), the memory \eqref{memtempapp} becomes
\begin{equation}
\label{memlinapp}
\bm{M}^{\rm ss}(t,t') = \bm{K}^{\rm sb} e^{\bm{K}^{\rm bb}(t-t')}\bm{K}^{\rm bs} 
\end{equation}
(precisely expression \eqref{memlin}), while the covariance \eqref{EffNoitem} of the noise in the subnetwork simplifies to 
\begin{equation}
\label{EffNoi}
\begin{split}
 \bm{N}_0^{\rm ss}(t,t') &=
 \bm{\Sigma}^{\rm ss}\delta(t-t')+\bm{K}^{\rm sb}e^{\bm{K}^{\rm bb}t}
 \bm{C}^{\rm bb}(0,0)e^{(\bm{K}^{\rm bb})^{T}t'}(\bm{K}^{\rm sb})^{T}\\
 &+\bm{K}^{\rm sb}\int_0^{\text{min}(t,t')} dt'' e^{\bm{K}^{\rm bb}(t-t'')}
 \bm{\Sigma}^{\rm bb}e^{(\bm{K}^{\rm bb})^{T}(t'-t'')}(\bm{K}^{\rm sb})^{T}
 \end{split}
\end{equation}

\section{Nonlinear corrections by perturbation theory}
\label{pert}
We give details here of the effective action including the first order nonlinear corrections to the Gaussian approximation.
The effective action can be treated as a field theoretic free energy and can be expressed perturbatively \cite{kleinert}. Let us decompose the action into a ``non-interacting'' part (containing only terms quadratic in all the variables) and 
an interacting one (containing higher powers), which can be seen as a small perturbation
\begin{equation}
 \mathcal{H}=\mathcal{H}_0+\Delta\mathcal{H}
\end{equation}
The expression for the effective action \eqref{EffAct}, applying the idea of a perturbative expansion, becomes
\begin{eqnarray}
\label{effactpert}
&&\mathcal{H}_{\text{eff}}=\ln {\int D\bm{x}^{\rm b}D\bm{\hat{x}}^{\rm b} e^{\mathcal{H}_0 + \Delta \mathcal{H}}}= \\
&&=\ln {\int D\bm{x}^{\rm b}D\bm{\hat{x}}^{\rm b} e^{\mathcal{H}_0}\left(1+\frac{\int D\bm{x}^{\rm b}D\bm{\hat{x}}^{\rm b} 
e^{\mathcal{H}_0}\Delta \mathcal{H}}{\int D\bm{x}^{\rm b}D\bm{\hat{x}}^{\rm b} e^{\mathcal{H}_0}}+ 
\mathcal{O}(\Delta \mathcal{H})^2\right)}=\notag\\
&&=\ln{\int D\bm{x}^{\rm b}D\bm{\hat{x}}^{\rm b} e^{\mathcal{H}_0}+\int D\bm{x}^{\rm b}D\bm{\hat{x}}^{\rm b}
Q_0(\bm{x}^{\rm b},\bm{\hat{x}}^{\rm b}|\bm{x}^{\rm s},\bm{\hat{x}}^{\rm s})\Delta \mathcal{H}+
\mathcal{O}(\Delta \mathcal{H})^2}\notag
\end{eqnarray}
where $e^{\mathcal{H}_0}=Q_0(\bm{x}^{\rm b},\bm{\hat{x}}^{\rm b},\bm{x}^{\rm s},\bm{\hat{x}}^{\rm s})$ and 
$\int  D\bm{x}^{\rm b}D\bm{\hat{x}}^{\rm b} e^{\mathcal{H}_0}=Q_0(\bm{x}^{\rm s},\bm{\hat{x}}^{\rm s})$,
therefore $Q_0(\bm{x}^{\rm b},\bm{\hat{x}}^{\rm b}|\bm{x}^{\rm s},\bm{\hat{x}}^{\rm s}) =
e^{\mathcal{H}_0}/\int D\bm{x}^{\rm b}D\bm{\hat{x}}^{\rm b} e^{\mathcal{H}_0}$. The reference distribution for the perturbative
expansion $Q_0(\bm{x}^{\rm b},\bm{\hat{x}}^{\rm b}|\bm{x}^{\rm s},\bm{\hat{x}}^{\rm s})$ is chosen as the LNA, i.e.\ 
a Gaussian conditional distribution centered around the deterministic steady states.
To define the effective action we fix 
subnetwork variables, i.e.\ condition on them, and need the corresponding conditional bulk probabilities. The perturbative corrections arise from cubic terms in the action, which explicitly read
\begin{equation}
\label{deltah}
\begin{split}
\Delta \mathcal{H} = \int_0^{T} dt\Bigg\lbrace&\sum_{i}\text{i}\hat{x}_i(t) \Big[\sum_{j,l,j \neq l} k^{+}_{ij,l}
\delta x_i(t) \delta x_j(t)-\frac{1}{2}\sum_{j,l,j\neq l}k^{+}_{jl,i}\delta x_j(t)\delta x_l(t) + \\
&\sum_{l}k^{+}_{ii,l}\delta x_i(t) \delta x_i(t) -\frac{1}{2}\sum_{j}k^{+}_{jj,i}\delta x_j(t)\delta x_j(t)\Big]\Bigg\rbrace
\end{split}
\end{equation}
The indices can belong to bulk or subnetwork, giving different combinations; we introduce a vectorial notation
\begin{eqnarray}
\label{deltaH}
&&-\Delta\mathcal{H}= \int_0^{T} dt 
\bigg\lbrace\text{i}\hat{\bm{x}}^{\rm s\,\it{T}}\bigg[\bm{K}^{\rm  s, ss}\big(\delta\bm{x}^{\rm s}\circ\delta\bm{x}^{\rm s}\big)+ 
\bm{K}^{\rm  s,sb}\big(\delta\bm{x}^{\rm s}\circ\delta\bm{x}^{\rm b}\big) +\bm{K}^{\rm  s, bb}\big(
\delta\bm{x}^{\rm b}\circ\delta\bm{x}^{\rm b}\big)\bigg]+\notag\\
&&\qquad\qquad\quad\text{i}\hat{\bm{x}}^{\rm b\,\it{T}}
\bigg[\bm{K}^{\rm  b, ss}\big(\delta\bm{x}^{\rm s}\circ\delta\bm{x}^{\rm s}\big)+
\bm{K}^{\rm  b, sb}\big(\delta\bm{x}^{\rm s}\circ\delta\bm{x}^{\rm b}\big)+
\bm{K}^{\rm  b,bb}\big(\delta\bm{x}^{\rm b}\circ\delta\bm{x}^{\rm b}\big)\bigg]\bigg\rbrace
\end{eqnarray}
where $\circ$ indicates a ``flattened'' outer product (see main text) and 
$\bm{K}^{\rm s,sb}$, $\bm{K}^{\rm b,sb}$, $\bm{K}^{\rm b,bb}, \bm{K}^{\rm  b, ss}, \bm{K}^{\rm  s, bb}$ are matrices collecting the relevant nonlinear couplings.
As explained in the main text, we do not consider all the possible reactions occurring between subnetwork and bulk,
thus we set $\bm{K}^{\rm  b, ss}=\bm{K}^{\rm  s, bb}\equiv 0$ (though these would be easy to re-instate). We have 
also already dropped $\delta$ from $\hat{\bm{x}}^{\rm s}$ and $\hat{\bm{x}}^{\rm b}$ as
the marginal auxiliary means satisfy $\bm{\hat{\mu}}^{\rm s}=\bm{\hat{\mu}}^{\rm b} \equiv 0$. 

The first order correction (as clear from the second term in \eqref{effactpert})
reduces to an average w.r.t.\ the conditional distribution $Q_0(\bm{x}^{\rm b},\bm{\hat{x}}^{\rm b}|\bm{x}^{\rm s},\bm{\hat{x}}^{\rm s})$. This average
is equivalent to replacing combinations of bulk variables in \eqref{deltah} by conditional Gaussian moments which, by Wick's theorem,
can be expressed in terms of means and correlations
\begin{eqnarray}
\label{dsCompact}
 -\langle\Delta \mathcal{H}\rangle &=&
 \int_0^{T} dt \bigg\lbrace\text{i}\hat{\bm{x}}^{\rm s \,\it{T}}
 \bigg[\bm{K}^{\rm  s, ss}\big(\delta\bm{x}^{\rm s}\circ\delta\bm{x}^{\rm s}\big)
 + \bm{K}^{\rm  s,sb}\big(\delta\bm{x}^{\rm s}\circ\delta\bm{\mu}^{\rm b|s}\big)\bigg]+\\
&&\quad \quad \quad\quad\text{i}\hat{\bm{\mu}}^{\rm b|s \,\it{T}}
\bigg[\bm{K}^{\rm  b, sb}\big(\delta\bm{x}^{\rm s}\circ\delta\bm{\mu}^{\rm b|s}\big)+
\bm{K}^{\rm  b,bb}\big(\delta\bm{\mu}^{\rm b|s}\circ\delta\bm{\mu}^{\rm b|s}+\bm{C}^{\rm bb|s}(t,t)\big)\bigg]\bigg\rbrace\notag
\end{eqnarray}
where one uses that the conditional average $\langle\delta \bm{x}^{\rm b}(t)\delta \bm{x}^{\rm b\, \it{T}}(t)\rangle=
\delta \bm{\mu}^{\rm b|s}(t)\delta \bm{\mu}^{\rm b|s\, \it{T}}(t)+\bm{C}^{\rm bb|s}(t,t)$.
In developing the Wick's theorem for moments of the schematic form $\langle \bm{\hat{x}}^{\rm b} \delta \bm{x}^{\rm b} \delta \bm{x}^{\rm b}\rangle$ 
or $\langle \bm{\hat{x}}^{\rm b} \delta \bm{x}^{\rm b} \delta \bm{x}^{\rm s}\rangle$
we have used that the conditional equal time response functions vanish, $\bm{R}^{\rm bb|s}(t,t)=0$. We next substitute the expression for $\text{i}\hat{\bm{\mu}}^{\rm b|s}$ \eqref{condaux} 
\begin{eqnarray}
\label{dsCompact0}
\fl  &&-\langle\Delta \mathcal{H}\rangle =
 \int_0^{T} dt \,\text{i}\hat{\bm{x}}^{\rm s \,\it{T}}(t)
 \bigg[\bm{K}^{\rm  s, ss}\big(\delta\bm{x}^{\rm s}(t)\circ\delta\bm{x}^{\rm s}(t)\big)
 + \bm{K}^{\rm  s,sb}\big(\delta\bm{x}^{\rm s}(t)\circ\delta\bm{\mu}^{\rm b|s}(t)\big)\bigg]+\\
\fl && \int_0^{T} dt'\int_{t'}^{T} dt\, \text{i}\hat{\bm{x}}^{\rm s\,\it{T}}(t)\bm{K}^{\rm sb}e^{\bm{K}^{\rm bb}(t-t')}
\bigg[\bm{K}^{\rm  b, sb}\big(\delta\bm{x}^{\rm s}(t')\circ\delta\bm{\mu}^{\rm b|s}(t')\big)+
\bm{K}^{\rm  b,bb}\big(\delta\bm{\mu}^{\rm b|s}(t')\circ\delta\bm{\mu}^{\rm b|s}(t')+
\bm{C}^{\rm bb|s}(t',t')\big)\bigg]\notag
\end{eqnarray}
We have re-instated the explicit time dependences here, which are important for our derivation of the memory functions. One can extract the nonlinear reduced dynamics of $\delta\bm{x}^{\rm s}(t)$ from this action by considering the terms multiplying $\text{i}\hat{\bm{x}}^{\rm s}(t)$ ($\bm{C}^{\rm bb|s}(t',t')$ gives an additional deterministic
contribution that we will analyze in \ref{projvsgva}). 
First, we note that $\bm{K}^{\rm s, ss}$ appearing in the first line of \eqref{dsCompact0} gives the nonlinear subnetwork dynamics, which is local in time.
The reduced dynamics for $\delta\bm{x}^{\rm s}(t)$ includes a memory
$\bm{M}^{\rm ss}(t,t')$ and a coloured noise $\bm{\chi}(t)$ with covariance 
\be
\langle\bm{\chi}(t)\bm{\chi}^{T}(t')\rangle=\bm{N}^{\rm ss}(t,t')
\ee
to be determined. One can write $\bm{N}^{\rm ss}(t,t')=\bm{N}^{\rm ss}_{0}(t,t')+ \bm{N}^{\rm ss}_{1}(t,t')$ as the sum of a
purely Gaussian term and a first order correction. Similarly the memory comprises two terms: $\bm{M}^{\rm ss}(t-t')$, which like
$\bm{N}^{\rm ss}_{0}(t,t')$ can be calculated starting from the quadratic part of the action (as shown in \ref{MemNo}) and $\bm{M}^{\rm s,ss}(t,t',t'')$, which like 
$\bm{N}^{\rm ss}_{1}(t,t')$ contains the contributions of cubic terms treated by
means of perturbation theory. To evaluate these first order corrections, we insert into \eqref{dsCompact0} the expressions for 
$\bm{\mu}^{\rm b|s}(t)$ \eqref{bulkmean}: to illustrate the procedure we focus on the first term in the second line of \eqref{dsCompact0}, which becomes
\begin{eqnarray}
\label{dsCompact1}
\fl  &&\int_0^Tdt'\int_{t'}^{T} dt\, \text{i}\hat{\bm{x}}^{\rm s\,\it{T}}(t)\bm{K}^{\rm sb}e^{\bm{K}^{\rm bb}(t-t')}
\bm{K}^{\rm  b, sb}\big(\delta\bm{x}^{\rm s}(t')\circ\delta\bm{\mu}^{\rm b|s}(t')\big)=\\
\fl  \qquad&&\int_0^Tdt'\int_{t'}^{T} dt\, \text{i}\hat{\bm{x}}^{\rm s\,\it{T}}(t)\bm{K}^{\rm sb}e^{\bm{K}^{\rm bb}(t-t')}
\bm{K}^{\rm  b, sb}\bigg(\delta\bm{x}^{\rm s}(t')\circ \int_0^{t'} dt'' e^{\bm{K}^{\rm bb}(t'-t'')}\bm{K}^{\rm bs}\delta\bm{x}^{\rm s}(t'')\notag\\
&&\qquad \qquad \qquad\qquad \qquad \qquad \qquad -\delta\bm{x}^{\rm s}(t')\circ \int_0^{T}dt''\bm{C}^{\rm bb|s}(t',t'')(\bm{K}^{\rm sb})^{T}\text{i}\bm{\hat{x}}^{\rm s}(t'')\bigg)\notag
\end{eqnarray}
The first of the resulting two terms contributes to the reduced subnetwork dynamics via a temporal integral defining a nonlinear memory term given by
\be
\label{memfirst}
\int_0^tdt'\int_{0}^{t'} dt'' \bm{K}^{\rm sb} e^{\bm{K}^{\rm bb}(t-t')}\bm{K}^{\rm b,sb}\big(\delta \bm{x}^{\rm s}(t')\circ 
 e^{\bm{K}^{\rm bb}(t'-t'')}\bm{K}^{\rm bs}\delta \bm{x}^{\rm s}(t'')\big)
\ee
This describes how \emph{products} of subnetwork concentrations feed back into the evolution of a single concentration. More generally, one can 
see that terms $\sim \bm{\hat{x}}^{\rm s}\delta\bm{x}^{\rm s}\delta\bm{x}^{\rm s}$ define the nonlinear 
memory function. Extracting all of these terms from \eqref{dsCompact0} gives the result \eqref{eq:effMemoryNLtimes} stated in the main text 
(where the kernel of \eqref{memfirst} appears in the second line).
In the second term of \eqref{dsCompact1}, the factor multiplying $\text{i}\hat{\bm{x}}^{\rm s\,\it{T}}(t)$ and $\text{i}\hat{\bm{x}}^{\rm s}(t'')$ 
\be
\int_0^t dt' \bm{K}^{\rm sb} e^{\bm{K}^{\rm bb}(t-t')}\bm{K}^{\rm b,sb}
\big(\delta\bm{x}^{\rm s}(t')\circ\bm{C}^{\rm bb|s}(t',t'')(\bm{K}^{\rm sb})^{T}\big)
\ee
contributes to the covariance of the effective noise (which always enters the MSRJD formalism via quadratic terms in the auxiliary variables, as is clear from \eqref{action}) 
with an additional $\bm{x}^{\rm s}$-dependence.
Treating the other terms in \eqref{dsCompact0} in the same way and dropping third order terms $\sim \bm{\hat{x}}^{\rm s}\bm{\hat{x}}^{\rm s}\bm{\hat{x}}^{\rm s}$ that 
encode non-vanishing higher cumulants of the noise distribution one obtains 
the perturbative contribution to the effective noise covariance, given by all terms of the form $\sim \bm{\hat{x}}^{\rm s}\bm{\hat{x}}^{\rm s}\delta\bm{x}^{\rm s}$; explicitly
\begin{eqnarray}
\label{covNL}
\fl \bm{N}_1^{\rm ss}(t,t'')&=&\bm{K}^{\rm s, sb}
\big(\delta\bm{x}^{\rm s}(t)\circ \bm{C}^{\rm bb|s}(t,t'')(\bm{K}^{\rm sb})^{T}\big)+
\int_0^t dt' \bm{K}^{\rm sb} e^{\bm{K}^{\rm bb}(t-t')}\bm{K}^{\rm b,sb}
\big(\delta\bm{x}^{\rm s}(t')\circ\bm{C}^{\rm bb|s}(t',t'')(\bm{K}^{\rm sb})^{T}\big)\notag\\
\fl &+&\int_0^t ds\int_0^{s}dt'\bm{K}^{\rm sb} e^{\bm{K}^{\rm bb}(t-s)}\bigg[\bm{K}^{\rm b,bb}\big(e^{\bm{K}^{\rm bb}(s-t')}\bm{K}^{\rm bs}
\delta\bm{x}^{\rm s}(t')\circ\bm{C}^{\rm bb|s}(s,t'')(\bm{K}^{\rm sb})^{T}\big)+\\
\fl &&\qquad \qquad \qquad \qquad \qquad
\bm{K}^{\rm b,bb}\big(\bm{C}^{\rm bb|s}(s,t'')(\bm{K}^{\rm sb})^{T} \circ
e^{\bm{K}^{\rm bb}(s-t')}\bm{K}^{\rm bs}\delta\bm{x}^{\rm s}(t')\big)\bigg]+ \ldots^T \notag
\end{eqnarray}
where $\ldots^T$ indicates that to each written term its transpose has to be added, as is required to make $\bm{N}_1^{\rm ss}(t,t'')$ symmetric. 
We have extended our circle product notation here to products of vectors and matrices, which are defined with the same ordering restrictions as the original product. 
For example, in the second line of \eqref{covNL} we have a term of the form
$\bm{K}^{\rm b,bb}(\bm{v}^{\rm b} \circ \bm{A}^{\rm bs})$ where $\bm{v}^{\rm b}$ is a vector and $\bm{A}^{\rm bs}$ a matrix. 
This term is to be read as a matrix with $ij$-element $\sum_{k \leq l} K_{i,kl}\big(v_k A_{lj}\big)$.
For the sake of completeness, one should also include corrections stemming from the fact 
that the white noise covariance $\bm{\Sigma}(\bm{x})$ as given by \eqref{fluct} \emph{depends} on concentrations.
We defer an analysis of these to a dedicated future paper on stochastic effects in subnetwork modelling.

\section{Effective equations solver}
\label{eff_solver}
It turns out that the numerical solution of the \emph{integro-differential} equations  \eqref{nonlingva} can be simplified by mapping them to a system of ordinary \emph{differential} equations. The idea here is that every memory integral term can be seen as the 
solution of a differential equation. 
Let us start from the subnetwork reduced dynamics in the form
\be
\label{eqsub}
\frac{d \delta\bm{x}^{\rm s}(t)}{dt}  = \bm{K}^{\rm ss}\delta\bm{x}^{\rm s}(t) + 
\bm{K}^{\rm s,ss}\big(\delta\bm{x}^{\rm s}(t)\circ\delta\bm{x}^{\rm s}(t))
+\mathcal{\bm{M}}(t)
\ee
where we have taken the large volume limit in order to neglect the intrinsic noise.
By stepping back through the derivation of the memory vector $\mathcal{\bm{M}}(t)$ in terms of bulk conditional means
$\delta\bm{\mu}^{\rm b|s}(t)= \bm{\p}(t)+\bm{\hat{\p}}(t)$, we can rewrite it as in \eqref{memfullcond}, i.e.\
\be
\label{intimpl}
\begin{split}
\mathcal{\bm{M}}(t) = &\bm{K}^{\rm sb}\bm{\p}(t)+ \bm{K}^{\rm s,sb}\big(\delta\bm{x}^{\rm s}(t)\circ\bm{\p}(t)\big)+\\
&\int_0^t dt' \bm{K}^{\rm sb} e^{\bm{K}^{\rm bb}(t-t')}\left[\bm{K}^{\rm b,sb}\big(\delta\bm{x}^{\rm s}(t')\circ\bm{\p}(t')\big)+ 
\bm{K}^{\rm b,bb}\big( \bm{\p}(t')\circ\bm{\p}(t')\big)\right]
\end{split}
\ee 
with $\bm{\p}(t')= \int_{0}^{t'} dt'' \big[e^{\bm{K}^{\rm bb}(t'-t'')}\bm{K}^{\rm bs}\big]\delta \bm{x}^{\rm s}(t'')$,
a vector of size $N^{\rm b}$ (number of bulk species). The $\bm{\p}(t)$ can be obtained equivalently by solving the $N^{\rm b}$ \emph{linear} differential equations
\be
\label{eqnu}
\frac{d\bm{\p}(t)}{dt}= \bm{K}^{\rm bb} \bm{\p}(t) + \bm{K}^{\rm bs} \delta\bm{x}^{\rm s}(t)
\ee
The time integral left in the last term of \eqref{intimpl} appears as a result of the insertion of $\bm{\hat{\mu}}^{\rm b|s}(t)$ in the effective action. 
This term can be rearranged and evaluated by introducing \emph{ad hoc} auxiliary variables. 
We need to decompose the exponential kernel $e^{\bm{K}^{\rm bb}(t-t')}$ into a superposition of pure exponentials (by diagonalizing $\bm{K}^{\rm bb}$)
\be
\begin{split}
&\int_0^t dt' \bm{K}^{\rm sb} e^{\bm{K}^{\rm bb}(t-t')}\left[\bm{K}^{\rm b,sb}\big(\delta\bm{x}^{\rm s}(t')\circ\bm{\p}(t')\big)+ 
\bm{K}^{\rm b,bb}\big( \bm{\p}(t')\circ\bm{\p}(t')\big)\right]=\\
&=\int_0^t dt' \bm{K}^{\rm sb}\sum_{c=1}^{N^{\rm b}} \bm{r}^{c}\bm{l}^{c\, \it{T}}\,e^{\lambda_{c}(t-t')}
\left[\bm{K}^{\rm b,sb}\big(\delta\bm{x}^{\rm s}(t')\circ\bm{\p}(t')\big)+ 
\bm{K}^{\rm b,bb}\big( \bm{\p}(t')\circ\bm{\p}(t')\big)\right]
\end{split} 
 \ee 
 where the $\lambda_{c}$ are the $N^{\rm b}$ eigenvalues of $\bm{K}^{\rm bb}$ (with negative real part to ensure stability),
 $\bm{r}^{c}$ and $\bm{l}^{c}$ are respectively the right and left eigenvectors of $\bm{K}^{\rm bb}$, playing the role of 
 the coefficients of this decomposition. We consider a vector $\bm{z}(t)=\lbrace z^c(t) \rbrace$, $c=1,...,N^{\rm b}$ whose
 components, one for each eigenvalue $\lambda_{c}$, satisfy
\be
\label{eqz}
\frac{d z^{c}(t)}{dt} = \lambda_{c} \,z^{c}(t) + \bm{l}^{c\, \it{T}}\left[\bm{K}^{\rm b,sb}\big(\delta\bm{x}^{\rm s}(t)\circ\bm{\p}(t)\big)+ 
\bm{K}^{\rm b,bb}\big( \bm{\p}(t)\circ\bm{\p}(t)\big)\right] \qquad z^{c}(0)=0
\ee 
Then \eqref{eqsub} can be translated into
\be
\frac{d \delta\bm{x}^{\rm s}(t)}{dt} = \bm{K}^{\rm ss}\delta\bm{x}^{\rm s}(t) + \bm{K}^{\rm s,ss}\big(\delta\bm{x}^{\rm s}(t)
\circ\delta\bm{x}^{\rm s}(t)\big)+
\bm{K}^{\rm sb}\bm{\p}(t)+ \bm{K}^{\rm s,sb}\big(\delta\bm{x}^{\rm s}(t)\circ\bm{\p}(t)\big) + \bm{K}^{\rm sb}\sum_{c=1}^{N^{\rm b}}\bm{r}^{c}z^{c}(t) 
\ee 
To summarize, solving the $N^{\rm s}$ subnetwork equations with integral memory terms is equivalent to solving 
a system with $2 N^{\rm b}$ additional equations, the $N^{\rm b}$ ones describing $\bm{\p}(t)$ (see \eqref{eqnu}) and 
the $N^{\rm b}$ ones for $\bm{z}(t)$ (see \eqref{eqz}). For projection methods, the additional variables one needs to introduce in order to
express memory integrals via differential equations is given by the
dimension of the bulk variable space (see \cite{katy}) including all the possible concentration products, and hence scales 
as $(N^{\rm b}\times N^{\rm b})+(N^{\rm s}\times N^{\rm b})$.

\section{Projection methods vs GVA}
\label{projvsgva}
The application of projection methods to protein-protein interaction networks has been studied by Rubin et al. in \cite{katy}:
here we summarize the basic aspects needed to develop a comparison with the GVA. 

One is interested in deriving a closed-form expression for the memory function and the random force
both for the linearized and the fully nonlinear dynamics: in the second case, one has to appeal to the limit of vanishing noise.
As the variance of copy number fluctuations scales as $\epsilon=1/V$, i.e.\ the inverse of the reaction volume, the contribution of the noise becomes negligible for $\epsilon \rightarrow 0$, i.e.\ 
for suitably large reaction volumes (see Sec.~\ref{set_up} in the main text). While one needs a nonzero $\epsilon$ for initially applying 
the Zwanzig-Mori formalism, the authors always then take the small $\epsilon$ limit. Let us stress that the small $\epsilon$ limit 
is not necessary for the linear dynamics, as the noise drops out from the equations for conditionally averaged 
concentrations whatever the value of $\epsilon$.

In addition, to evaluate the memory and the random force from projection operators, one needs the steady state distribution for the
deviations $\delta \bm{x}$. If the noise is negligible ($\epsilon \rightarrow 0$), $\delta \bm{x}$ is small and one can find 
the steady state distribution by linearizing the Langevin equation around $\delta \bm{x}=0$; this gives a Gaussian
steady state distribution of $\delta \bm{x}$ with covariance satisfying the Lyapunov equation. 
The structure of the covariance is not unique, but fixed 
by the type of fluctuations. The choice of independent Poisson fluctuations for each species, thus
a covariance with a diagonal structure, produces the simplest projected equations. These describe the evolution for the subnetwork 
conditionally on the available knowledge of the initial conditions which are
specified, for the bulk, via some probability distribution. The solution is the mean trajectory w.r.t.\ this initial distribution, from which stochastic 
fluctuations are thus averaged out: we will use for it the same notation as for the single instances $\bm{x}^{\rm s}(t)$ as they
coincide in the limit $\epsilon \rightarrow 0$.
As a consequence, we expect that the equations of motion given by projection methods for such conditionally averaged concentrations 
agree with the noise averaged subnetwork equations in the GVA. We next look closely at this comparison.

\subsection{Linearized projected equations}
\label{projlin}
The starting point is the Langevin dynamics \eqref{eq:steq} and its corresponding Fokker-Planck equation, 
encoded by $\mathcal{L}$, the so-called adjoint Fokker-Planck operator. Referring to \cite{katy} for the entire
derivation, we shall directly provide the final form of the linearized dynamics
\begin{equation}
 \frac{d\delta x_i(t)}{dt}=\sum_{j=1}^{N^{\rm s}}\delta x_j(t)\bm{\Omega}^{\rm ss}_{ji}+ 
 \sum_{j=1}^{N^{\rm s}}\int_0^t dt'\delta x_j(t')\bm{M}^{\rm ss}_{ji}(t-t')+r_i(t)
\end{equation}
This is obtained by applying two operators projecting either onto the subspace of subnetwork d.o.f.\ (degrees of freedom) or onto the orthogonal one; in the linearized
dynamics, it is found that they can be represented by matrices with a simple block structure, $\bm{P}$ and $\bm{Q}$ respectively
 \[\bm{P}=\begin{pmatrix}
\mathbb{1}   & \mathbb{0} \\
\mathbb{0}   & \mathbb{0} \\
\end{pmatrix}\]
 \[\bm{Q}=\begin{pmatrix}
\mathbb{0}   & \mathbb{0} \\
\mathbb{0}   & \mathbb{1} \\
\end{pmatrix}\]
The adjoint Fokker-Planck operator $\mathcal{L}$ can also be cast in matrix form and is then
denoted as $\bm{L}$. Note that $\bm{L}=\bm{K}^{T}$, i.e.\ it is equivalent to the transpose of the dynamical matrix of the GVA; like $\bm{K}$ 
it can be partitioned into blocks for the bulk and the subnetwork part as
\[\bm{L}=\begin{pmatrix}
\bm{L}^{\rm ss}   & \bm{L}^{\rm sb} \\
\bm{L}^{\rm bs}   & \bm{L}^{\rm bb} \\
\end{pmatrix}\] 
Exploiting this correspondence between operators and matrices one has for the linear dynamics:
\begin{itemize}
\item $\bm{\Omega}$ is the top left block (related to subnetwork variables) of $\bm{PL}$, thus $\bm{\Omega}=\bm{L}^{\rm ss}$
\item $\bm{M}(t-t')$ is the top left block of $\bm{P}\bm{L}\bm{Q} e^{\bm{Q}\bm{L}\bm{Q}(t-t')} \bm{Q}\bm{L}$, i.e.\
\be
\label{memlin0}
\bm{M}(t-t')=\bm{L}^{\rm sb} e^{\bm{L}^{\rm bb}(t-t')}\bm{L}^{\rm bs}
\ee
\item The random force is given by the $\rm s$-entries of $\delta \bm{x}^{\rm b\, \it{T}}(0) e^{\bm{Q}\bm{L}\bm{Q} t} \bm{Q}\bm{L}$, i.e.\ 
\be
\label{rf0}
\bm{r}^{T}_0(t)=\delta \bm{x}^{\rm b\, \it{T}}(0)e^{\bm{L}^{\rm bb}t}\bm{L}^{\rm bs}
\ee
\end{itemize}
Note that with the index conventions used in the projection method~\cite{katy}, the memory function above is the matrix \emph{transpose} of the one in our GVA approach. To avoid introducing more notation, we use the same symbol $\bm{M}$ nonetheless. 

\subsection{Nonlinear projected equations}
\label{nonlinproj}
For the nonlinear projected equations one again starts from suitable matrix representations of the operators involved.
Note that in the projection method, even once a subnetwork has been selected, there is freedom in choosing the space of observables onto which to project.
Rubin et al.~show \cite{katy} that the best option is to project onto concentrations and products of concentrations.

The nonlinearity is represented as a linear coupling between a concentration and products of two concentrations; 
the projected equations have the form
\begin{equation}
\label{nonlinprojeq}
\begin{split}
\frac{d\delta x_i(t)}{dt}&=\sum_{j=1}^{N^{\rm s}}\delta x_j(t)\bm{\Omega}^{\rm ss}_{ji}+
\sum_{1\leq j\leq k\leq N^{\rm s}}\delta x_j(t)\delta x_k(t)\bm{\Omega}^{\rm ss,s}_{(jk),i}+
\int_0^t dt'\bigg( \sum_{j=1}^{N^{\rm s}}\delta x_j(t')\bm{M}^{\rm ss}_{ji}(t-t')\\
&+\sum_{1\leq j\leq k\leq N^{\rm s}}\delta x_j(t')\delta x_k(t')
\bm{M}^{\rm ss,s}_{(jk),i}(t-t')\bigg)+r_i(t)
\end{split}
\end{equation}
Focussing on an observable $z_{\alpha}$ (summarizing both simple concentrations and products),
one can write the adjoint Fokker-Planck operator $\mathcal{L}$ in a matrix form such that
\be
\partial_t z_{\alpha} = \sum_{\beta} z_{\beta} L_{\beta \alpha} + \delta x^3 + \mathcal{O}(\epsilon) 
\ee
$\delta x^3$ represents cubic terms, which are not captured at this order of accuracy,
while $\mathcal{O}(\epsilon)$-terms vanish in the small $\epsilon$ limit. The matrix representation therefore mirrors the choice of an enlarged space containing also products, and it is
valid for small $\epsilon$. It reads explicitly as follows
\[\bm{L}=\begin{pmatrix}
\bm{L}^{\rm ss} & \bm{L}^{\rm sb}  & \mathbb{0}  & \mathbb{0}  & \mathbb{0}  \\
\bm{L}^{\rm bs} & \bm{L}^{\rm bb}  & \mathbb{0}  & \mathbb{0}  & \mathbb{0}  \\
\bm{L}^{\rm ss,s} & \bm{L}^{\rm ss,b}  & \bm{L}^{\rm ss,\rm ss}  & \bm{L}^{\rm ss,\rm sb}  & \bm{L}^{\rm ss,\rm bb}  \\
\bm{L}^{\rm sb,s} & \bm{L}^{\rm sb,b}  & \bm{L}^{\rm sb,\rm ss}  & \bm{L}^{\rm sb,\rm sb}  & \bm{L}^{\rm sb,\rm bb}  \\
\bm{L}^{\rm bb,s} & \bm{L}^{\rm bb,b}  & \bm{L}^{\rm bb,\rm ss}  & \bm{L}^{\rm bb,\rm sb}  & \bm{L}^{\rm bb,\rm bb}  \\
\end{pmatrix}\]
$\bm{L}$ consists of 5 block rows and columns, 
referring to linear subnetwork concentrations ($\rm s$), subnetwork products ($\rm ss$), mixed subnetwork-bulk products 
($\rm sb$), which are considered part of the bulk subspace, and products of bulk concentrations ($\rm bb$). The dynamics for the 
products, contained in the bottom right blocks ($\bm{L}^{\rm ss,\rm ss}$, $\bm{L}^{\rm ss,\rm sb}$ etc.), is simply derived from linearized
dynamics, thus
\be
\partial_t(\delta x_i \delta x_j)= \delta x_j \partial_t \delta x_i + \delta x_i \partial_t \delta x_j 
\ee
In other words, in the evolution of a product only products appear: this explains why
the top right block vanishes, i.e.\ applying $\bm{L}$ to quadratic observables does not give linear terms.
All the coefficients in the bottom right blocks are simply ``imported'' from the linearized dynamics.
Genuine nonlinearities enter via the bottom left blocks ($\bm{L}^{\rm ss,s}$, $\bm{L}^{\rm ss,b}$ etc.). These contain the coefficients 
multiplying products in the equations for linear observables. For the projection matrices one has
 \[\bm{P}=\begin{pmatrix}
\mathbb{1}   & \mathbb{0}  & \mathbb{0}  & \mathbb{0}  & \mathbb{0}  \\
\mathbb{0}   & \mathbb{0}  & \mathbb{0}  & \mathbb{0}  & \mathbb{0} \\
\mathbb{0}  & \mathbb{0}  &\mathbb{1}   & \mathbb{0}  & \mathbb{0} \\
\mathbb{0}   & \mathbb{0}  & \mathbb{0}  & \mathbb{0}  & \mathbb{0} \\
\mathbb{0}   & \mathbb{0}  & \mathbb{0}  & \mathbb{0}  & \mathbb{0} \\
\end{pmatrix}\]
while $\bm{Q}$ has the roles of $\mathbb{1}$ and $\mathbb{0}$ along the diagonal interchanged
 \[\bm{Q}=\begin{pmatrix}
\mathbb{0}   & \mathbb{0}  & \mathbb{0}  & \mathbb{0}  & \mathbb{0}  \\
\mathbb{0}   & \mathbb{1}  & \mathbb{0}  & \mathbb{0}  & \mathbb{0} \\
\mathbb{0}  & \mathbb{0}  &\mathbb{0}   & \mathbb{0}  & \mathbb{0} \\
\mathbb{0}   & \mathbb{0}  & \mathbb{0}  & \mathbb{1}  & \mathbb{0} \\
\mathbb{0}   & \mathbb{0}  & \mathbb{0}  & \mathbb{0}  & \mathbb{1} \\
\end{pmatrix}\]
The combinations of matrices one needs, to generalize the formulas of the previous section, are $\bm{PL}$, $\bm{QLQ}$ and $\bm{QL}$, see \cite{katy} for explicit expressions. We emphasize that $\bm{QLQ}$ has a lower triangular block structure, thus 
$\bm{E}(t)=e^{\bm{QLQ}t}$ has the same structure with diagonal 
blocks that are the exponentials of those in $\bm{QLQ}$, i.e.\ for example $[e^{\bm{QLQ}t}]_{\rm bb}=e^{\bm{L}^{\rm bb}t}$. 
Another property that can be deduced from such a structure is
 \begin{subequations}\label{eq:iden1}
 \begin{align}
 \bm{E}_{\rm sb,b}(t)=&\int_0^t dt' \bm{E}_{\rm s b, s b}(t-t')\bm{L}^{ \rm sb, \rm b}\bm{E}_{\rm bb}(t')\\
 \bm{E}_{\rm bb,b}(t)=&\int_0^t dt' \bm{E}_{\rm b b, b b}(t-t')\bm{L}^{\rm bb, \rm b}\bm{E}_{\rm bb}(t')
 \end{align}
 \end{subequations}

The rate constants, the memory function and the random force for the full nonlinear dynamics can now be found in analogy with the linear case provided that we consider
the above enlarged matrices; therefore:
\begin{itemize}
\item The nonlinear rate matrix for the internal subnetwork dynamics is $\bm{\Omega}^{\rm ss,s}=\bm{L}^{\rm ss,s}$
\item The memory is the $\rm ss$ block of $\bm{P}\bm{L}\bm{Q} e^{\bm{Q}\bm{L}\bm{Q} t} \bm{Q}\bm{L}$, whose nonlinear part is
\begin{equation}
 \label{memProj}
    \bm{M}^{\rm ss,s}(t-t')=\bm{L}^{\rm ss,b}\bm{E}_{\rm bb}(t-t')\bm{L}^{\rm bs}+\bm{L}^{\rm ss,\cdot b}\bm{E}_{\cdot \rm b,b}(t-t')
    \bm{L}^{\rm bs} + \bm{L}^{\rm ss,\cdot b} \bm{E}_{\rm \cdot b, \cdot b}(t-t') \bm{L}^{\cdot \rm b, s} 
\end{equation} 
where ``$\cdot \rm b$'' indicates the union of the ``$\rm sb$'' and ``$\rm bb$'' product ranges.
Assumptions about what reactions between the bulk and the subnetwork can occur imply that some $\bm{L}$ blocks are zero, namely:
 $\bm{L}^{\rm ss,\rm bb}=\bm{L}^{\rm bb,\rm ss}=\bm{L}^{\rm ss,b}=\bm{L}^{\rm bb,s}=\mathbb{0}$~\cite{katy}.
 These constraints then simplify the expression for the memory considerably
 \begin{equation}
 \bm{M}^{\rm ss,s}(t-t')=\bm{L}^{\rm ss,\rm sb}\bm{E}_{\rm sb,b}(t-t')\bm{L}^{\rm bs} + \bm{L}^{\rm ss,sb} \bm{E}_{\rm sb,\rm sb}(t-t') 
 \bm{L}^{\rm sb,s}  
 \end{equation}
 We shall nevertheless stick to the most general case, to keep our arguments general. The memory function in the dynamics for 
 $\delta\bm{x}^{\rm s}(t)$ is then embedded within a time integral over the past history of all possible 
 subnetwork products $\delta\bm{x}^{\rm ss}(t')$, thus the general nonlinear memory for the projected equations is 
 \be
 \label{mproj}
 \bm{\mathcal{M}}_{\rm proj}^T(t)= m_1 + m_2 + m_3
 \ee
   with
   \be
   \label{m1}
   m_1=\int_0^t dt' \delta\bm{x}^{\rm ss\,\it{T}}(t')\bm{L}^{\rm ss,b}\bm{E}_{\rm bb}(t-t')\bm{L}^{\rm bs}
   \ee
   \be
   \label{m2}
   m_2=\int_0^t dt' \delta\bm{x}^{\rm ss\,\it{T}}(t')\bm{L}^{\rm ss,\cdot b}\bm{E}_{\cdot \rm b,b}(t-t')\bm{L}^{\rm bs}
   \ee
   \be
   \label{m3}
   m_3=\int_0^t dt' \delta\bm{x}^{\rm ss\,\it{T}}(t')\bm{L}^{\rm ss,\cdot b} \bm{E}_{\rm \cdot b, \cdot b}(t-t') \bm{L}^{\cdot \rm b, s}
   \ee
 \item The random force cannot be calculated  in closed form from the matrix representations introduced, but an expansion can still be provided up to quadratic order in deviations form the steady state: one may hope that higher order terms not captured in this way are small or negligible. We shall write $\bm{r}(t)=\bm{r}_0(t)+\bm{r}_1(t)$,
 $\bm{r}_1(t)$ being the nonlinear random force given by
 \begin{equation} 
 \label{RF}
 \begin{split}  
 \bm{r}_1^{T}(t)= &\,\delta\bm{x}^{\rm sb\,\it{T}}(0)
 \left[\bm{E}_{\rm sb,b}(t)\bm{L}^{\rm bs} + \bm{E}_{\rm sb,\rm sb}(t)\bm{L}^{\rm sb,s}\right]+\delta\bm{x}^{\rm bb\,\it{T}}(0)\left[\bm{E}_{\rm bb,b}(t)\bm{L}^{\rm bs}+\bm{E}_{\rm bb,\rm sb}(t)\bm{L}^{\rm sb,s}\right]
 \end{split}  
 \end{equation}  
\end{itemize}   
From these results, we see that the projection method divides the non-Markovian contribution from the bulk into the random
force, which depends only on initial conditions, and the memory, containing all the single time points in the past for single species and 
products. This structural feature is important for the comparison to the GVA, which does not separate  
the non-local-in-time terms so neatly but still gives an \emph{equivalent} approximation \emph{up to the second order and in the limit
$\epsilon \to 0$}. To provide the tools for this comparison, we first recall the perturbative expansion allowing us 
to derive nonlinear corrections in the Gaussian variational approach (see \ref{gva_full})
and then explicitly derive this equivalence in \ref{comparison}.

\subsection{Full nonlinear memory and random force in the GVA}
\label{gva_full}
As we mentioned, the GVA and the projection method are expected to agree only for $\epsilon \rightarrow 0$: 
we then restrict ourselves to this case, i.e.\ the deterministic dynamics. 
Let us translate the perturbative approach of \ref{pert} to a 
notation analogous to the one in the previous section;
in particular the cubic part of the action \eqref{deltaH} then reads 
 \begin{eqnarray}
 \label{deltaH_new}
&&-\Delta\mathcal{H}= \int_0^{T} dt\bigg\lbrace\bigg[\big(\delta\bm{x}^{\rm s}\circ \delta\bm{x}^{\rm s}\big)^T\bm{L}^{\rm  ss, s}+ 
\big(\delta\bm{x}^{\rm s}\circ\delta\bm{x}^{\rm b}\big)^T\bm{L}^{\rm  sb, s}
+\big(\delta\bm{x}^{\rm b}\circ \delta\bm{x}^{\rm b}\big)^T\bm{L}^{\rm  bb, s}\bigg]\text{i}\hat{\bm{x}}^{\rm s}+\notag\\
&&\qquad\qquad\qquad\bigg[\big(\delta\bm{x}^{\rm s}\circ \delta\bm{x}^{\rm s}\big)^T\bm{L}^{\rm  ss, b}+
\big(\delta\bm{x}^{\rm s}\circ\delta\bm{x}^{\rm b}\big)^T\bm{L}^{\rm  sb, b}+\big(\delta\bm{x}^{\rm b}\circ
\delta\bm{x}^{\rm b}\big)^T\bm{L}^{\rm  bb, b}\bigg]\text{i}\hat{\bm{x}}^{\rm b}\bigg\rbrace
 \end{eqnarray}
 In light of the fact that we allow only certain processes, after \eqref{deltaH} we had invoked
 the simplification $\bm{L}^{\rm ss,\rm b}=\bm{L}^{\rm bb,\rm s}\equiv0$, while here, for the sake of a general comparison,
 we keep all the nonlinear couplings.
The nonlinear memory and effective noise covariance are evaluated by taking the average w.r.t.\ the Gaussian conditional 
distribution over bulk variables, as given by \eqref{dsCompact0}. In the new notation it becomes
\begin{eqnarray}
\label{dsCompact_new}
\fl &&-\langle\Delta \mathcal{H}\rangle=\int_0^{T} dt\big(\delta\bm{x}^{\rm s}(t)\circ\delta\bm{x}^{\rm s}(t)\big)^{T}\bm{L}^{\rm ss, s}
 \text{i}\hat{\bm{x}}^{\rm s}(t)+\\
\fl &&\int_0^{T} dt \bigg\lbrace\bigg[\big(\delta\bm{x}^{\rm s}(t)\circ\delta\bm{\mu}^{\rm b|s}(t)\big)^{T}\bm{L}^{\rm sb, s}+
 \big(\delta\bm{\mu}^{\rm b|s}(t)\circ\delta\bm{\mu}^{\rm b|s}(t)+
 \bm{C}^{\rm bb|s}(t,t)\big)^{T}\bm{L}^{\rm bb, \rm s}\bigg]\text{i}\hat{\bm{x}}^{\rm s}(t)+\notag\\
\fl &&\int_{t}^{T} dt'\bigg[\big(\delta\bm{x}^{\rm s}(t)\circ\delta\bm{\mu}^{\rm b|s}(t)\big)^{T}
 \bm{L}^{\rm sb, \rm b}+ \big(\delta\bm{\mu}^{\rm b|s}(t)\circ\delta\bm{\mu}^{\rm b|s}(t)+\bm{C}^{\rm bb|s}(t,t)\big)^{T}\bm{L}^{\rm bb, \rm b}
 \bigg]\bm{E}_{\rm bb}(t'-t)\bm{L}^{\rm bs}\text{i}\hat{\bm{x}}^{\rm s}(t')\bigg\rbrace\notag
\end{eqnarray}
given that the auxiliary conditional mean, from \eqref{condaux}, can be rewritten as 
\begin{equation} 
\label{condL}
\ii\hat{\bm{\mu}}^{\rm b|s}(t)= \int_{t}^{T} dt' \,\bm{E}_{\rm bb}(t'-t)\bm{L}^{\rm bs}\text{i}\hat{\bm{x}}^{\rm s}(t') 
\end{equation} 
Then one substitutes the expressions for the conditional means \eqref{bulkmean}, which now read
\be
\label{condmea}
\delta\bm{\mu}^{\rm b|s} (t)=\bm{\mu}^{\rm b|s} (t)-\bm{\mu}^{\rm b} (t)=\int_0^t dt' e^{(\bm{L}^{\rm bb})^{T}(t-t')}(\bm{L}^{\rm sb})^{T}\delta\bm{x}^{\rm s}(t')-
\int_0^{T}dt'\bm{C}^{\rm bb|s}(t,t')\bm{L}^{\rm bs}\text{i}\bm{\hat{x}}^{\rm s}(t')
\ee
into \eqref{dsCompact_new} and by isolating the terms multiplying $\text{i}\hat{\bm{x}}^{\rm s}(t)$ one
can read the nonlinear reduced dynamics of $\delta\bm{x}^{\rm s}(t)$.
The memory contribution of the GVA comes from the terms that are quadratic in $\delta\bm{x}^{\rm s}$ and non-local in time
  \begin{eqnarray} 
  \label{mem1}
\fl   \bm{\mathcal{M}}_{\rm GVA}^{T}(t)&=&\int_0^t dt'\bigg\lbrace 
  \big(\delta\bm{x}^{\rm s \, \it{T}}(t')\bm{L}^{\rm sb}\bm{E}_{\rm bb}(t-t') \circ \delta\bm{x}^{\rm s \, \it{T}}(t)\big)
   \bm{L}^{\rm sb, s}+\\
\fl &&\int_{t'}^{t} dt'' \big(\delta\bm{x}^{\rm s \, \it{T}}(t') \bm{L}^{\rm sb}\bm{E}_{\rm bb}(t''-t')\circ
   \delta\bm{x}^{\rm s \, \it{T}}(t'')\big)\bm{L}^{\rm sb, \rm b}\bm{E}_{\rm bb}(t-t'')\bm{L}^{\rm bs} +\notag\\
\fl  &&  \int_0^t dt''\int_{\text{max}(t',t'')}^t ds\, \big(\delta\bm{x}^{\rm s \, \it{T}}(t') \bm{L}^{\rm sb}\bm{E}_{\rm bb}(s-t')\circ
    \delta\bm{x}^{\rm s \, \it{T}}(t'')
    \bm{L}^{\rm sb}\bm{E}_{\rm bb}(s-t'')\big)\bm{L}^{\rm bb, \rm s}+\notag\\
\fl && \int_0^t dt''\int_{\text{max}(t',t'')}^t ds\, \big(\delta\bm{x}^{\rm s \, \it{T}}(t') \bm{L}^{\rm sb}\bm{E}_{\rm bb}(s-t')
    \circ\delta\bm{x}^{\rm s \, \it{T}}(t'')\bm{L}^{\rm sb}\bm{E}_{\rm bb}(s-t'')\big)\bm{L}^{\rm bb, \rm b}
    \bm{E}_{\rm bb}(t-s)\bm{L}^{\rm bs}\bigg\rbrace\notag
   \end{eqnarray}
   This corresponds to the double time integral of \eqref{eq:effMemoryNLtimes} in the modified notation and keeping all the couplings.
As before, by switching to the projection method notation, we are taking the transposes of the memory functions in the main text, 
   equations \eqref{memlin} and \eqref{eq:effMemoryNLtimes}.
   
Similarly from the terms $\sim\bm{\hat{x}}^{\rm s}\delta\bm{x}^{\rm s}\bm{\hat{x}}^{\rm s}$ we can deduce the effective noise covariance
\begin{eqnarray}
\label{cov1}
\fl &&\bm{N}_1^{\rm ss}(t,t'')=\big((\bm{L}^{\rm bs})^{T}\bm{C}^{\rm bb|s}(t,t'')\circ\delta\bm{x}^{\rm s \, \it{T}}(t)\big)\bm{L}^{\rm sb, s}+\int_0^t dt' 
\big((\bm{L}^{\rm bs})^{T}\bm{C}^{\rm bb|s}(t',t'')\circ\delta\bm{x}^{\rm s \, \it{T}}(t')\big)\bm{L}^{\rm sb, \rm b}\bm{E}_{\rm bb}(t-t')\bm{L}^{\rm bs}\notag\\
\fl &&+\int_0^t dt'\bigg[\big((\bm{L}^{\rm bs})^{T}\bm{C}^{\rm bb|s}(t,t'')\circ\delta\bm{x}^{\rm s \, \it{T}}(t') 
\bm{L}^{\rm sb}\bm{E}_{\rm bb}(t-t')\big)+\big(\delta\bm{x}^{\rm s \, \it{T}}(t') \bm{L}^{\rm sb}\bm{E}_{\rm bb}(t-t')\circ(\bm{L}^{\rm bs})^{T}
\bm{C}^{\rm bb|s}(t,t'')\big)\bigg]\bm{L}^{\rm bb, \rm s}\notag\\
\fl &&+\int_0^t ds\int_0^{s}dt'\bigg[\big((\bm{L}^{\rm bs})^{T}\bm{C}^{\rm bb|s}(s,t'')\circ\delta\bm{x}^{\rm s \, \it{T}}(t')
\bm{L}^{\rm sb}\bm{E}_{\rm bb}(s-t')\big)+\\
\fl &&\qquad \qquad\qquad\big(\delta\bm{x}^{\rm s \, \it{T}}(t') \bm{L}^{\rm sb}\bm{E}_{\rm bb}(s-t')\circ(\bm{L}^{\rm bs})^{T}
\bm{C}^{\rm bb|s}(s,t'')\big) \bigg]\bm{L}^{\rm bb, \rm b}
\bm{E}_{\rm bb}(t-s)\bm{L}^{\rm bs}+\ldots^T\notag
\end{eqnarray}
which generalizes \eqref{covNL}. For $\epsilon \to 0$, the bulk conditional correlator is 
simply
   \begin{equation}
   \label{simple_corr}
   \bm{C}^{\rm bb|s}(t',t'')= \bm{E}_{\rm bb}^T(t')\bm{C}^{\rm bb}(0,0)\bm{E}_{\rm bb}(t'')
   \end{equation}
as can be deduced from \eqref{bulkcov}.

Let us stress that projection methods expand the random force, while our perturbative approach expands the correlator of a coloured noise.
 For the sake of comparison, it is therefore convenient to think in terms of the effective coloured noise with covariance
 $\bm{N}_0^{\rm ss}+\bm{N}_1^{\rm ss}$. $\bm{N}_1^{\rm ss}$ can be read off from \eqref{cov1}, while $\bm{N}_0^{\rm ss}$, 
   the effective noise covariance of the linearized dynamics, is given by \eqref{EffNoi}; we rewrite it as
\begin{equation}
\label{cov2}
 \bm{N}_0^{\rm ss}(t,t')=(\bm{L}^{\rm bs})^T\bm{E}_{\rm bb}^T(t)
 \bm{C}^{\rm bb}(0,0)\bm{E}_{\rm bb}(t')\bm{L}^{\rm bs}
\end{equation}
Note that we have dropped the intrinsic noise contribution $\bm{\Sigma}^{\rm ss}$, $\bm{\Sigma}^{\rm bb}$
as we focus on $\epsilon\to 0$.
The Gaussian noise $\bm{\chi}_0(t)$ such that $\langle\bm{\chi}_0(t)\bm{\chi}_0^{\it{T}}(t')\rangle=\bm{N}_0^{\rm ss}(t,t')$ is therefore
\be
\label{chi_0pr}
\bm{\chi}_0^{\it{T}}(t) = \delta\bm{x}^{\rm b\, \it{T}}(0)\bm{E}_{\rm bb}(t)\bm{L}^{\rm bs}
\ee
If we define $\bm{\chi}_1(t)$ as follows 
   \begin{eqnarray}    
   \label{chi1}    
    \fl &\bm{\chi}_1^{T}(t)= \,\big(\delta\bm{x}^{\rm b \, \it{T}}(0)\bm{E}^{\rm bb}(t)\circ\delta\bm{x}^{\rm s \, \it{T}}(t)\big)\bm{L}^{\rm sb,s}+
 \int_0^t dt' \big(\delta\bm{x}^{\rm b \, \it{T}}(0)\bm{E}_{\rm bb}(t') \circ\delta\bm{x}^{\rm s \, \it{T}}(t') \big)\bm{L}^{\rm sb,b} 
 \bm{E}_{\rm bb}(t-t')\bm{L}^{\rm bs}+\notag\\
\fl& \int_0^{t} dt'\bigg[\big(\delta\bm{x}^{\rm b \, \it{T}}(0)\bm{E}_{\rm bb}(t) \circ
\delta\bm{x}^{\rm s \, \it{T}}(t') \bm{L}^{\rm sb}
 \bm{E}_{\rm bb}(t-t')\big) +\big(\delta \bm{x}^{\rm s \, \it{T}}(t')\bm{L}^{\rm sb}\bm{E}_{\rm bb}(t-t')\circ
 \delta\bm{x}^{\rm b \, \it{T}}(0) \bm{E}_{\rm bb}(t)\big)\bigg]\bm{L}^{\rm bb,s} \notag\\
\fl &+\int_0^t ds\int_0^{s} dt'\bigg[\big(\delta\bm{x}^{\rm b \, \it{T}}(0)\bm{E}_{\rm bb}(s) \circ
 \delta\bm{x}^{\rm s \, \it{T}}(t') \bm{L}^{\rm sb}
 \bm{E}_{\rm bb}(s-t')\big)+\notag\\
\fl &\qquad \qquad \qquad\big(\delta\bm{x}^{\rm s \, \it{T}}(t')\bm{L}^{\rm sb}\bm{E}_{\rm bb}(s-t')
 \circ\delta\bm{x}^{\rm b \,\it{T}}(0)\bm{E}_{\rm bb}(s)\big)\bigg]\bm{L}^{\rm bb,b} \bm{E}_{\rm bb}(t-s)\bm{L}^{\rm bs}
 \end{eqnarray}   
then the correlation function of $\bm{\chi}_0+\bm{\chi}_1$ 
is given by $\bm{N}_0^{\rm ss}+\bm{N}_1^{\rm ss}$ at the linear order in
$\delta\bm{x}^{\rm s}$ (this is the order in $\delta\bm{x}^{\rm s}$ to which $\bm{N}_1^{\rm ss}$ is calculated).
   For example, let us take the first term in \eqref{cov1}: it is derived 
from cross-correlations of $\bm{\chi}_0$ \eqref{chi_0pr} and the first term of $\bm{\chi}_1$ \eqref{chi1} as follows
\be
\bigg\langle \big((\bm{L}^{\rm bs})^{T}\bm{E}_{\rm bb}^{T}(t)\delta\bm{x}^{\rm b}(0)
\delta\bm{x}^{\rm b \, \it{T}}(0)\bm{E}_{\rm bb}(t'')\circ\delta\bm{x}^{\rm s \, \it{T}}(t'')\big)\bm{L}^{\rm sb,s}\bigg\rangle
= \big((\bm{L}^{\rm bs})^{T}\bm{C}^{\rm bb|s}(t,t'')\circ\delta\bm{x}^{\rm s \, \it{T}}(t'')\big)\bm{L}^{\rm sb, s}
\ee
 by using the definition \eqref{simple_corr}. 
 
From expression \eqref{dsCompact_new} (see also \eqref{dsCompact}), 
it can be seen that the Gaussian effective dynamics of $\delta\bm{x}^{\rm s}(t)$ exhibits also the $\bm{x}^{\rm s}$-independent term
\begin{equation}
 \label{addTerm}
\bm{C}^{\rm bb|s}(t,t)\bm{L}^{\rm bb, \rm s}+\int_0^t dt' \bm{C}^{\rm bb|s}(t',t')\bm{L}^{\rm bb, \rm b}\bm{E}_{\rm bb}(t-t')\bm{L}^{\rm bs}=\langle\bm{\psi}_1^{T}(t)\rangle
\end{equation}
which can be written as the conditional average of a vector $\bm{\psi}_1(t)$ defined as follows
\be
\label{psi1}
\begin{split}
\bm{\psi}_1^{T}(t) = &\big(\delta\bm{x}^{\rm b \, \it{T}}(0)\bm{E}_{\rm bb}(t)\circ\,\delta\bm{x}^{\rm b \, \it{T}}(0)\bm{E}_{\rm bb}(t)\big)
\bm{L}^{\rm bb,s}+\\
&\int_0^t dt'  \big(\delta\bm{x}^{\rm b \, \it{T}}(0)\bm{E}_{\rm bb}(t')\circ\,\delta\bm{x}^{\rm b \, \it{T}}(0)\bm{E}_{\rm bb}(t')\big)
\bm{L}^{\rm bb,b}\bm{E}_{\rm bb}(t-t')\bm{L}^{\rm bs}
\end{split}
\ee
The conditional average of $\bm{\psi}_1$ produces in the action a cubic term, $\sim \hat{\bm{x}}^{s}\delta\bm{x}^{\rm b} \delta\bm{x}^{\rm b}$, 
while its variance would appear in the action via a term of $6$th order, thus it is not present at the order we have kept in our calculation. 
From \eqref{psi1} we see that $\bm{\psi}_1(t)$ is a temporally correlated term in the reduced
subnetwork dynamics that depends quadratically on 
$\delta\bm{x}^{\rm b}(0)$; it can be regarded as a further contribution to the nonlinear ``random force'' of the GVA, as by definition 
the random force summarizes the uncertainty on the bulk initial conditions. 
As a result, the systematic perturbative expansion up to the cubic order in the action gives it as
\be
\bm{\tilde{r}}_{1\,\rm GVA}(t)= \bm{\chi}_1(t)+ \langle\bm{\psi}_1(t)\rangle
\ee
If we consider the fluctuating version of this quantity, namely
\be
\label{rfgva}
\bm{r}_{1\,\rm GVA}(t)= \bm{\chi}_1(t)+ \bm{\psi}_1(t)
\ee
we can prove the equivalence with the projection formalism.

\subsection{Proof of the equivalence}  
\label{comparison}
Let us assume that the initial deviations from the means are proportional to some factor $\delta$, 
$\delta \bm{x}^{\rm s}(0)\sim \delta$ and $\delta \bm{x}^{\rm b}(0)\sim \delta$. The solution 
$\delta \bm{x}^{\rm s}(t)$ can be expanded in powers of $\delta$ and we want to find out to what extent 
projection and GVA give the same results for this expansion in the limit $\epsilon \to 0$. We expect that the two descriptions are
equivalent at $\mathcal{O}(\delta^2)$ as in the GVA we keep cubic terms in the effective action 
while in projection methods we keep all quadratic observables.
\subsubsection{Linear order in $\delta$.}
To work out the solution of $\delta \bm{x}^{\rm s}(t)$ in this case, we need to include 
in its dynamics the expressions for memory and random force at $\mathcal{O}(\delta)$. 
By re-writing the linear memory \eqref{memlin} of the GVA 
in the notation of \ref{nonlinproj} one has
\be
 \label{memlinpr}
\bm{M}^{\rm ss}(t-t') = \bm{L}^{\rm sb} \bm{E}_{\rm bb}(t-t') \bm{L}^{\rm bs} 
\ee
where we have used $e^{\bm{L}^{\rm bb}t}=\bm{E}_{\rm bb}(t)$. The effective coloured noise at this order
for the GVA is given by \eqref{chi_0pr}.
Expressions \eqref{memlinpr} and \eqref{chi_0pr}
are the same as from projection methods \eqref{memlin0} and \eqref{rf0}, thus the resulting expansion of $\delta \bm{x}^{\rm s}(t)$ is
identical to $\mathcal{O}(\delta)$.

\subsubsection{Quadratic order in $\delta$.}
In this case, we need to include in the
dynamics of $\delta \bm{x}^{\rm s}(t)$ the memory and the effective noise/random force at $\mathcal{O}(\delta^2)$.
This includes linear memory terms acting on the second order part of $\delta \bm{x}^{\rm s}(t)$, 
which will be the same for projection and GVA. The remaining terms are the 
nonlinear memory and nonlinear effective noise/random force evaluated to order $\mathcal{O}(\delta^2)$.

Expressions for nonlinear memory and random force from the two approaches (respectively \eqref{mproj} and
\eqref{mem1}, \eqref{RF} and \eqref{rfgva}) do not coincide if taken 
\emph{separately}; we nevertheless expect that their \emph{combination} is actually equivalent (because
there can be different ways of writing expansions that are correct at the same order, namely $\mathcal{O}(\delta^2)$).
   
Before verifying this expectation, we briefly outline the reasoning. First of all, the memory function \eqref{mem1} differs from the one in the projection methods \eqref{mproj}:
in particular, the two subnetwork species in quadratic terms are calculated at two past times instead of the same time;
additional integrals over time also appear. One can however eliminate the intermediate times, and thus match the one-time and two-times structures in the algebra,
by explicitly expressing everywhere the dependence on subnetwork initial values.
Next we observe that in the structure of the enlarged (i.e.\ including quadratic observables) $\bm{L}$, the non-linearity of interactions 
   enters genuinely only in the bottom left off-diagonal block. The bottom right (diagonal) block, 
   which represents the coupling between products, does not provide any additional information with respect to the linear dynamics: it simply
   describes how 
   the products propagate under the linear dynamics. 
   Note that this is precisely what the perturbative expansion implements, i.e.\ it describes products
   under the linear evolution. Therefore, we can drop the purely non-linear block and write also the products of subnetwork 
   variables and 
   conditional means of the perturbative expansion as functions of the initial conditions in the subnetwork via the exponential of a 
   matrix consisting of a linear and a quadratic block.
   
   We need to substitute into the nonlinear memory and random force the first order 
   deterministic solution for $\delta \bm{x}^{\rm s}(t)$ and  $\delta \bm{x}^{\rm b}(t)$ to evaluate them consistently to 
   $\mathcal{O}(\delta^2)$: let us perform the substitution in both approaches and compare the results. The projection memory is evaluated at $\mathcal{O}(\delta^2)$ using the expression
\begin{equation}
\label{product}
\delta \bm{x}^{\cdot \cdot\,\it{T}}(t) = \delta \bm{x}^{\cdot \cdot\,\it{T}}(0) e^{\bm{L}^{\cdot \cdot, \cdot \cdot }t} 
\end{equation}
where ``$\cdot \cdot$'' refers to the block obtained by joining the ``ss'',``sb'' and ``bb'' ranges, here
and everywhere below.
Also in the GVA we have to write the products as in \eqref{product} but with the following
caveat. The GVA is obtained via the inclusion of the dynamics of conditional \emph{averages}, where the average is taken also 
over bulk initial conditions. This, importantly, implies that $\delta \bm{x}^{\rm b}(0)$ should not be treated 
as a fluctuating quantity as happens in \eqref{product}.
More explicitly, we recall that the conditional bulk variables can be written 
as $\delta\bm{\mu}^{\rm b|s \,\it{T}}(t)= \bm{\p}^{\it{T}}(t)+\bm{\hat{\p}}^{\it{T}}(t)$,
where only the deterministic part $\bm{\p}(t)$ contributes to the memory
\begin{equation}
\label{detpart}
\bm{\p}^{\it{T}}(t)=\int_0^t dt'\delta\bm{x}^{\rm s \, \it{T}}(t')\bm{L}^{\rm sb}\,\bm{E}_{\rm bb}(t-t')
\end{equation}
(from expression \eqref{bulkmean}). On the other hand, the evolution of the products introduced by \eqref{product} 
accounts for a solution for bulk variables as follows
\begin{equation}
\label{addition}
 \delta\bm{x}^{\rm b \, \it{T}}(t) =  \delta\bm{x}^{\rm b \, \it{T}}(0)\bm{E}_{\rm bb}(t) +
 \int_0^t dt' \delta\bm{x}^{\rm s \, \it{T}}(t')\bm{L}^{\rm sb}\,\bm{E}_{\rm bb}(t-t')
\end{equation}
with also deviations in the initial conditions $\delta\bm{x}^{\rm b}(0)$.
  The nonlinear memory for the variational approach \eqref{mem1} can be written as
\begin{eqnarray}
\label{memvar}
\bm{\mathcal{M}}_{\rm GVA}^{T}(t)&=&\int_0^t dt'\bigg\langle\delta\bm{x}^{\cdot \cdot\,\it{T}}(0) e^{\bm{L}^{\cdot \cdot, \cdot \cdot }t'}
\bigg\rangle_{\cdot \cdot}
\,\bm{L}^{\cdot \cdot, \rm b}\bm{E}_{\rm bb}(t-t')\bm{L}^{\rm bs} + 
 \bigg\langle\delta\bm{x}^{\cdot \cdot\,\it{T}}(0) e^{\bm{L}^{\cdot \cdot, \cdot \cdot }t}\bigg\rangle_{\cdot \rm b} \bm{L}^{\cdot \rm b, s}=\notag\\
&=& \int_0^t dt'\big(\delta\bm{x}^{\cdot \cdot\,\it{T}}(0) e^{\bm{L}^{\cdot \cdot, \cdot \cdot }t'}\big)_{\cdot \cdot}\,\bm{L}^{\cdot \cdot, \rm b}
\bm{E}_{\rm bb}(t-t')\bm{L}^{\rm bs} + 
 \big(\delta\bm{x}^{\cdot \cdot\,\it{T}}(0) e^{\bm{L}^{\cdot \cdot, \cdot \cdot }t}\big)_{\cdot \rm b}\bm{L}^{\cdot \rm b, s} 
 -m'_4 =\notag\\
&=&m'_1+m'_2+m'_3 - m'_4
\end{eqnarray}
We have 
\begin{eqnarray}
m'_1&=&\int_0^t dt'(\delta\bm{x}^{\cdot \cdot\,\it{T}}(0) e^{\bm{L}^{\cdot \cdot, \cdot \cdot }t'})_{\rm ss} \bm{L}^{\rm ss, \rm b}\bm{E}_{\rm bb}
(t-t')\bm{L}^{\rm bs} \label{m1p}\\
m'_2&=&\int_0^t dt'(\delta\bm{x}^{\cdot \cdot\,\it{T}}(0) e^{\bm{L}^{\cdot \cdot, \cdot \cdot }t'})_{\cdot \rm b} \bm{L}^{\cdot \rm b, \rm b}
\bm{E}_{\rm bb}(t-t')\bm{L}^{\rm bs} \label{m2p}\\
m'_3&=&(\delta\bm{x}^{\cdot \cdot\,\it{T}}(0) e^{\bm{L}^{\cdot \cdot, \cdot \cdot }t})_{\cdot \rm b} \bm{L}^{\cdot \rm b, s} \label{m3p}
\end{eqnarray}
$m'_4$ has minus sign as it is a correction term needed for the substitution in the second line of \eqref{memvar} to be valid
(in other words, to compensate for the use of \eqref{addition} instead of \eqref{detpart}); namely
\begin{eqnarray}
\label{correction}
\fl &m'_4 = \,\big(\delta\bm{x}^{\rm b \, \it{T}}(0)\bm{E}^{\rm bb}(t)\circ\delta\bm{x}^{\rm s \, \it{T}}(t)\big)\bm{L}^{\rm sb,s}+
 \int_0^t dt' \big(\delta\bm{x}^{\rm b \, \it{T}}(0)\bm{E}_{\rm bb}(t') \circ\delta\bm{x}^{\rm s \, \it{T}}(t') \big)\bm{L}^{\rm sb,b} 
 \bm{E}_{\rm bb}(t-t')\bm{L}^{\rm bs}+\notag\\
\fl& \int_0^{t} dt'\bigg[\big(\delta\bm{x}^{\rm b \, \it{T}}(0)\bm{E}_{\rm bb}(t) \circ
\delta\bm{x}^{\rm s \, \it{T}}(t') \bm{L}^{\rm sb}
 \bm{E}_{\rm bb}(t-t')\big) +\big(\delta \bm{x}^{\rm s \, \it{T}}(t')\bm{L}^{\rm sb}\bm{E}_{\rm bb}(t-t')\circ
 \delta\bm{x}^{\rm b \, \it{T}}(0) \bm{E}_{\rm bb}(t)\big)\bigg]\bm{L}^{\rm bb,s} +\notag\\
 \fl &\big(\delta\bm{x}^{\rm b \, \it{T}}(0)\bm{E}_{\rm bb}(t)\circ\delta\bm{x}^{\rm b \, \it{T}}(0)
 \bm{E}_{\rm bb}(t)\big)\bm{L}^{\rm bb,s}+\int_0^t dt'  \big(\delta\bm{x}^{\rm b \, \it{T}}(0)\bm{E}_{\rm bb}(t')\circ\,\delta\bm{x}^{\rm b \, \it{T}}(0)\bm{E}_{\rm bb}(t')\big)
\bm{L}^{\rm bb,b}\bm{E}_{\rm bb}(t-t')\bm{L}^{\rm bs}\notag\\ 
\fl &+\int_0^t ds\int_0^{s} dt'\bigg[\big(\delta\bm{x}^{\rm b \, \it{T}}(0)\bm{E}_{\rm bb}(s) \circ
 \delta\bm{x}^{\rm s \, \it{T}}(t') \bm{L}^{\rm sb}
 \bm{E}_{\rm bb}(s-t')\big)+\notag\\
\fl &\qquad \qquad \qquad\big(\delta\bm{x}^{\rm s \, \it{T}}(t')\bm{L}^{\rm sb}\bm{E}_{\rm bb}(s-t')
 \circ\delta\bm{x}^{\rm b \,\it{T}}(0)\bm{E}_{\rm bb}(s)\big)\bigg]\bm{L}^{\rm bb,b} \bm{E}_{\rm bb}(t-s)\bm{L}^{\rm bs}
\end{eqnarray}
Recalling equations \eqref{chi1}, \eqref{psi1} and \eqref{rfgva}, one can write
\be
\label{m4}
m'_4 = \bm{r}_{1\,\rm GVA}^T(t)
\ee
From \eqref{memvar} and \eqref{m4} we have
\be
\label{mrvar}
m'_1 + m'_2 + m'_3=\bm{\mathcal{M}}_{\rm GVA}^T(t)+ \bm{r}_{1\,\rm GVA}^T(t)
\ee

Let us now compare term by term the nonlinear memory from
projection methods \eqref{mproj} and from the GVA \eqref{memvar}: one has immediately $m_1=m'_1$ (see \eqref{m1} and \eqref{m1p}). 
We apply next the identity
\begin{equation}
\label{iden}
(\delta\bm{x}^{\cdot \cdot\,\it{T}}(0) e^{\bm{L}^{\cdot \cdot, \cdot \cdot }t'})_{\rm ss} \bm{L}^{\rm ss,\cdot b}=
 \partial_{t'}(\delta\bm{x}^{\cdot \cdot\,\it{T}}(0) e^{\bm{L}^{\cdot \cdot, \cdot \cdot }t'})_{\cdot \rm b}-(\delta\bm{x}^{\cdot \cdot\,\it{T}}(0) e^{\bm{L}^{\cdot \cdot, \cdot \cdot }t'})_{\cdot \rm b}\bm{L}^{\cdot \rm b, \cdot \rm b}   
\end{equation}
to manipulate the sum of $m_2$ \eqref{m2} and $m_3$ \eqref{m3}, as follows
\begin{equation}
\label{m2m3}
\begin{split}
&m_2 + m_3= \int_0^t dt' (\delta\bm{x}^{\cdot \cdot\,\it{T}}(0) e^{\bm{L}^{\cdot \cdot, \cdot \cdot }t'})_{\rm ss}\bm{L}^{\rm ss,\cdot b}
[\bm{E}_{\cdot \rm b,b}(t-t')\bm{L}^{\rm bs} + \bm{E}_{\rm \cdot b, \cdot b}(t-t') \bm{L}^{\cdot \rm b, s}]=\\
&= \int_0^t dt'\big[\partial_{t'}(\delta\bm{x}^{\cdot \cdot\,\it{T}}(0) e^{\bm{L}^{\cdot \cdot, \cdot \cdot }t'})_{\cdot \rm b}-
(\delta\bm{x}^{\cdot \cdot\,\it{T}}(0) e^{\bm{L}^{\cdot \cdot, \cdot \cdot }t'})_{\cdot \rm b}\bm{L}^{\cdot \rm b, \cdot \rm b}\big]
\big[\bm{E}_{\cdot \rm b,b}(t-t')\bm{L}^{\rm bs}+\bm{E}_{\rm \cdot b, \cdot b}(t-t') \bm{L}^{\cdot \rm b, s}\big]=\\
&= - \int_0^t dt'(\delta\bm{x}^{\cdot \cdot\,\it{T}}(0) e^{\bm{L}^{\cdot \cdot, \cdot \cdot }t'})_{\cdot \rm b}\partial_{t'}[\bm{E}_{\cdot \rm b,b}
(t-t')\bm{L}^{\rm bs}+\bm{E}_{\rm \cdot b, \cdot b}(t-t')\bm{L}^{\cdot \rm b, s}\big]\\
&\quad+(\delta\bm{x}^{\cdot \cdot\,\it{T}}(0) e^{\bm{L}^{\cdot \cdot, \cdot \cdot }t'})_{\cdot \rm b}[\bm{E}_{\cdot \rm b,b}(t-t')\bm{L}^{\rm bs}+
\bm{E}_{\rm \cdot b, \cdot b}(t-t')\bm{L}^{\cdot \rm b, s}]\bigg|_0^{t}\\
&\quad -\int_0^t dt'(\delta\bm{x}^{\cdot \cdot\,\it{T}}(0) e^{\bm{L}^{\cdot \cdot, \cdot \cdot }t'})_{\cdot \rm b}\bm{L}^{\cdot \rm b, \cdot \rm b}
[\bm{E}_{\cdot \rm b,b}(t-t')\bm{L}^{\rm bs}+\bm{E}_{\rm \cdot b, \cdot b}(t-t')\bm{L}^{\cdot \rm b, s}]
\end{split}
\end{equation}
In the third line, an integration by parts has been implemented and $-\partial_{t'}$ can be further substituted by $\partial_{t}$ as it acts 
on the difference $(t-t')$. Let us the consider the boundary term from this integration (fourth line)
\begin{equation}
\begin{split}
(\delta\bm{x}^{\cdot \cdot\,\it{T}}(0) e^{\bm{L}^{\cdot \cdot, \cdot \cdot }t})_{\cdot \rm b} \bm{L}^{\cdot \rm b, s} &- 
\delta\bm{x}^{\cdot \rm b\,\it{T}}(0)\big(\bm{E}_{\cdot \rm b,b}(t)\bm{L}^{\rm bs}+\bm{E}_{\rm \cdot b, \cdot b}(t)
\bm{L}^{\cdot \rm b, s}\big)\\
&=m'_3 - \delta\bm{x}^{\cdot \rm b\,\it{T}}(0)\big(\bm{E}_{\cdot \rm b,b}(t)\bm{L}^{\rm bs}+\bm{E}_{\rm \cdot b, \cdot b}(t)\bm{L}^{\cdot \rm b, s}\big)
\end{split}
\end{equation}
where we used \eqref{m3p}, 
$\bm{E}_{\rm sb,b}(0)=\bm{E}_{\rm bb, \rm b}(0)= \mathbb{0}$,
$\bm{E}_{\rm sb,\rm sb}(0)=\bm{E}_{\rm bb,\rm bb}(0)=\mathbb{1}$
and the fact that all the terms in $e^{\bm{L}^{\cdot \cdot, \cdot \cdot }t'}$ at $t'=0$ give $\mathbb{1}$. Equation \eqref{m2m3} can be rewritten
\begin{equation}
\label{m2m3n}
\begin{split}
m_2 + m_3 =&\, m'_3 - \delta\bm{x}^{\cdot \rm b\,\it{T}}(0)(\bm{E}_{\cdot \rm b,b}(t)\bm{L}^{\rm bs}+\bm{E}_{\rm \cdot b, \cdot b}(t)\bm{L}^{\cdot \rm b, s})\\
& + \int dt'(\delta\bm{x}^{\cdot \cdot\,\it{T}}(0) e^{\bm{L}^{\cdot \cdot, \cdot \cdot }t'})_{\cdot \rm b}\partial_{t}[\bm{E}_{\cdot \rm b,b}(t-t')
\bm{L}^{\rm bs}+\bm{E}_{\rm \cdot b, \cdot b}(t-t')\bm{L}^{\cdot \rm b, s}]\\
& - \int dt'(\delta\bm{x}^{\cdot \cdot\,\it{T}}(0) e^{\bm{L}^{\cdot \cdot, \cdot \cdot }t'})_{\cdot \rm b}\bm{L}^{\cdot \rm b, \cdot \rm b}
[\bm{E}_{\cdot \rm b,b}(t-t')\bm{L}^{\rm bs}+\bm{E}_{\rm \cdot b, \cdot b}(t-t')\bm{L}^{\cdot \rm b, s}]
\end{split}
\end{equation}
In the second line of \eqref{m2m3n}, we can apply the identity
\begin{equation}
\partial_{t}\bm{E}_{\cdot \rm b, \rm b}(t-t')= \bm{L}^{\cdot \rm b, \cdot \rm b}\bm{E}_{\cdot \rm b, \rm b}(t-t')+ \bm{L}^{\cdot \rm b, \rm b}\bm{E}_{\rm bb}(t-t') 
\end{equation}
which can be deduced from the properties \eqref{eq:iden1} and from 
\begin{equation}
\partial_{t}\bm{E}_{\rm \cdot b, \cdot b}(t-t')= \bm{L}^{\cdot \rm b, \cdot \rm b}\bm{E}_{\rm \cdot b, \cdot b}(t-t') 
\end{equation}
As a consequence, also by recalling \eqref{m2p}, we obtain
\begin{equation}
m_2 + m_3 = m'_3 - \delta\bm{x}^{\cdot \rm b\,\it{T}}(0)(\bm{E}_{\cdot \rm b,b}(t)\bm{L}^{\rm bs}+\bm{E}_{\rm \cdot b, \cdot b}(t)\bm{L}^{\cdot \rm b, s})+m'_2
\end{equation}
In light of \eqref{mproj} and \eqref{mrvar} we thus have that, symbolically,
\begin{equation}
 \bm{\mathcal{M}}_{\rm GVA}^T(t)+\bm{r}_{1\,\rm GVA}^T(t)= 
 \bm{\mathcal{M}}_{\rm proj}^T(t) + \delta\bm{x}^{\cdot \rm b\,\it{T}}(0)(\bm{E}_{\cdot \rm b,b}(t)\bm{L}^{\rm bs}+
 \bm{E}_{\rm \cdot b, \cdot b}(t)\bm{L}^{\cdot \rm b, s})
\end{equation}
Here the second term on the r.h.s.\  is exactly the nonlinear part of the projection random force \eqref{RF}, which we here denote 
$\bm{r}_{1\,\rm proj}^T(t)$. Finally one obtains the equivalence at $\mathcal{O}(\delta^2)$ we aimed to prove, i.e.
 \begin{equation}
  \bm{\mathcal{M}}_{\rm GVA}(t)+ \bm{r}_{1\,\rm GVA}(t) = \bm{\mathcal{M}}_{\rm proj}(t)+ \bm{r}_{1\,\rm proj}(t)
 \end{equation}
 The comparison becomes even simpler if the bulk is assumed initially at steady state, i.e.\ $\delta\bm{x}^{\rm b}(0) \equiv 0$, as all the random force terms then vanish and 
 $\bm{\mathcal{M}}_{\rm GVA}(t) = \bm{\mathcal{M}}_{\rm proj}(t)$.

\end{document}